# Mining user reviews of COVID contact-tracing apps: An exploratory analysis of nine European apps

| Vahid Garousi<br>Queen's University Belfast, UK<br>Bahar Software Engineering Consulting Corporation, UK<br>v.garousi@qub.ac.uk | David Cutting<br>Queen's University Belfast, UK<br>d.cutting@qub.ac.uk | Michael Felderer<br>University of Innsbruck, Austria<br>Blekinge Institute of Technology, Sweden<br>michael.felderer@uibk.ac.at |
|---|---|---|

**Abstract:**

*Context:* More than 50 countries have developed COVID contact-tracing apps to limit the spread of coronavirus. However, many experts and scientists cast doubt on the effectiveness of those apps. For each app, a large number of reviews have been entered by end-users in app stores.

*Objective:* Our goal is to gain insights into the user reviews of those apps, and to find out the main problems that users have reported. Our focus is to assess the "software in society" aspects of the apps, based on user reviews.

*Method:* We selected nine European national apps for our analysis and used a commercial app-review analytics tool to extract and mine the user reviews. For all the apps combined, our dataset includes 39,425 user reviews.

*Results:* Results show that users are generally dissatisfied with the nine apps under study, except the Scottish ("Protect Scotland") app. Some of the major issues that users have complained about are high battery drainage and doubts on whether apps are really working.

*Conclusion:* Our results show that more work is needed by the stakeholders behind the apps (e.g., app developers, decision-makers, public health experts) to improve the public adoption, software quality and public perception of these apps.

## TABLE OF CONTENTS

## 1 Introduction

As of October 2020, more than 50 countries and regions have developed so far (or are developing) COVID contact-tracing apps to limit the spread of coronavirus[1]. The list is quickly growing, and as of this writing, 19 of those apps are open source.

Contact-tracing apps generally use Bluetooth signals to log when smartphones, and hence their owners, are close to each other, so if someone develops COVID symptoms or tests positive, an alert can be sent to other users they may have infected. An app can be developed using two different approaches: centralized or decentralized. Under the centralized model, the data gathered is uploaded to a remote server where matches are made with other contacts should a person start to develop COVID symptoms. This is the method that countries such as the UK were initially pursuing.

By contrast, the decentralized model gives users more control over their information by keeping it on the phone. It is there that matches are made with people who may have contracted the virus. This is the model promoted by Google, Apple and an international consortium, advised in part by the MIT-led Private Automated Contact Tracing (PACT) project (pact.mit.edu) [1]. Both types have their pros and cons. Since early summer 2020, a split emerged between the two approaches. However, privacy and platform support issues have pushed countries to use the decentralized model.

The apps have been promoted as a promising tool to help bring the COVID outbreak under control. However, there are many discussions in the media, the academic (peer-reviewed) literature [2], and also the grey literature about the 'efficacy of contact-tracing apps' (try a Google search for the term inside quotes). A systematic review [3] of 15 studies, which had studied the efficacy of contact-tracing apps, found that "*there is relatively limited evidence for the impact of contact-tracing apps*". A French news article[2] reported that, as of mid-August 2020, "*StopCovid [the French app] had over 2.3 million downloads [of a population of 67 million people] and only 72 notifications were sent [by the app]*".

One cannot help but wonder the reasons behind low efficacy and low adoption of the apps by the general public in many countries. The issue is a multi-faceted, complex, and interdisciplinary issue, as it relates to fields such as public health, behavioral science [4], epidemiology, and software engineering.

The software engineering aspect of contact-tracing apps is quite diverse in itself, e.g., whether different apps developed by different countries will cooperate/integrate (when people travel across counties/borders), and whether the app software would work as intended (e.g., will it record the nearby phone IDs properly, and will it send the alerts to all the recorded persons?). The decentralized nature of the system makes such a verification a challenging task. Some other related developments include a news article reporting that a large number of developers worldwide have found a large number of defects in one of the contract-tracing apps (England's open-source app)[3].

Another software engineering angle of the issue is the availability of a high number of user reviews in the two major app stores: the Google Play Store for Android apps and the Apple App Store for the iOS apps. A user review often contains information about the user's experience with the app and opinion of it, feature requests, or bug reports [5]. Many insights can be mined by analyzing the user reviews of these apps to figure out what end-users think of COVID contact-tracing apps, and that is what we analyze and present in this paper. Studies have shown that reviews written by the users represent a rich source of information for the app vendors and the developers, as they include information about bugs and ideas for new features [6]. Mining of app store data and app reviews has become an active area of research in software engineering [5] to extract valuable insights. User ratings and reviews are user-driven feedback that may help improve software quality and address missing application features.

---

[1] www.xda-developers.com/google-apple-covid-19-contact-tracing-exposure-notifications-api-app-list-countries/
[2] www.lefigaro.fr/secteur/high-tech/stopcovid-2-3-millions-de-telechargements-et-seulement-72-notifications-envoyees-20200819
[3] eandt.theiet.org/content/articles/2020/05/developers-find-new-flaws-in-source-code-of-nhs-contract-tracing-app/

Among the insights that we aim at deriving in this study are the ratios of users which, as per their reviews, have been happy or unhappy with the contact-tracing apps, and the main issues (problems) that most users have reported about the apps. The nature of our analysis is "exploratory" [7] in nature, as we want to explore the app reviews and extract insights from them which could be useful for the different stakeholders, e.g., app developers, decision-makers, researchers, and the public, to benefit from or act upon.

Also, the focus of our paper is software engineering "in society" [8], since it is clear that contact-tracing apps are widely discussed in the public media and are used by millions of people worldwide, and also have the potential to have major influences on people's lives in the challenges that the COVID pandemics has brought upon all the people of the world. Furthermore, many resources have argued that "*these apps are safety-critical*"[1], since "*a faulty proximity tracing app could lead to false positives, false negatives, or maybe both*". It is thus very important that these apps and their user reviews be carefully studied to ensure that upcoming updates of existing apps or new similar apps have the highest software quality.

Another motivating factor for this study is ongoing research and consulting engagement of the first author in relation to the Northern Irish contact-tracing app (called "StopCOVID NI"[2]). Since May 2020, he has been a member of an Expert Advisory Committee for the StopCOVID NI app. Some of his activities so far have included peer review and inspection of various software engineering artifacts of the app, e.g., UML design diagrams, test plans, and test suites (see page 13 of an online report by the local Health Authority[3]). In that Expert Advisory committee, the members have felt the need to review and mine insights from user reviews in app stores to be able to provide a feedback loop to the committee and the software engineering team of the app. Thus, the current study will provide benefits in that direction (to the committee), and also, by analyzing other apps from other countries, we will provide insight for other stakeholders (researchers and practitioners) elsewhere too.

The remainder of this paper is structured as follows. In Section 2, as background information, we provide a review of contact-tracing apps, and then a review of related works. We discuss the research approach, research design, and research questions of our study in Section 3. Section 4 presents the results of our study. In Section 5, we discuss a summary of our results and their implications for various stakeholders (app developers, decision-makers, researchers, the public, etc.). Finally, Section 6 concludes the paper and discusses our ongoing and future works.

## 2 Background and related work

As the background and related work, we review the following topics in the next several sub-sections:

1. Usage of computing and software technologies in the COVID pandemic
2. A review of contact-tracing apps and how they work
3. Closely related work: Mining of COVID app reviews
4. Grey literature on software engineering of contact-tracing apps
5. Formal and grey literature on overall quality issues of contact-tracing apps
6. Behavioral science, social aspects, and epidemiologic aspects of the apps
7. Related work on mining of app reviews

### 2.1 Usage of computing and software technologies in the COVID pandemic

A number of digital, computing, and software technologies have been developed and are in use in the public health response to COVID-19 pandemic [9]. A survey paper in the Nature Medicine magazine [9] reviewed the breadth of digital innovations (computing and software systems) for the public-health response to COVID-19 worldwide, their limitations, and barriers to their implementation, including legal, ethical, and privacy barriers, as well as organizational and workforce barriers. The paper argued that the future of public health is likely to become increasingly digital. We adopt a summary table from that paper in Table 1 [9].

[1] www.eff.org/deeplinks/2020/04/challenge-proximity-apps-covid-19-contact-tracing
[2] covid-19.hscni.net/stop-covid-ni-mobile-app/
[3] covid-19.hscni.net/wp-content/uploads/2020/07/Expleo-StopCOVIDNI-Closure-Report-V1.0.pdf

As the table shows, there are various public-health needs and various digital tools/technologies to address those needs. Contact-tracing mobile apps are just one of the digital tools/technologies to address one of those needs, i.e., interruption of community transmission.

**Table 1-Digital technologies in the public-health response to COVID-19 pandemic (from [9])**

| Public-health need | Digital tool or technology | Example of use |
|---|---|---|
| **Digital epidemiological surveillance** | Machine learning | Web-based epidemic intelligence tools and online syndromic surveillance |
| | Survey apps and websites | Symptom reporting |
| | Data extraction and visualization | Data dashboard |
| **Rapid case identification** | Connected diagnostic device | Point-of-care diagnosis |
| | Sensors including wearables | Febrile symptoms checking |
| | Machine learning | Medical image analysis |
| **Interruption of community transmission** | Smartphone app, low-power Bluetooth technology | Digital contact tracing |
| | Mobile-phone-location data | Mobility-pattern analysis |
| **Public communication** | Social-media platforms | Targeted communication |
| | Online search engine | Prioritized information |
| | Chat-bot | Personalized information |
| **Clinical care** | Tele-conferencing | Telemedicine, referral |

Other than contact-tracing mobile apps, other types of software systems have also been developed and used, related to the COVID pandemic, e.g., a system named Dot2Dot[1] which "*is a software tool to help health authorities trace and isolate people carrying an infectious disease*", and a mobile app named *COVIDCare NI*[2], developed in Northern Ireland, by the regional healthcare authority. The app provides various features to users, e.g., accessing personalized advice based on user's answers to a number of symptom-check questions; deciding if the user needs clinical advice and how to access it; and easily finding links to trusted information resources on COVID-19 advice and mental health resources.

## 2.2 A REVIEW OF CONTACT-TRACING APPS AND HOW THEY WORK

As discussed in Section 1, more than 50 countries and regions have developed so far (or, are developing) COVID contact-tracing apps to limit the spread of coronavirus[3]. The list is quickly growing, and as of this writing, 19 of those apps are open source.

Almost all proximity-detecting contact-tracing apps use Bluetooth signals emitting from nearby devices to record contact events [9]. However, in August 2020, news[4] came out that a WiFi-Based contact-tracing app has been developed in the University of Massachusetts Amherst.

A contact-tracing app can be developed using either of two different approaches: centralized or decentralized. Centralized contact-tracing apps share information about contacts and contact events with a central server (often set up by the healthcare authority of a region or country). A centralized app uploads information when a user reports testing positive for COVID. Decentralized apps upload only an anonymous identifier of the user who reports testing positive for COVID. This identifier is then broadcast to all users of the app, which compares the identifier with on-phone contact-event records. If there is a match on the mobile app of a given user, that app gives a notification to the user. Taken from a paper in this area [9], Figure 1 depicts the process of how these apps work. Another paper [10] has modeled the tracing process of a decentralized app as a UML sequence diagram (also shown in Figure 1).

[1] www.dot2dot.app
[2] play.google.com/store/apps/details?id=net.hscni.covid19ni
[3] www.xda-developers.com/google-apple-covid-19-contact-tracing-exposure-notifications-api-app-list-countries/
[4] www.eurekalert.org/pub_releases/2020-08/uoma-cso080720.php

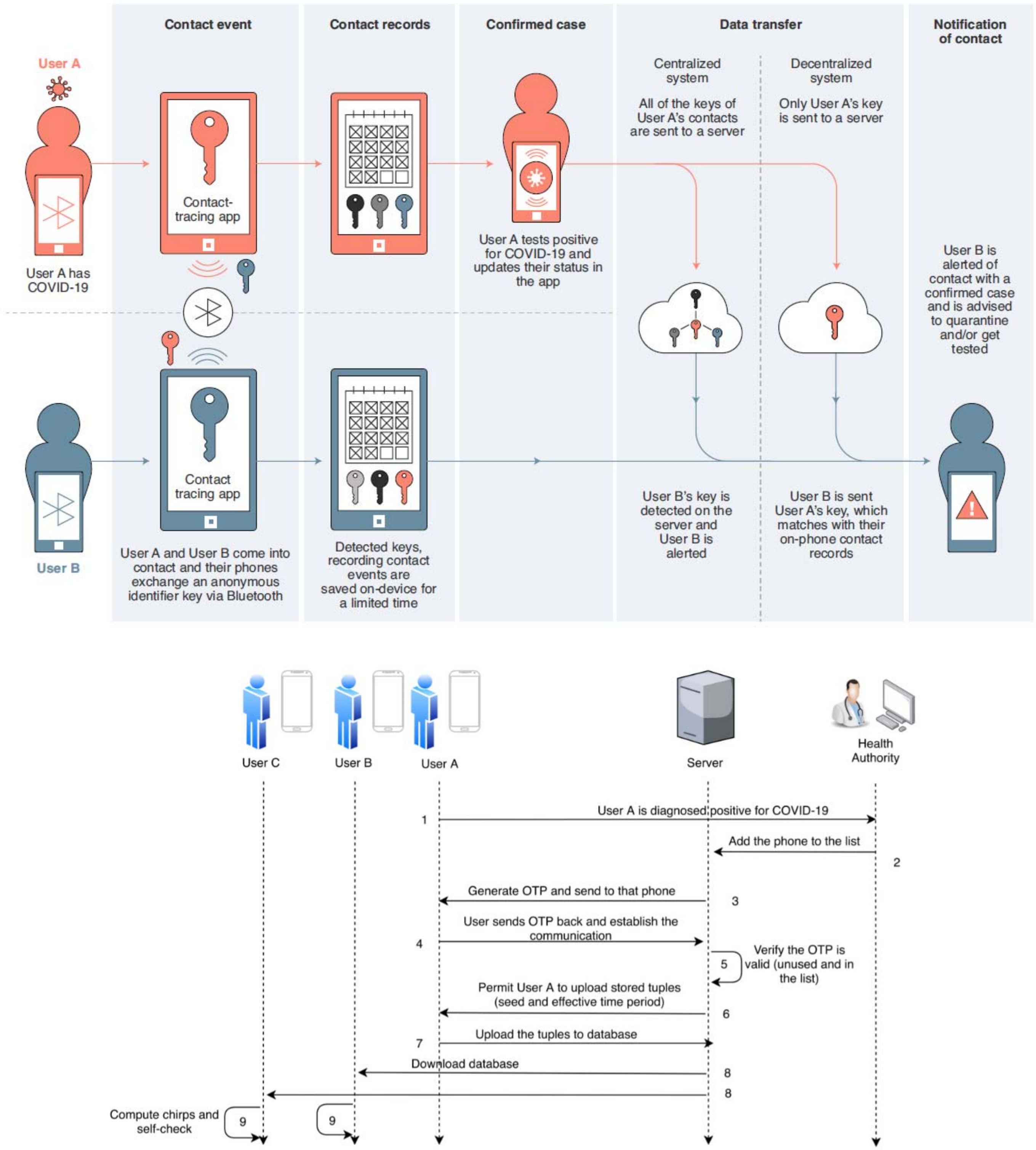

**Figure 1- How the COVID-19 contact-tracing apps work on Bluetooth-enabled smartphones [9]. Tracing process of a decentralized app (the sequence diagram is taken from [10]).**

The most widely used framework for developing decentralized contact-tracing apps is the "Exposure Notification"[1,2] framework/API, originally known as the "Privacy-Preserving Contact Tracing Project", which is a framework and protocol specification developed by Apple Inc. and Google to facilitate digital contact-tracing during the COVID-19 pandemic. The framework/API is a decentralized reporting-based protocol built on a combination of Bluetooth Low Energy (BLE) technology [11] and privacy-preserving cryptography.

[1] www.google.com/covid19/exposurenotifications/
[2] developer.apple.com/exposure-notification/

As of May 2020, at least 22 countries had received access to the protocol. Switzerland and Austria were among the first to back the protocol[1]. Shortly after, Germany announced it would back Exposure Notification, followed by Ireland and Italy.

More concretely, to know what features these apps provide, we show, as examples, several screenshots from the user interface of the *Protect Scotland* app and the *COVID Tracker Ireland* app in Figure 2. We have taken these screenshots from the apps' pages[2] in the Google Play Store. For the case of the *Protect Scotland* app, we can see that it only provides the "basic"/core contact-tracing features (use cases), i.e., tracing, adding test result, and sending notifications to recorded (traced) contacts. However, the *COVID Tracker Ireland* app provides some extra features in addition to the "core" features, e.g., showing the number of registrations since app's launch, COVID cases by county, etc.

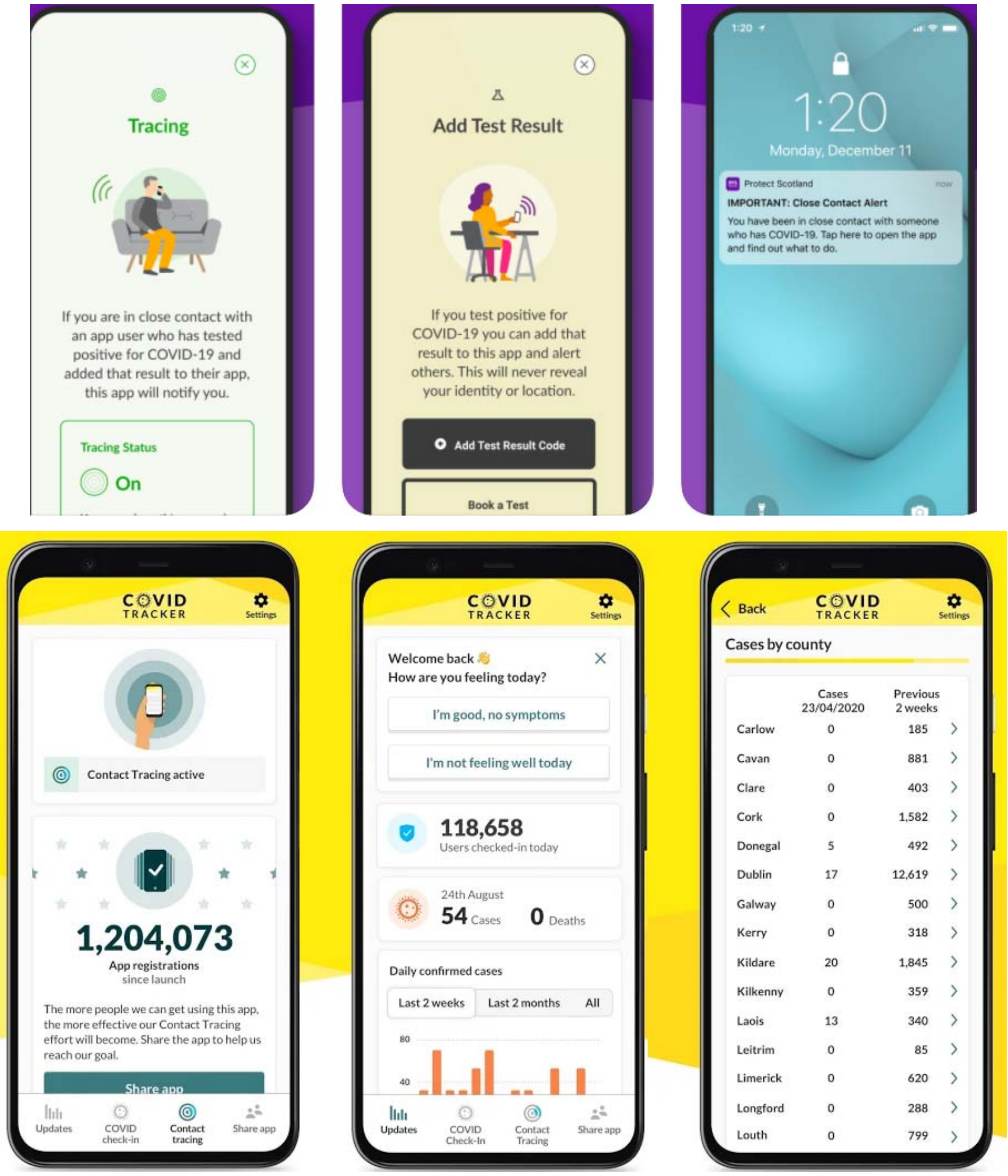

**Figure 2- Screenshots from the graphical user interface (GUI) of the *Protect Scotland* and *COVID Tracker Ireland* apps**

[1] www.reuters.com/article/health-coronavirus-europe-tech/switzerland-austria-align-with-gapple-on-corona-contact-tracing-idUSL3N2CA36L
[2] play.google.com/store/apps/details?id=gov.scot.covidtracker and play.google.com/store/apps/details?id=com.covidtracker.hse

Since early 2020, COVID has severely impacted the work and lives of almost everyone in the planet and contact-tracing apps have been widely discussed in online media, social media and news outlets. As of this writing (mid December 2020), a Google search for "contact-tracing app"[1] returned 2,110,000 hits on the web, many of which are news about these apps in the media.

Also, in the relatively short timeframe since early 2020, many research papers have been published about these apps. As of this writing (mid December 2020), a search in Google Scholar for "contact-tracing app"[2] returned 1,010 papers, which have been published in different research areas, e.g., public health, behavioral science [4], epidemiology and software engineering. We show a short list of a few interesting papers in the following, from that large set of papers:

- Contact tracing mobile apps for COVID-19: Privacy considerations and related trade-offs [12]
- One app to trace them all? Examining app specifications for mass acceptance of contact-tracing apps [13]
- A survey of covid-19 contact tracing apps [10]
- COVID-19 contact tracing apps: the 'elderly paradox' [14]
- On the accuracy of measured proximity of Bluetooth-based contact-tracing apps [15]
- Vetting Security and Privacy of Global COVID-19 Contact Tracing Applications [16]
- COVID-19 Contact-tracing Apps: A Survey on the Global Deployment and Challenges [17]

Also, various reports and news articles have discussed the high costs involved in engineering (development and testing) of contact-tracing apps. For example, for the Australian app, the cost was estimated to be 70 million Australian dollars ($49m USD)[3]. For the UK NHS contact-tracing app, the cost was reported to be more than £35 million pounds[4]. The development cost of the Irish app (*COVID Tracker Ireland*) was reported[5] to be about £773K pounds only.

### 2.3 Closely related work: Mining of COVID app reviews

In terms of related work, three insightful blog posts under a series entitled "*What went wrong with Covid-19 Contact Tracing Apps*" have recently appeared in the IEEE Software blog[6]. The articles reported analyses of user reviews of three such app: Australia's CovidSafe App, Germany's Corona-Warn App, and the Italian app. They presented thematic findings on what went wrong with the apps, e.g., lack of citizen involvement, lack of understanding of the technological context of Australian people, ambitious technical assumptions without cultural considerations, privacy and effectiveness concerns.

A recent paper [18] presented a sentiment analysis of user review on the Irish (*COVID Tracker Ireland*) app. While our current paper has some similarity in objectives with those articles, we look at more apps (nine) and also, the nature and scale of our analyses are different (more in-depth) compared to the analyses, reported in the above blog posts and paper.

### 2.4 Grey literature on software engineering of contact-tracing apps

Unlike academic (peer-reviewed), in the grey literature (such as news articles and technical reports), there are plenty of articles on the software engineering aspects of contact-tracing apps.

An interesting related news article was entitled: "*UK contact-tracing app launch shows flawed understanding of software development*"[7]. The article argued that: "*In a pandemic, speed is critical. When it comes to developing high-quality software at speed, using open-source is essential, which other nations were quick to recognize*". The article also criticized the approach taken by the UK healthcare authorities in developing their app from scratch: "*Countries such as Ireland, Germany, and Italy used open-source to build [develop] their own applications months ago. Sadly the UK did not follow suit, and wasted millions of pounds and hours of resources trying to build its own version.*"

Some other related developments include a news article[8] reporting that developers world-wide have found and reported a large number of defects in the England's open-source contract-tracing app. The peer reviews can be found on the GitHub page[1] of the app. The news article[2] went on to say that: "*… Developers have scrutinized every line of code and raised 27 issues on its Android version and 17 on the iOS version*", a summary of which can be found in Figure 3 (reproduced from the news article). The major concerns raised by software developers included: (1) the app storing the timestamps of contacts and every move of the user with GPS data on a central server; and (2) storage of Google Analytics tracking data which could help to identify users, and thus invalidate the entire idea of app usage being anonymous. In its privacy guidance, the NHSX app had promised: "*the app will not be able to track your location and it cannot be used for monitoring whether people are self-isolating or for any law enforcement purposes*". The news article argued that: "*New shortcomings in the NHSX contact-racing app could further limit effectiveness and scare away users*", which we think is a fair assessment.

[1] www.google.com/search?q=%22contact-tracing+apps%22
[2] scholar.google.com/scholar?%22contact-tracing+app%22
[3] www.bbc.co.uk/news/technology-53485569
[4] www.digitalhealth.net/2020/09/total-cost-of-nhs-contact-tracing-app-set-to-top-35-million/
[5] www.theguardian.com/world/2020/jul/20/cheap-popular-and-it-works-irelands-contact-tracing-app-success
[6] blog.ieeesoftware.org/2020/09/what-went-wrong-with-covid-19-contact.html
[7] www.verdict.co.uk/contact-tracing-app-launch/
[8] eandt.theiet.org/content/articles/2020/05/developers-find-new-flaws-in-source-code-of-nhs-contract-tracing-app

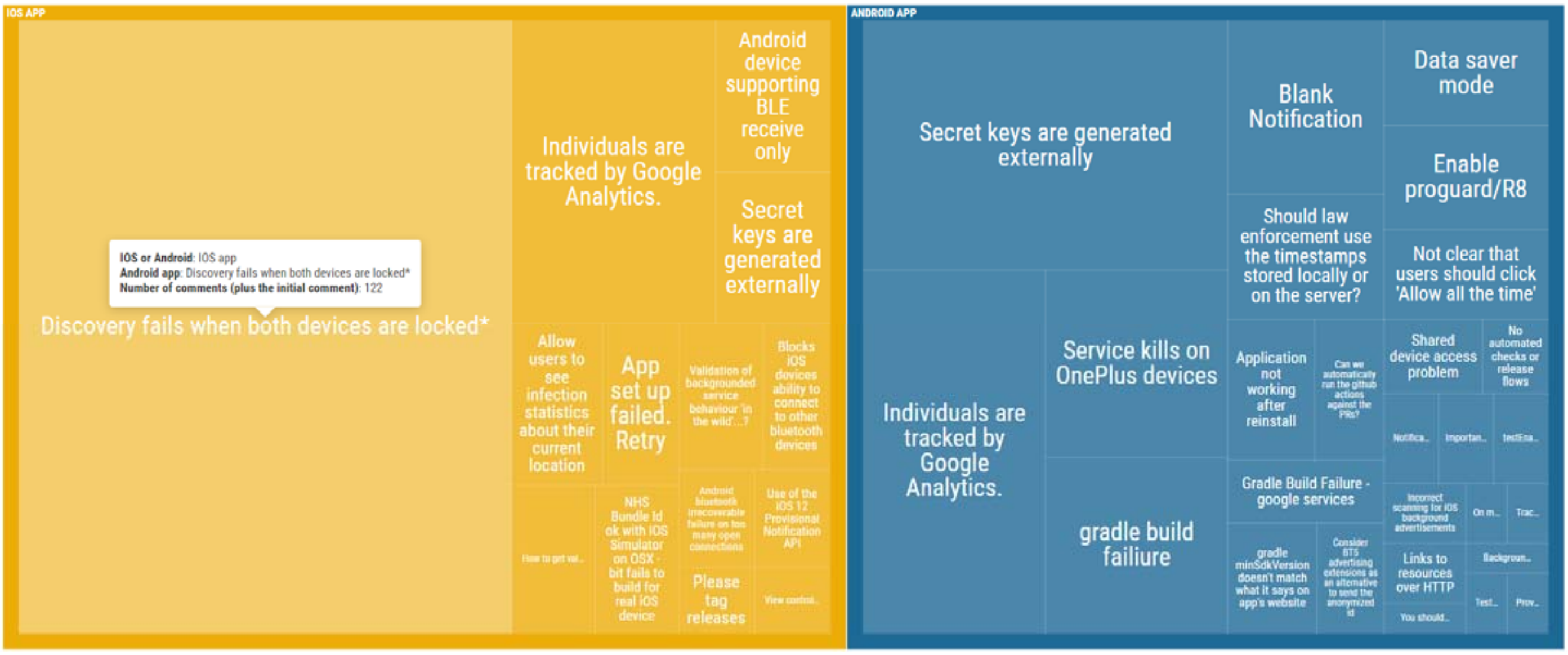

**Figure 3- Issues raised for the NHS COVID-19 iOS and Android BETA app versions on Github. Sized by number of comments posted (reproduced from: flo.uri.sh/visualisation/2515424/embed)**

Several companies, who have been involved in various software engineering aspects of contact-tracing apps, have also shared (published) grey literature materials (e.g., blog posts). For example, a large software company named ExpleoGroup (with presence in 25+ countries) was hired by the Irish Health Service Executive (i.e., Health Ministry) to conduct quality assurance and testing of the Irish "COVID Tracker" app[3]. The company published a blog post[4] on July 2020 about its test strategy for the app. The post discussed details of how the app was functionally tested, which was mostly manual, but in real-life settings, carrying out approximately 3,400 individual software tests, with work effort totaling 4,727 human-hours. In addition to functional testing, other types of testing were also conducted, according to the blog post: (1) Performance testing; (2) Exposure Notification Distance Testing: As devices have varying Bluetooth strengths, Expleo tested varying attenuation values that manage Bluetooth strength to maximize compliance with requirements; (3) Graphical user interface (GUI) testing: Expleo tested the look and feel of the app, ensuring user-friendly navigation; the correct and reliable function of all buttons; and that all content and text matched requirements.

The first two authors of this paper have also collaborated with the ExpleoGroup. For example, the first author provided consulting, for the StopCOVIDNI app, by conducting code review and inspection of test plans and test cases. Some of his contributions are discussed in an online technical report[5], serving as the testing "Closure Report" of the StopCOVIDNI app, published by ExpleoGroup.

The company which has developed both the apps for the Republic of Ireland and Northern Ireland is NearForm[6]. NearForm has published a blog post[7] in which it has discussed about bringing "privacy by design" to contact-tracing apps. The post includes many details about testing, including privacy- and security testing, e.g.: "*Intensive, repetitive testing is crucial when*

---

[1] github.com/nhsx/COVID-19-app-iOS-BETA/issues?q=
[2] eandt.theiet.org/content/articles/2020/05/developers-find-new-flaws-in-source-code-of-nhs-contract-tracing-app
[3] covidtracker.gov.ie
[4] expleogroup.com/news/expleo-announces-its-vital-role-in-ensuring-success-of-covid-19-contact-tracing-app
[5] covid-19.hscni.net/wp-content/uploads/2020/07/Expleo-StopCOVIDNI-Closure-Report-V1.0.pdf
[6] www.bbc.co.uk/news/uk-northern-ireland-53599514
[7] www.nearform.com/blog/bringing-privacy-by-design-to-contact-tracing-apps/

*it comes to privacy, and was a core element of the entire development process for this app. Because of the close collaboration among everyone involved, we were able to test the contact tracing app continually across multiple cohorts throughout both the design stage and app development".*

There have been many other news articles on software engineering aspects of the apps, e.g., for the case Australia's app, it was reported[1] that, "*… developers have highlighted ongoing problems with the [Australian] contact -racing app being able to exchange Bluetooth handshakes with iPhones if the iPhone screen is locked*".

### 2.5 Formal and grey literature on overall quality issues of contact-tracing apps

In addition to formal and grey literature on software engineering and software quality aspects of these apps, there are many sources (in both literature types) on "quality" issues (not specific to software). For example, a technical report [19] by two Irish researchers conducted an evaluation of Google/Apple Exposure Notification API for proximity detection in a commuter bus. The assessment focused on wireless networking aspects of the issue by measuring "attenuation", i.e., the loss of transmission signal strength measured in decibels (dB). Many media articles have reported various criticisms, using such reports as sources, e.g., Irish Times published an article entitled: "*Precision of tracing apps in doubt after TCD study*"[2].

German Deutsche Welle (DW) News agency published a video with the following title: "*Coronavirus tracing apps: False hope and hidden dangers?*"[3]. The Australian news agency, ZDNet, reported that: "*COVIDSafe's [Australia's contact-tracing app] problems aren't Google or Apple's fault despite government claims*"[4].

There are many reports and news articles in the US as well, e.g., a US-based nonprofit organization published a comprehensive article with the following title: "*The challenge of proximity apps for COVID-19 contact tracing*"[5]. Among the many arguments included in it was the following: "*Questions about quality, efficacy and accuracy may compound Americans' existing wariness toward tracking technologies like contact-tracing apps. Yet for theses apps to work, they need to be adopted by most of the population: Their benefit increases exponentially with the number of users. This presents a circular problem: The effectiveness of these apps will inevitably influence whether people are willing to install them, while the number of people who install the app will directly influence its effectiveness*". Another indeed insightful discussion was: "*Reliable applications of this sort typically go through many rounds of development and layers of testing and quality assurance, all of which takes time. And even then, new apps often have bugs. A faulty proximity tracing app could lead to false positives, false negatives, or maybe both, which stresses the fact that these apps are safety-critical*".

Furthermore, various metrics have been reported in support of efficacy of these apps: For the Irish app, a news article[6] reported that: "*A total of 308 users registered positive tests in the app's first seven weeks of operation, generating almost 700 close contact alerts, a proportion of whom subsequently tested positive for COVID-19*".

### 2.6 Behavioral science, social science and epidemiologic science of the apps

The use of contact tracing as a means of controlling infectious disease is long established [20] and has been seen recently for example in Ebola outbreaks [21]. The techniques used for contact tracing have however been largely centralized and focused on manual data collection by "contact tracers". Where technology such as apps have been previously used, they have been aids for the contact tracers to record data and/or systems for centralized use, analysis, and visualization of generated data [22, 23]. An actual contact-tracing app, one that actually performs rather than merely supports the process, measuring contacts and handling notifications is a novel innovation in public health.

While the *potential* benefits of such an app are generally agreed by epidemiologists, the critical success factor is adoption. For an app to be effective, it is estimated that half the population must both install it and have the app active, a significant challenge especially in countries where use of such an app is optional [13].

Some literature identifies that while contact-tracing apps face the same challenges as any technology platform, they also have additional challenges based around the sensitive health-related nature of their work and trust issues in governments. For example, Farronato et al. [24] identifies that platform failure is common place and most often because "*[the platform] never build[s] a critical mass of engaged users*" before citing some very public examples including Google+ and iTunes Ping.

[1] www.theguardian.com/australia-news/2020/jun/17/covid-safe-app-australia-covidsafe-contact-tracing-australian-government-covid19-tracking-problems-working

[2] www.irishtimes.com/news/ireland/irish-news/covid-19-precision-of-tracing-apps-in-doubt-after-tcd-study-1.4247865

[3] www.youtube.com/watch?v=pYFc5W8E91w

[4] www.zdnet.com/article/covidsafes-problems-arent-google-or-apples-fault-despite-government-claims/

[5] www.eff.org/deeplinks/2020/04/challenge-proximity-apps-covid-19-contact-tracing

[6] uk.reuters.com/article/us-health-coronavirus-ireland-apps/active-irish-covid-19-tracing-app-users-drop-on-battery-problem-hse-idUKKBN25N1PA

The study [24] hypothesized that most-contact tracing apps will also fail unless significant revision is made to their design and implementation arguing that the possible approach of mandating installation would be very poorly received in liberal democracies, the only possible successful outcome being the widespread optional use through clear demonstration of the value to the individual or community that could be provided. Generally, it seems the barrier to adoption is behavioral rather than technical, with the vast majority of the target audience having a device suitable to install the relevant app [25].

Throughout much of the literature is the common concept that potential users must be "sold" on using the app, either through seeing a clear individual benefit or encouragement to behave in a "pro-social" manner for societal benefit, while having their concerns clearly addressed [13, 24, 26].

The most common concern and therefore barrier to adoption raised repeatedly by potential users in different countries was around privacy [27-29]. While in most countries development was around the decentralized model a lack of information, clarity and transparency was seen to hinder public acceptance [26, 27].

In [30], a theory named the *Unified Theory of Acceptance and Use of Technology (UTAUT)* has been applied to evaluate the COVID tracing apps based on a survey with students from Germany and the Netherlands. Performance expectancy and perceived credibility have been determined to have a significant impact on the intention to use a contract-tracing app in the user base under study. Apart from receiving notifications about possible infections, current contract-tracing apps appear to not provide a clear benefit to the user and are perceived as somewhat privacy-invading. Furthermore, contact-tracing apps might turn out to be a failure, as the study [30] finds a low intention to use such apps. We apply a different approach, i.e., analyzing app reviews, to investigate the sentiment about COVID contract-tracing apps.

### 2.7 Related work on mining of app reviews

User feedback has long been an important component of understanding the successes or failures of software systems, traditionally in the form of direct feedback or focus groups and more recently through social media or the distribution channels themselves, i.e., feedback in app or software stores [31]. A systematic literature review (SLR) [5] of the approaches used to mine user opinion from app store reviews identified a number of approaches used to analyze such reviews and some interesting findings such as: correlation between app rating and downloads (apps rated as high-quality gain more users), but there are significant issues identifying the overall sentiment of many reviews through automated processing. Many of their reviewed studies identified the key difference in the ease with which ratings can be used numerically compared with the difficulties in "understanding" unstructured textual commentaries, especially in different societal and linguistic settings.

With ratings being seen as key to the success of apps [5], it is important to understand the concerns and issues that lead users to most commonly complain or leave poor reviews, work which is undertaken in [32]. A study was conducted using user reviews from 20 apps in the iTunes Store, which identified 12 specific types of common user complaints. The most common types of complaints are around functionality and include crashing, removal, or lack of features as well as functional defects in the app. Of particular note in the context of this paper and COVID apps is that privacy and ethics concerns are also a common type of complaint with an example review given of an unnamed app that it is *"yet another app that thinks your contacts are fair game"* [32]. Beyond the iOS focus of [32], most successful apps co-exist in at least two ecosystems (Apple and Google) and share the same brand even if they may not share the same codebase.

Hu et al. [33] seek to analyze reviews of the "same" app from both Android and iOS and compare the cross-platform results, finding that nearly half (32 out of 68) of hybrid apps (where the codebase is largely shared between platforms) "receive a significantly different distribution of star ratings across both studied platforms". The authors state that this shows a great deal of variability in how users perceive the apps even with the same fundamental features and interface depending on the users' platform.

Focusing on the societal and linguistic differences, another study [34] specifically considered country differences and how that can be a challenge for software engineering of mobile apps. The most obvious headline difference is the variation in the platform, and hence app store, adoption in different countries depending on market penetration. Other differences included the frequency of visits to app stores and the number of downloads as well as significant variation in the attitude to paid-for apps. Furthermore, large differences were seen in the willingness of users to rate apps and provide feedback in different countries based on a number of societal attributes.

Being able to take advantage of user feedback to learn lessons and improve current or future apps has also been studied. Several papers have taken advantage of this data, especially where the volume of reviews and ratings may make manual analysis impractical. Scherr et al. [35] presented a lightweight framework built on the use of emojis as representative of emotive feeling and expression of an app, building from their initial findings that large numbers of textual reviews also included emojis. Beyond general opinions, Guzman and Maalej [36] presented an approach to look at user sentiment with relation to specific features and use techniques such as Natural Language Processing (NLP) to gain this insight.

Within the health domain, Stoyanov et al. [37] defined a mobile app rating scale called "MARS" with a specific focus on descriptors aligned to health apps including mental health ranging from UX to quality perceptions and technical considerations. This approach has the potential to be widely applied to health-related apps and used as a base of comparison between them.

Mining of app-store data, including app reviews, has become an active area of research in software engineering [5]. Papers on this topic are typically published in the Mining Software Repositories (MSR) community. Authors in [5] provide a systematic literature review (SLR) on opinion mining studies from mobile app store user reviews. The SLR shows that mobile app ecosystems and user reviews contain a wealth of information about user experience and expectations. Furthermore, it is highlighted that developers and app store regulators can leverage the information to better understand their audience. This also holds for COVID contact-tracing apps as applied in this paper. However, the SLR also highlights that opinion spam or fake review detection is one of the largest problems in the domain. With respect to what users complain about in mobile apps, Khalid et al. [32] qualitatively studied 6,390 low-rated user reviews for 20 free-to-download iOS apps. They uncovered 12 types of user complaints. The most frequent complaints were functional errors, feature requests, and app crashes. Complaints about privacy and ethical issues and hidden app costs most negatively affected ratings and also play a role in the context of this paper. Further studies on app ratings cover topics on quality improvement through lightweight feedback analyses [35], sentiment analysis of app reviews [36], and consistency of star ratings and reviews of popular free hybrid Android and iOS apps [33].

When mining app store data also country-specific differences in mobile app user behavior were identified [34]. The authors collected data from more than 15 countries, including USA, China, Japan, Germany, France, Brazil, United Kingdom, Italy, Russia, India, Canada, Spain, Australia, Mexico, and South Korea. Analysis of data provided by 4,824 participants showed significant differences in app user behaviors across countries, for example, users from USA are more likely to download medical apps, users from the United Kingdom and Canada are more likely to be influenced by price, users from Japan and Australia are less likely to rate apps. Also, in this paper we analyze app reviews from several countries and therefore should be aware of country-specific differences when analyzing the data.

## 3 Research context, method and questions

We discuss next the context of our research, the research method and the research questions of our study. We then discuss our dataset, and the tool that we have used to extract and mine the app reviews.

### 3.1 Context of research

It is important to understand and clarify the research context (scope) of our work. To best present that, we have designed a context diagram as shown in Figure 4.

In the center of the study are the contact-tracing apps and the reviews entered by users in the app stores, who are citizens of a given country, and their number is often in the millions. A team of software engineers develops and releases the app in stores.

A team of public-health experts and decision-makers who work for a country's Health Authority (e.g., ministry) manage the entire work (project), set strategies and policies related to the apps and their release to the public. Software engineers are either direct employees of the Health Authority or employees of an independent software company, which is contracted by the Health Authority to develop and maintain the app.

The focus of this paper is to mine the user reviews and gather insights, with the aim of providing benefits for various stakeholders: the software engineering teams of the apps, public-health experts, decision-makers, and also the research community in this area.

We should mention that the involved teams of software engineers may already read and analyze the user reviews, sometimes replying to them in the app stores, and make improvements in their apps accordingly. However, those software engineers often only focus on reviews of their own apps. Our study extends and takes a different angle on the issue by considering nine European apps and analyzes the various trends in the reviews of those apps.

While our focus in this work positions this work in the area of software and software engineering in society [8], we also show in Figure 4 two related fields (behavioral science and epidemiology), which we reviewed for relevant literature, related to contact-tracing apps, in Section 2.5. Our analysis in this paper (Section 4) could provide potential benefits to researchers and practitioners in those fields as well.

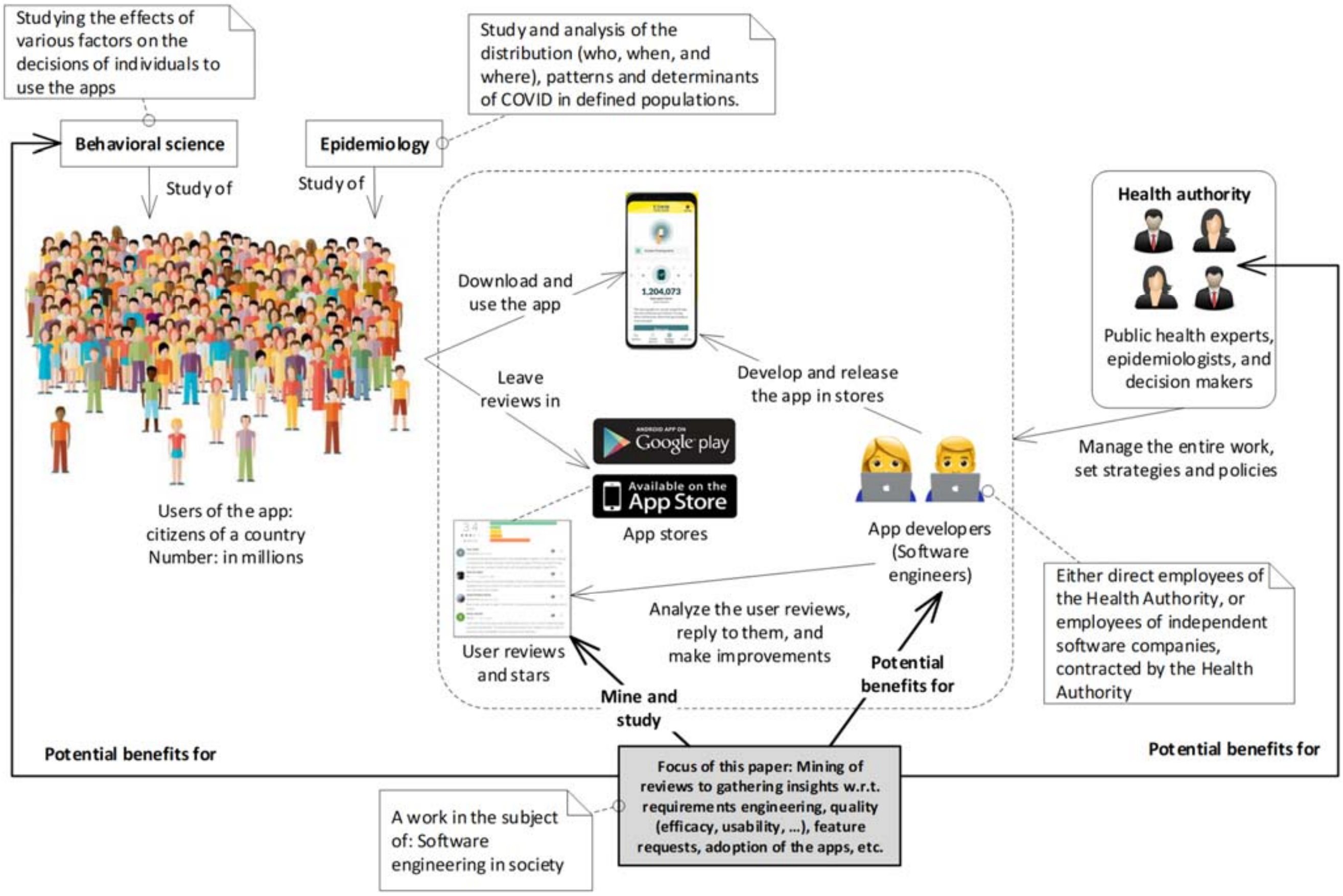

**Figure 4- Research context of this study, including the key stakeholders: app users (the public), app software engineers, public-health experts, and decision-makers**

### 3.2 RESEARCH METHOD AND RESEARCH QUESTIONS

The research method applied in this paper is an "exploratory" case study [7]. As defined in a widely-cited guideline paper for conducting and reporting case study research in software engineering [7], the goals of exploratory studies are "*finding out what is happening, seeking new insights and generating ideas and hypotheses for new research*", and those have been the goals of our study.

For data collection and measurement, we used the Goal-Question-Metric (GQM) approach [38]. Stated using the GQM's goal template [38], the goal of the exploratory case study reported in this paper is to understand and to gain insights into the user reviews (feedback) of a subset of COVID contact-tracing apps from the point of view of stakeholders of these apps (e.g., app developers, decision-makers, and public health experts).

Based on the above goal and also given the types of user review data available in app stores, we derived the following research questions (RQs):

- RQ1: What ratios of users are satisfied/dissatisfied (happy/ unhappy) with the apps?
- RQ2: What level of diversity/variability exists among different reviews and their informativeness?
- RQ3: What are the key problems reported by users about the apps?
- RQ4: By looking at the "positive" reviews, what aspects have users liked about the apps?
- RQ5: What feature requests have been submitted by users in their reviews?
- RQ6: When comparing the reviews of Android versus the iOS versions of a given app, what similarities and differences could be observed?
- RQ7: Is there a correlation between the number of app downloads and the country's population size?
- RQ8: Are there correlations between the number of reviews and the country's population or the number of downloads? And also, what ratio of app users has provided reviews?

- RQ9: What insights can be observed from the trends of review volumes and their sentiments over time?

An important aspect of our research method is the data analysis technique, which is mainly data mining. As we discuss in-depth in Section 3.4, we have selected and used a widely used commercial app-review data mining and analytics tool.

### 3.3 DATASET: SAMPLING A SUBSET OF ALL WORLD-WIDE CONTACT-TRACING APPS

As discussed in Section 1, according to a regularly-updated online article[1] in the grey literature, more than 50 countries and regions have developed so far (or are developing) contact-tracing apps. At least five other open-source contact-tracing implementations have been developed, based on the Apple-Google Exposure Notification API[2], e.g., the apps by MIT and MITRE Corporation. They could be, in principle, reused and adapted by any country/region's healthcare agency. We show in Figure 5 the 35 countries and 15 US states that have developed and published contact-tracing apps. These data have been taken from the above online article[1] (as of mid-September 2020).

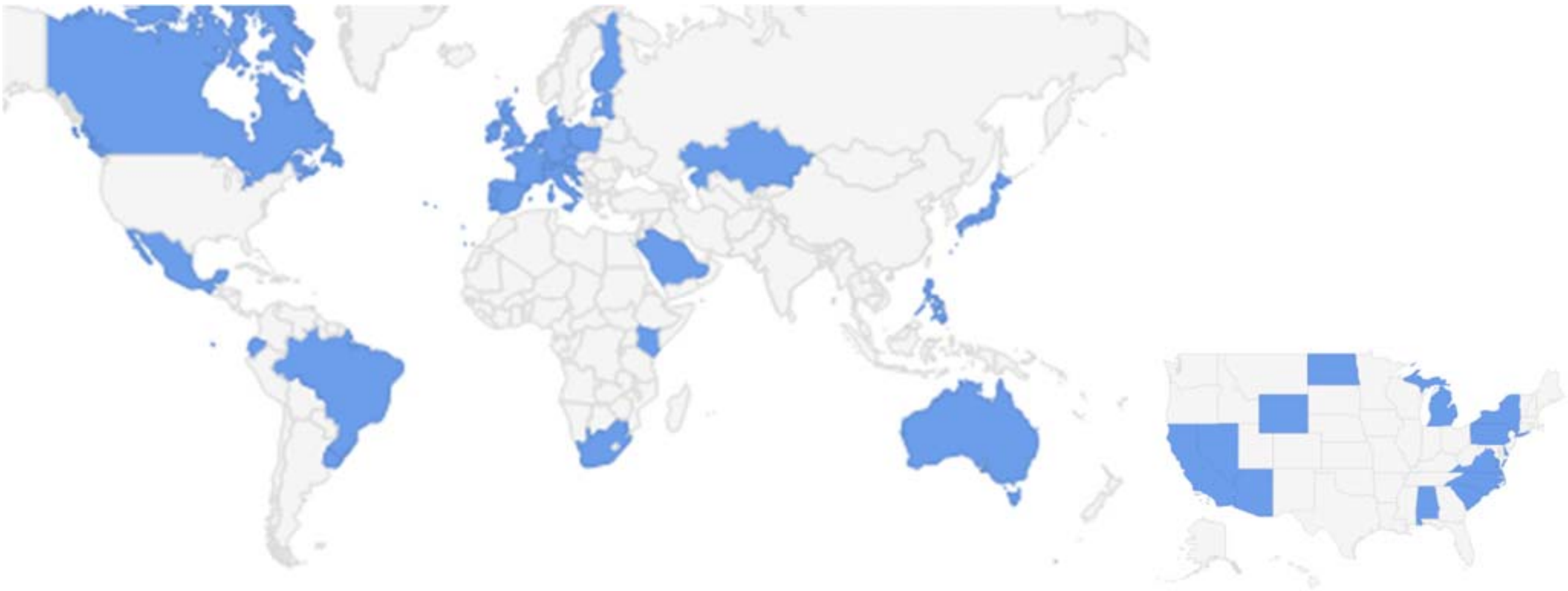

**Figure 5- The 35 countries and 15 US states which have developed contact-tracing apps**

Analyzing user reviews of "all" those 50+ apps would have been a major undertaking, and thus, instead, we decided to sample a subset of all worldwide contact-tracing apps including nine apps. Also, to make the assessments more comparable, we limited the sampling to European countries by selecting the four apps developed in the British Isles and five apps from mainland Europe. We selected the apps developed for England and Wales (parts of the UK), Republic of Ireland, Scotland, Northern Ireland, Germany, Switzerland, France, Austria, and Finland.

Table 2 lists the names, key information (such as first release dates and versions since the first release), and descriptive statistics of both Android and iOS versions of the eight selected apps. Each app can easily be found in each of the two app stores by searching for its name. Let us note that all data used for our analysis in this paper was gathered on September 17, 2020. We discuss in the next section the tools we used to extract and mine the data in this paper (including those shown in Table 2).

In terms of the number of downloads, we did not find any publicly-available exact metrics in the app stores. Google Play Store provides approximate download counts in the form of, for example, 100,000+ (meaning 100,001–500,000). Apple App Store does not provide any exact nor estimate of download counts for the iOS apps.

An interesting point in Table 2 is that some apps have had many versions since the first release, and some only had a few. Each app has received anywhere between only 63 (NHS COVID) to 20,972 reviews (Corona-Warn Germany), and counting. It is interesting to see that, in all cases, the Android apps have received more reviews compared to iOS apps. This seems to align with the general trend in the app industry, as reported in the grey literature: "*Android users tend to participate more in reviewing their apps*"[3] and "*Android apps get way more reviews than iOS apps*"[4].

We will conduct and report some correlations analysis in Section 4.7 on some of the metrics shown in Table 2.

---

[1] www.xda-developers.com/google-apple-covid-19-contact-tracing-exposure-notifications-api-app-list-countries/
[2] google.com/covid19/exposurenotifications
[3] medium.com/@takuma.kakehi/we-need-app-reviews-but-we-need-to-ask-at-the-right-time-e2916b126c8e
[4] medium.com/@chiragpinjar/why-android-apps-get-way-more-reviews-than-ios-apps-30c5b9e7ee71

**Table 2-The sampled apps and their descriptive statistics (*: As discussed in the text, all data used for our analysis in this paper were gathered on Sept. 17, 2020)**

| App | OS | First release date | # of downloads | Versions since first release | Reviews* | | |
|---|---|---|---|---|---|---|---|
| | | | | | # of reviews (as of our analysis) | Avg. stars of reviews | Ratio of Android to iOS reviews |
| 1-StopCOVID NI | Android | July 28, 2020 | 100,000+ | 2 | 195 | 3 | 2.01 |
| | iOS | | - | | 97 | 2.5 | |
| 2-NHS COVID (ENG) | Android | August 13, 2020 | 100,000+ | 2 | 174 | 1.9 | 2.76 |
| | iOS | | - | | 63 | 2.3 | |
| 3-Protect Scotland (SCO) | Android | September 10, 2020 | 100,000+ | 3 | 573 | 4 | 5.21 |
| | iOS | | - | | 110 | 4 | |
| 4-COVID Tracker Ireland (IE) | Android | June 19, 2020 | 500,000+ | 3 | 1,463 | 2.9 | 5.34 |
| | iOS | | - | | 274 | 3.1 | |
| 5-Corona-Warn Germany (DE) | Android | June 25, 2020 | 5,000,000+ | 8 | 20,972 | 2.7 | 3.10 |
| | iOS | | - | | 6,772 | 2.3 | |
| 6-SwissCovid (CH) | Android | June 18, 2020 | 500,000+ | 9 | 1,370 | 3.1 | 2.10 |
| | iOS | | - | | 652 | 3.1 | |
| 7-StopCovid France (FR) | Android | June 6, 2020 | 1,000,000+ | 10 | 2,397 | 2.6 | 9.95 |
| | iOS | | - | | 241 | 2.1 | |
| 8- Stopp Corona Austria | Android | Mar 27, 2020 | 100,000+ | 10 | 1,961 | 2.4 | 3.27 |
| | iOS | | - | | 599 | 2 | |
| 9-Finland Koronavilkku (FI) | Android | August 31, 2020 | 1,000,000+ | 3 | 1,276 | 3.4 | 5.41 |
| | iOS | | - | | 236 | 3.3 | |

We should mention that in the writing phase of this paper (in November 2020), we heard the news that France launched[1] a new contact-tracing app, named *TousAntiCovid* (literally translates to: "All Anti Covid") in late October 2020, which replaced the previous app, StopCovid, in the app stores. However, the review data that we had fetched using our chosen analytics tool (AppBot, as discussed in the next section) was until mid-September, so our analysis is on the France' StopCovid app, and the dataset had integrity w.r.t. that app.

### 3.4 Tool used to extract and mine the app reviews

We wanted to use an automated approach to extract, mine, and analyze the apps' user reviews. We came across the Google Play API[2], which provides a set of functions (web services) to get such data. At the same time, we found that there are many powerful online tools that do the job of fetching the review data from app stores and even include useful advanced features such as text mining, topic analysis and sentiment analysis [36] on review texts. The large number of such tools indicate the fact that an active market for app review "analytics" is emerging.

We came across a high-quality candidate tool to extract and mine the app reviews, i.e., a commercial tool named AppBot (appbot.co). The tool provides a large number of data-mining and sentiment analysis features. For example, as we will use in Section 4.1, AppBot uses an advanced method, based on AI and Natural Language Processing (NLP), to assign one of the four types of sentiments for each given review: positive, neutral, mixed, and negative sentiment. Also, as we will discuss in 4.5, another feature of AppBot is to automatically distinguish reviews that contain "Feature requests" submitted by users among all reviews of an app.

To do the above analysis, there have been specific papers that have proposed (semi-) automated techniques, which could be somewhat seen as the competitors for commercial App-analytics tools, such as AppBot. For example, a paper by Maalej and Nabil [39] introduced several probabilistic techniques to classify app reviews into four types: bug reports, feature requests, user experiences, and ratings. The approach uses review metadata such as the star rating and the tense, as well as, text classification, NLP, and sentiment analysis techniques.

Other papers have proposed or used sentiments techniques to classify each review, e.g., into positive or negative review, just like what the AppBot tool does. For example, authors of [36] used NLP techniques to extract the user sentiments about apps' features.

In summary, to make our choice of tools/techniques to extract and mine the app reviews, we could either use the approaches presented in the above papers, or the commercial tool AppBot. To make our tool choice, we tried the AppBot tool on several apps in our selected pool, and observed that the tool works well, and its outputs are precise. Also, the fact that “24 of Fortune-100 companies” (according to the tool’s website: appbot.co), e.g., Microsoft, Tweeter, BMW, LinkedIn, Expedia and New York Times are among the users of the tool, were strong motivations for us in favor of the AppBot tool over the techniques presented in the above papers. In addition, almost all techniques presented in the above papers had no publicly-available tool support and if we had to choose them, we had to develop new tools, which was clearly extra work, for which we saw no reason. Thus, we selected and used AppBot for all the data extraction and data mining.

[1] www.healthcareitnews.com/news/emea/france-launches-new-contact-tracing-app-tousanticovid
[2] developers.google.com/android-publisher/api-ref/rest/v3/reviews

However, we were still curious about the precision of the analyses (e.g., sentiment-analysis algorithm) done by AppBot. We initiated personal email communication with co-founder of AppBot, asking about the precision of the analyses by the tool. The reply that we received, was: “*We [have] trained our own sentiment analysis so it worked well with app reviews. Here's the details of our algorithm:*

- *Developed specifically for short forms of user feedback, like app reviews*
- *Understands the abbreviations, nuanced grammar and emoji*
- *Powered by machine learning*
- *Over 93% accuracy*
- *Trained on over 400 million records*”

We thus were quite satisfied that the tool that we were going to use has high quality and high precision in the analyses and results that it produces.

From another perspective, we are followers of the “open science” philosophy and reproducible research, especially in empirical software engineering [40], and we believe that empirical data generated and analyzed in any empirical software engineering study should be provided online (when possible) for possible use by other researchers, e.g., for replication and transparency. By following that principle, we provide all the data extracted and synthesized for this paper in the following online repository: www.doi.org/10.5281/zenodo.4059087. Since to download the raw review data and analyze them using the commercial tool AppBot, we acquired a paid license for it, we cannot share the raw dump of all review data for all the apps in the above online repository, but instead we share in there the aggregated statistics that we have gathered from the raw review data. Interested readers can easily acquire a license for the tool (AppBot) and download the raw data.

We also think that some readers may be interested to explore the dataset and reviews by their own and possibly conduct further studies like ours. To help with those, we have recorded and provide a brief (10-minute) video of live interaction with the dataset (to be analyzed in this paper) using AppBot, which can be found in: youtu.be/qXZ_8ZTr8cc.

## 4 RESULTS

We present next the results of our analysis by answering the RQs of our study.

### 4.1 RQ1: WHAT RATIOS OF USERS ARE SATISFIED/ DISSATISFIED (HAPPY/UNHAPPY) WITH THE APPS?

Our first exploratory RQ (analysis) was to assess the ratios of users, which, as per their reviews, have been happy or unhappy with the apps.

“Stars” (a value between 1-5) are the built-in rubric of app stores (both the Google Play and the Apple App Store) which let users mention their level of satisfaction or dissatisfaction with an app, when they submit their review. This feature is also widely used in many other online software systems, such as online shopping (e-commerce) web applications including Amazon. For the case of online shopping and also paid mobile apps, the number “stars” on a product (or app), often strongly impacts the choice of other users whether to buy a product (or app) or not [33], a relationship also seen in the levels of adoption of apps in mobile app stores [5].

A user can choose between 1 to 5 stars, when s/he submits a review. Another more sophisticated way to derive users’ satisfaction with an app is to look at the semantic tone of the review text, e.g., when a user mentioned in her/his review: "*I really like this app!*", that would clearly mean her/his satisfaction with the app. On the other hand, a review text like: "*the app crashed on my phone several times. Thus, it is not a usable app.*", implies the user's dissatisfaction with the app. Making broad use of this, especially on longer textual reviews, can have some limitations when automatically analyzed, but the majority of reviews can have sentiment successfully detected [5].

In the NLP literature, automatic identification of the semantic tone of a given text is referred to as *sentiment* analysis [41]. Sentiment analysis refers to the use of NLP to systematically quantify the affective state of a given text. A given text can have four types of sentiments [41]: positive, negative, neutral, and mixed. A positive sentiment denotes that the text has a positive tone in its message. “Neutral” sentiment implies that there is no strong sentiment in the text, e.g., "*I have used this app*". A text is given the “mixed” sentiment when it is conflicting sentiments (both positive and negative).

Our chosen data-mining tool (AppBot) supports the above four types of sentiments for each given review: positive, neutral, mixed, and negative sentiment. To classify the sentiment for a given review, AppBot calculates and provides a sentiment score of each review (a value between 0-100%).

We show in Figure 6 the distribution of stars as entered by the users in reviews, and also the distribution of reviews' sentiment categories. We show both a 100% stacked-bar and a stacked bar of absolute values for the stars. As we can see, since the German Corona-Warn app has received many more reviews compared to the others in the set, it has overshadowed the others in the stacked-bar figure.

We can see from these charts and also the average stars of each app (Table 2) that the users are generally dissatisfied with the apps under study, except the Scottish app. We furthermore averaged the stars from the mean score of Android and iOS versions of each app, e.g., for *StopCOVID NI*, this resulted in 2.75 (average of 3 and 2.5). Based on this metric, the *Protect Scotland* app is the highest starred (4/5), and *NHS COVID* is the least starred (2.1/5). The average of stars for all the other apps range between these two values. We should note that we have not installed nor tried any of the apps, and thus all our analyses are purely based on mining user reviews.

One very interesting consideration is what factors have led to the Scottish app be ranked the highest in terms of stars. Reviewing a subset of its reviews revealed that the app seems easy to use and is quite effective, e.g., one user said: "*Brilliant app. It collects zero personal data, no sign ups, no requirement to turn on location, nothing! All you have to do is turn on Bluetooth, that's it.*"[1]. Of course, more in-depth assessment and comparison of the apps are needed to be done.

We were expecting that stars and the reviews' sentiments would have correlations, i.e., if a user has left a 1 star for an app, s/he has most probably has also left a negative (critical) comment in the review, and vice versa. We show in Figure 6 a scatter-plot of those two metrics, in which 16 dots correspond to the 16 apps under study. The Pearson correlation coefficient of the two measures is 0.93, showing a strong correlation.

When comparing the average stars with the positive reviews sentiment percentages in Figure 6-(d), the sentiment percentages seem to be consistently more negative (have lower values in the Y-axis). By analyzing a subset of the dataset (reviews), we observed that many reviews are similar to the following phrase/tone: "*I like the app, but rant rant rant*", and then the user has entered 4 stars, for example. The "rant" (complain) part could be quite harsh, thus causing the textual reviews sentiment score to fall down.

It is also interesting to see in Figure 6-(d) that, generally, the dots of the two OS versions of each app, are relatively close to each other in this scatter-plot, meaning that users have independently scored both versions of each app in quite similar levels. In some cases, the iOS version of a given app has a slightly higher average star value than the Android version, and it was the other way around for the other apps. A uniform relationship could not be observed.

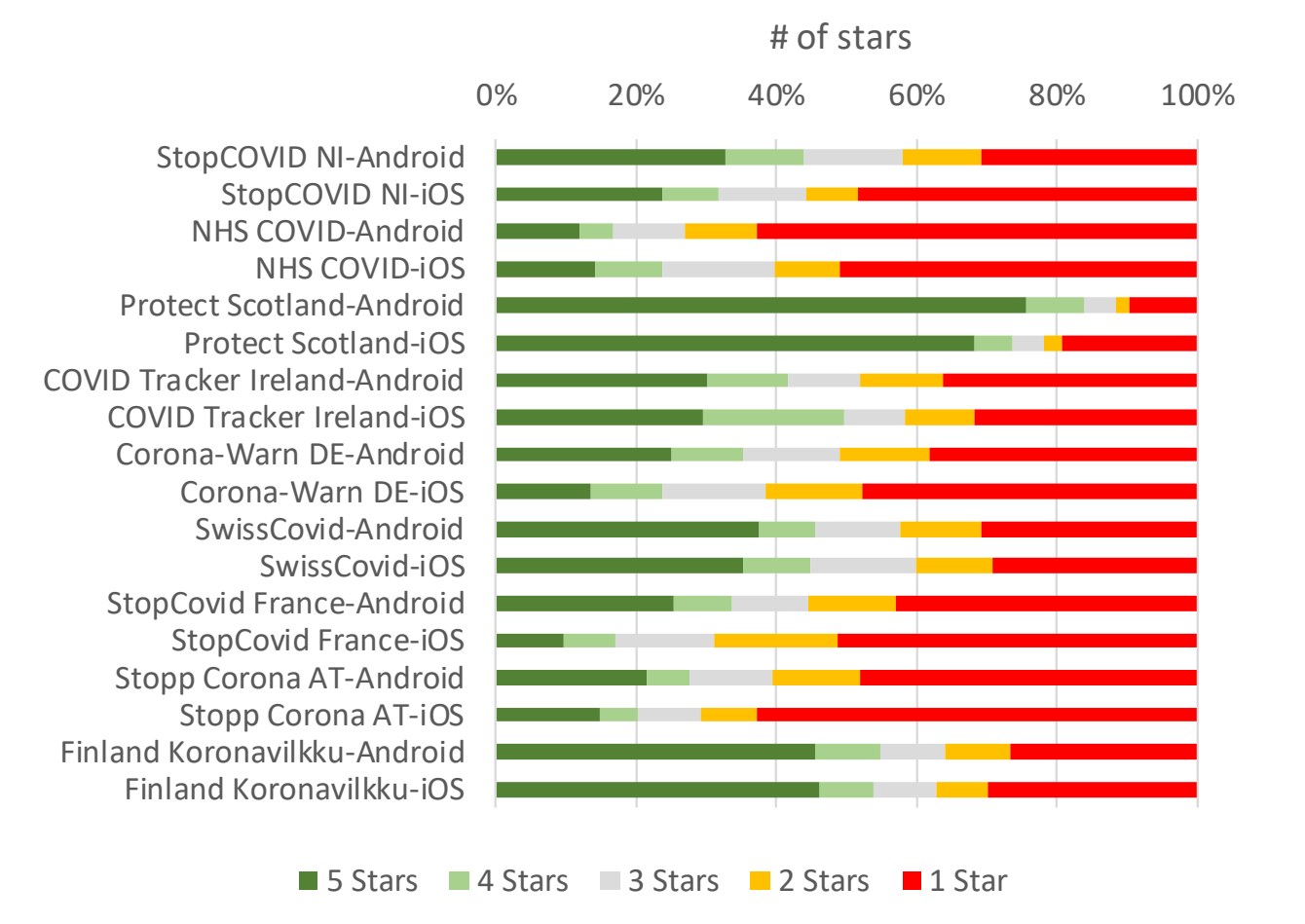

(a):Stacked bar-chart (percentages) of review stars

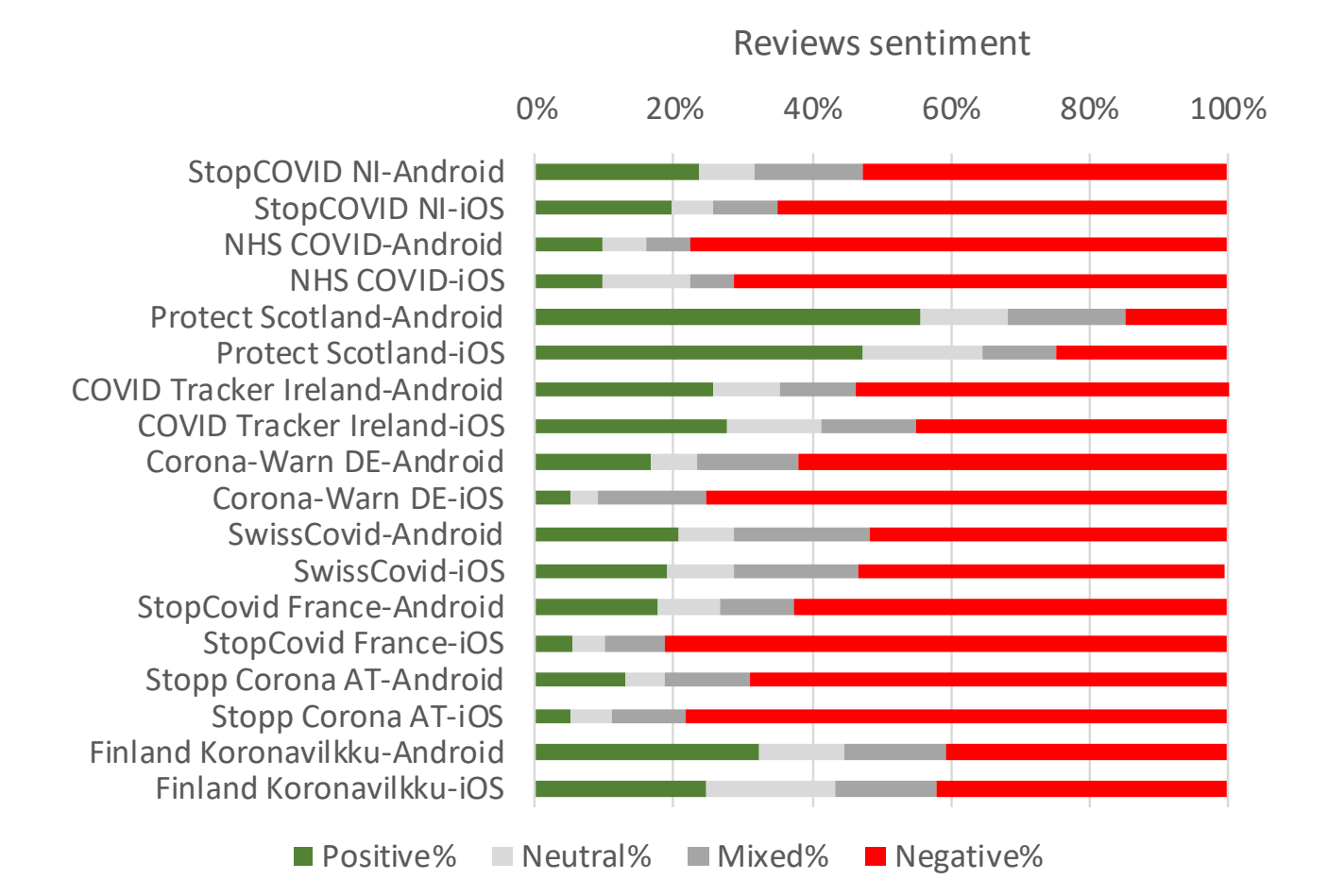

(b):Stacked bar-chart of sentiment categories

[1] bit.ly/ScottishAppAPositiveReview

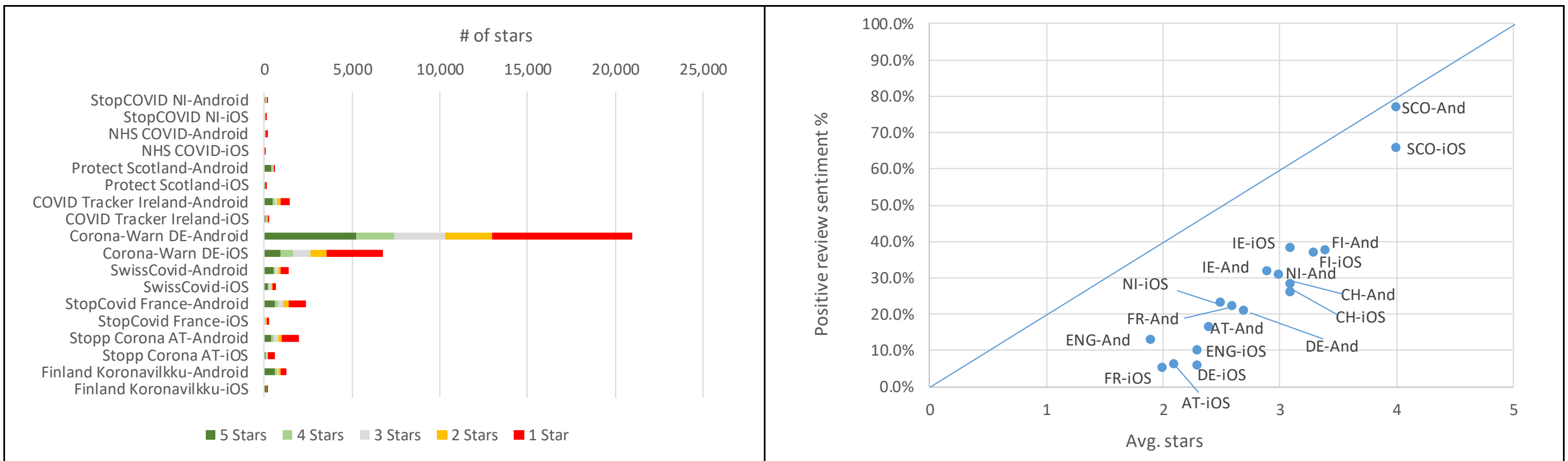

(c):Stacked bar-chart (absolute values)

(d):Correlations of review stars with review sentiments

**Figure 6- Distribution of review stars, and review sentiment categories**

**Lesson learned/recommendation**: The users are generally dissatisfied with the apps under study, except the Scottish app. Future studies could look into what factors has made the Scottish app be different than others in the pool of apps under study. That could a research question (RQ) to be studied by researchers in future works.

### 4.2 RQ2: A LARGE DIVERSITY/VARIABILITY IN REVIEWS AND THEIR INFORMATIVENESS

In addition to using the AppBot tool for automated text mining and sentiment analysis of the large set of reviews, it was important to read a subset of reviews, to actually get a sense of the dataset. For example, we browsed through the large list of 27,000+ reviews of the German Corona-Warn app. Thanks to a useful feature of app stores, other users can "like" a certain review and thus, one can order reviews by the number of likes (often denoting how informative they are). We found a few such reviews, such as the following[1]:

> *"Solid user interface and good explanation of the data privacy concept. Surprisingly well done, I was expecting it to be more cumbersome. Edit: It would be good if we could see how many tokens the app has collected in the last 14 days. This would make the app more attractive to open and raise the confidence in that it actually works. Also interesting metric would be to know how many users have been warned by the app. This has been released to the public (I believe 300 notifications so far). Unfortunately, I cannot find the Android system settings which apparently shows the number of contacts collected. Either its not available easily or I just can't find it. Anyway - I think it would be great if the app could show this information rather than asking the user to search for information in the settings."* (translated automatically from German by AppBot, which uses the Google Translate API).

Many reviews, including the above one, were feature requests, and some of them could indeed be useful for the development team for improving the app. Many reviews were also replied by the development team in a careful way, which was refreshing to see. For example, there was the following thread in one of the reviews for the German app:

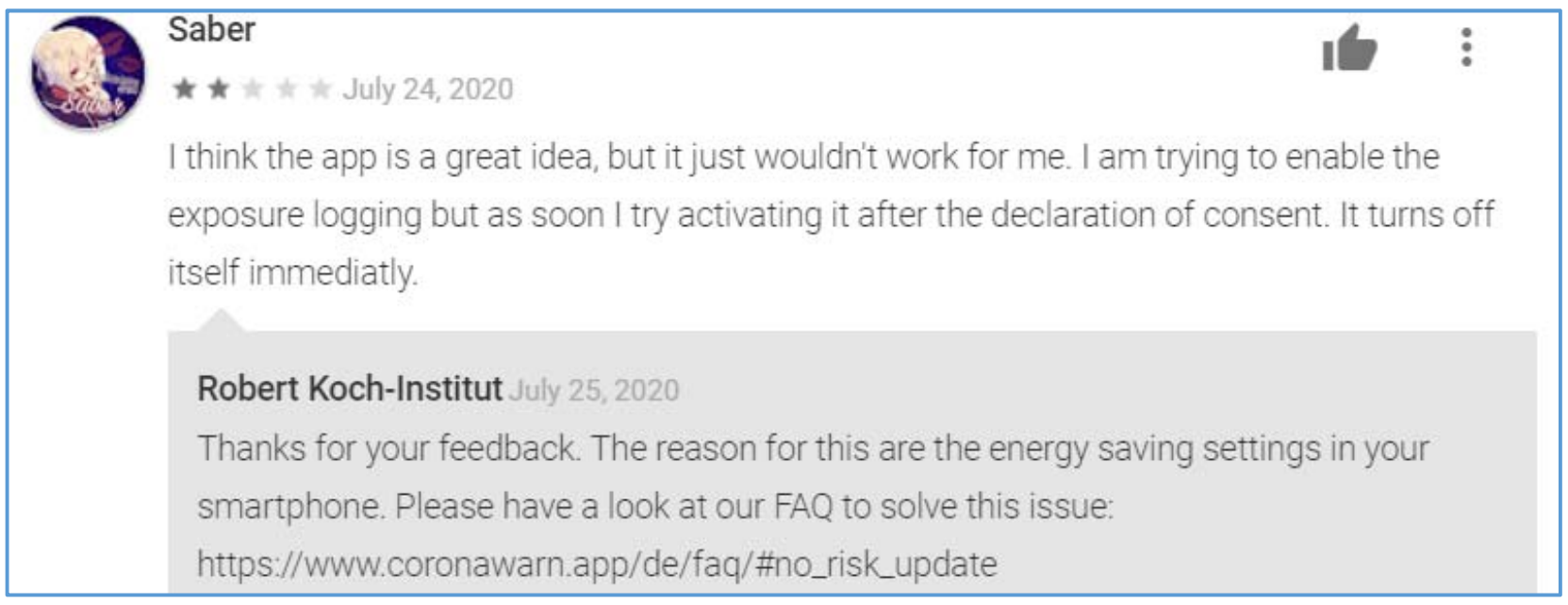

**Figure 7- A user review and the reply by the development team, for the German app**

From the above example review, we can realize that the apps should be designed as simply as possible, since typical citizens ("laymen") are not often "technical" people and we cannot assume that they will review the online FAQ pages of the app to properly configure it.

[1] bit.ly/ADetailedReviewInGermanApp

**Lesson learned/recommendation**: Contact-tracing apps should be designed as simple as possible (for usability), as we cannot expect layperson citizens to review the online FAQ pages of the app to properly configure it, especially for a safety-critical health-related app.

**Lesson learned/recommendation**: Developers of the apps can and should engage directly with reviews and reply, not only gaining insight into the most commonly raised concerns but also to answer the questions in public view. This can even provide a positive "image" of the software engineering team behind the app, in public view (in terms of accountability, responsiveness and being open to feedback).

Essentially, similar to any other mobile app, reviews could range from short phrases such as "*Not working. Weird privacy settings*"[1], which are often not useful nor insightful for any stakeholder, to detailed objective reviews (like the one discussed above), which are often useful.

One way of analyzing diversity/variability of reviews was to measure each review's length in words. We gathered those data for five of the nine apps (as examples) and provide the boxplots of both OS versions of those five example apps in Figure 8. Since we observed that there are many "outlier" data points in the box-plots, we provide the plots with and without outliers. For the readers who are less familiar with boxplots, we provide a conceptual example in Figure 8 about the meaning of the boxes in boxplots and lines in it. More details about boxplots and their terminology can be found in the statistics literature [42].

As we can see in Figure 8, for all five apps, the bulk of reviews are relatively short in length. The German app, on both Android and iOS platforms, has a more noticeable collection of longer comments which is particularly evident for the iOS version. The slight increase of the median may be due to linguistic differences but the relatively large number of longer (>200 word) reviews implies a great degree of user engagement in commenting on these apps, especially on iOS, or could have cultural/social root causes, e.g., it could be German users often tend to provide "detailed" (extensive) feedbacks.

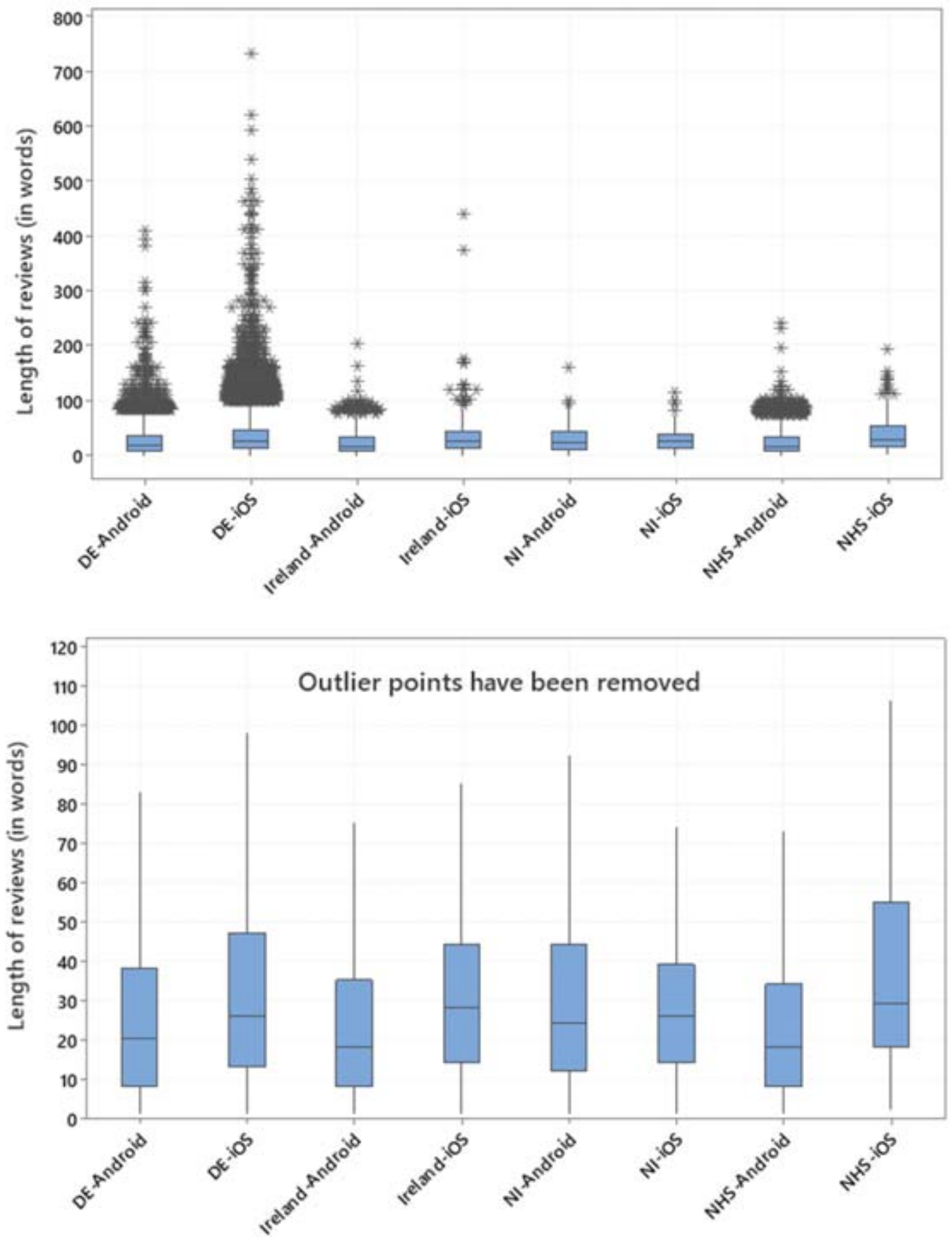

---

[1] bit.ly/AShortReviewInGermanApp

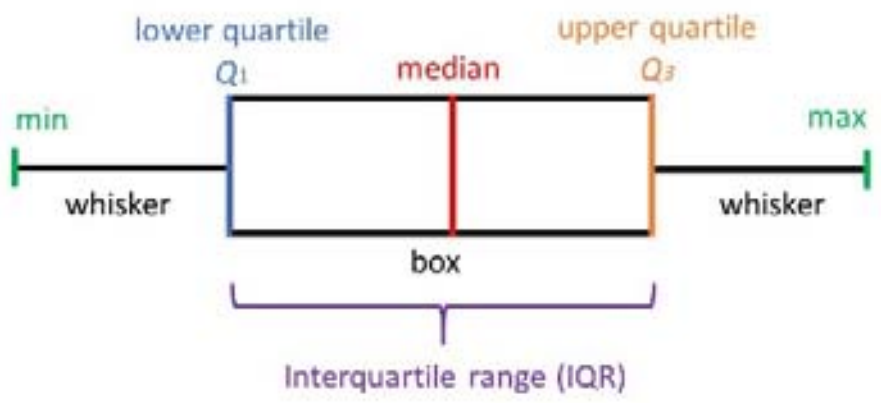

Q1: 25% of the data. Q3: 75% of the data

**Figure 8- Boxplot showing the distribution of textual "length" of reviews (in words) for five example apps. With and without "outlier" data points**

As another insight, we found that a proportion of reviews included error messages or crash reports. For example, for the German app (Corona-Warn) again, a user mentioned in her/his review[1]: *"After a few days the app stopped working. Several error messages appeared including 'cause: 3' and 'cause: 9002'. Tried to troubleshoot it by checking the Google services version, deleting the cache, reinstalling the app etc."*

When we interpret this review, it is logical to conclude that it is quite impossible for a layperson to deal with such errors and error messages, given the nature and full public outreach of the app. Thus, we wonder whether such error messages and crashes have been one of the several reasons why the apps under study have been rated quite low in reviews, overall. By reading more reviews, we observed that many users (citizens) with some technical (IT) background have taken various steps to make the apps work, e.g., reinstalling them, etc. However, we believe that, for a layperson, taking such troubleshooting steps is out of the question, and such a person would usually ignore and remove the app, and possibly would leave a harsh review for it in the app store, and submit a low score for the app.

**Lesson learned/recommendation**: Just like any other mobile app, user reviews for contact-tracing apps range from short phrases such as "*Not working*", often not that useful nor insightful, to detailed objective reviews that could be useful for various stakeholders. Thus, if any stakeholder (e.g., the app's development team) wants to benefit in a qualitative way from the reviewers, they need to filter and analyze the "informative" reviews.

### 4.3 RQ3: PROBLEMS REPORTED BY USERS ABOUT THE APPS

As another "core" RQ of our study, we wanted to identify the main issues (problems) that users have reported about. Having received anywhere between 63 and 20,972 reviews (and counting) as of this writing for each app, the nine apps had in total 39,425 review comments. Of course, manual analysis of such a large and diverse textual feedback was not an option. The AppBot tool provides various features such as sentiment analysis [36] and critical reviews to make sense of large review text datasets. We show the outputs of word-cloud visualization for all the nine apps in Figure 9. We also include the AppBot tool's user-interface in Figure 9, as a glimpse into how it works.

For generating word-clouds based on reviews, AppBot provides six types of options to filter review subsets: interesting reviews, popular reviews, critical reviews, trending up reviews, trending down reviews, and new reviews. AppBot has a sophisticated NLP engine to tag views under those six categories, for example: "*The Popular tab shows you the 10 words that are most common in your reviews. This helps you to identify the most common themes in your app reviews*"[2]; and "*the Critical tab is a quick way to find scary stuff in your reviews. This can help isolate bugs and crashes, so you can quickly locate and fix problems in your app faster*"[2]. Since we are interested to know about the problems reported by users about the apps, to generate the word-cloud visualization of Figure 9, we have filtered by "critical" reviews.

For apps of non-English-speaking nations, e.g., Germany and France, unsurprisingly, almost all reviews were in their official languages, and we used the Chrome browser's built-in translate feature to see the review texts in English. For readers wondering about the original reviews in the original languages, we also show the word-clouds of two example apps (StopCovid France, and Stopp-Corona Austria) based on their original review data.

Let us consider the *COVID Tracker Ireland* app as an example. As we can see in its word-cloud, "battery, "draining" and "uninstall" are among the most "critical" words. The fact that these apps make regular usage of Bluetooth signals leads to high battery usage, and this issue has been widely discussed in many online sources[3]. Furthermore, the terms "work" and "update" appear prominently with negative sentiment in the word-cloud of app reviews from Germany and Finland. By reading a subset of those reviews, we observed that it has not been obvious for many users of those apps how to use those apps properly (how to get them to "work").

[1] play.google.com/store/apps/details?id=de.rki.coronawarnapp&hl=en_CA&reviewId=gp%3AAOqpTOHW5kfcmZxpbP-qYBNTBbd-_P6n7boLJC-jAYDzt1I-1FTzkSNtd-IY-vea7mf3Ki2c_6MaqHVUAhy8Og

[2] support.appbot.co/help-docs/using-words-page/

[3] www.lancasterguardian.co.uk/health/coronavirus/nhs-test-and-trace-app-shouldnt-drain-your-battery-or-affect-your-privacy-2982165

In Figure 9, words in a word-cloud are colored according to their sentiments in reviews. AppBot provides four types of sentiments: positive (green labels in the word-cloud), negative (red), neutral (grey), and mixed (orange). There is another useful feature in AppBot: when we click on each word in the cloud, all the reviews containing that word are listed.

Lots of insights can be gained from the word-clouds, word sentiments, and also by live interaction with the dataset in the AppBot tool (we invite the interested readers to do so). As discussed in Section 3.4, we have posted an online video of live interaction with the dataset in: youtu.be/qXZ_8ZTr8cc

Of course, comparing these data and findings for two different contact-tracing apps should be done with a "grain of salt", since their contexts (users' demographics, software features and requirements) are quite different. For example, England's NHS COVID app has a feature to allow users to scan a QR code, as the government has asked shops to request shoppers to do so when entering shops. Many of the reviews for this app are about issues with that feature (see the word "code" in the word-cloud), a feature which apparently does not exist in the other apps.

Among the word-clouds, we can visually notice that the *Protect Scotland* app's word-cloud shows an overall positive picture, with lots of green (=positive) sentiments. In the rest of the word-clouds, red (negative) sentiments are the majority.

**Lesson learned**: Problems reported by users about each of the apps are quite different from one another. However, there are still some issues reported by users for most apps, e.g., high battery usage.

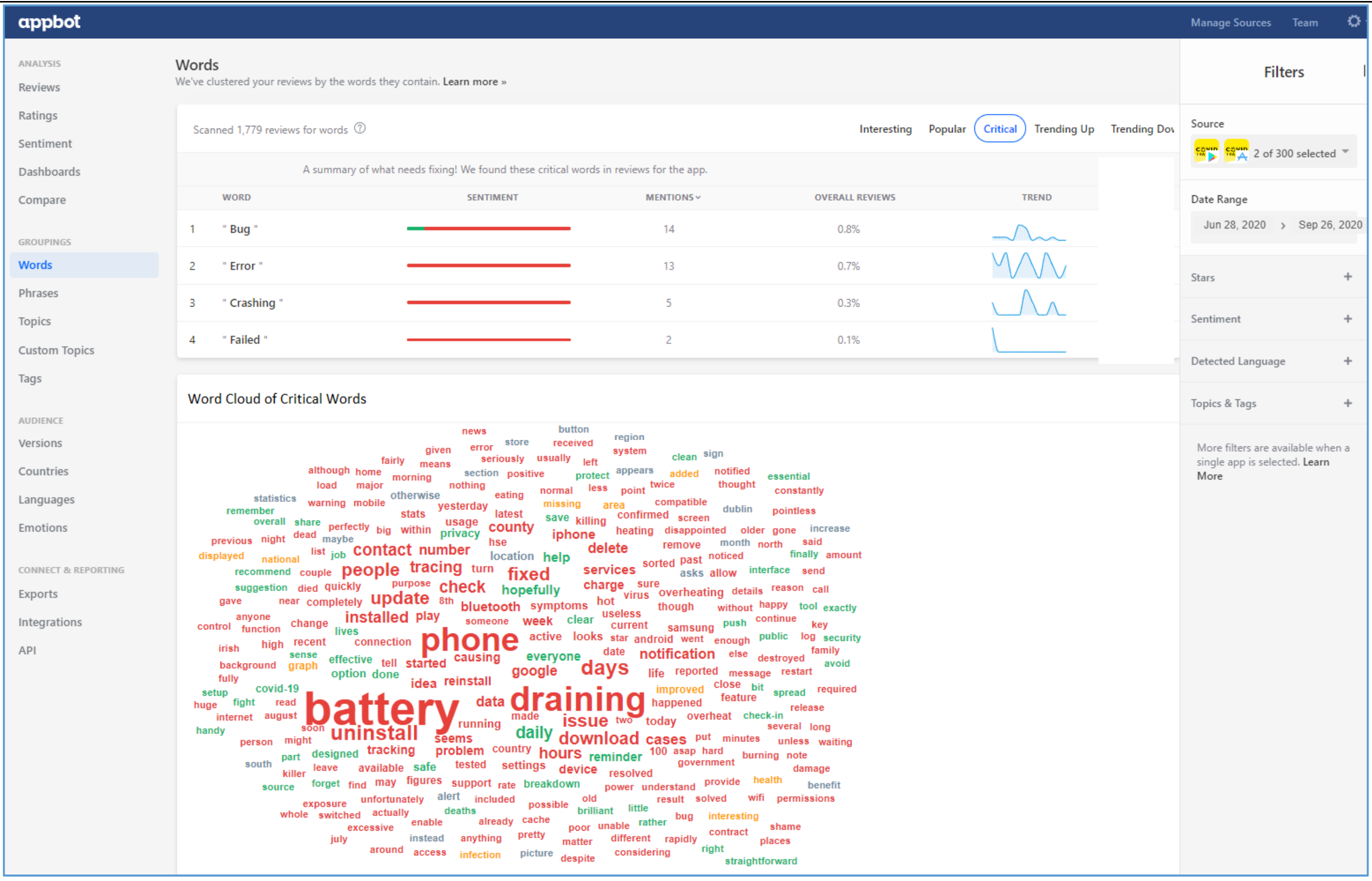

COVID Tracker Ireland

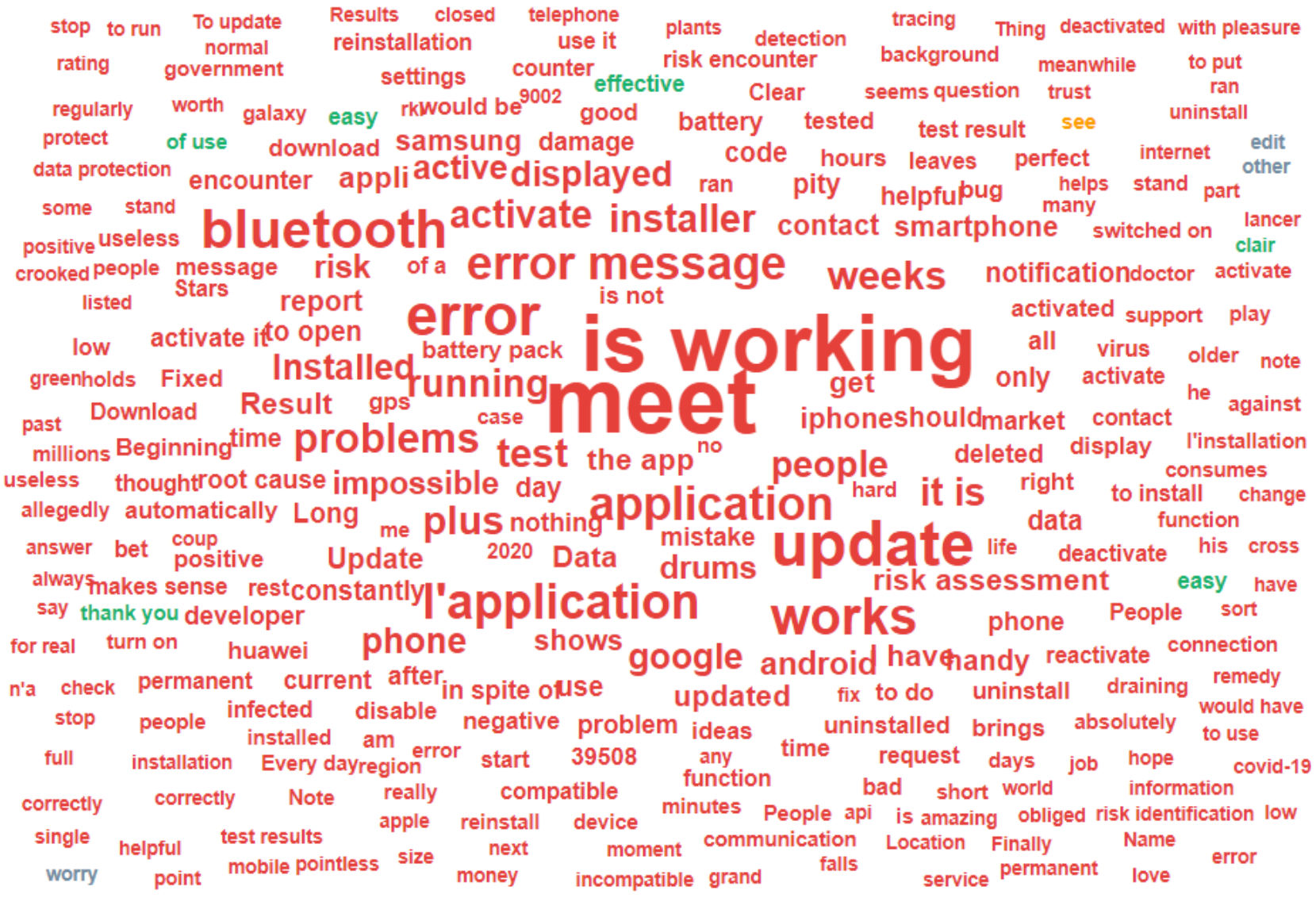

When all the reviews of all nine apps are analyzed (n=39,425 reviews)

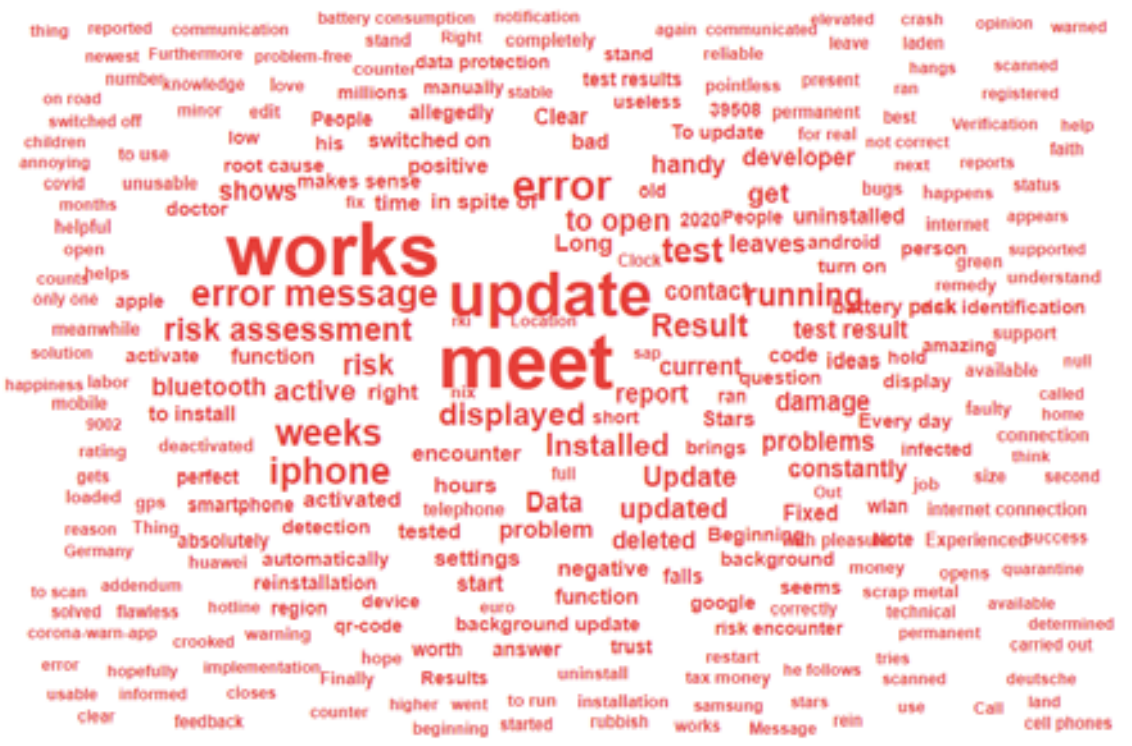

Corona-Warn Germany

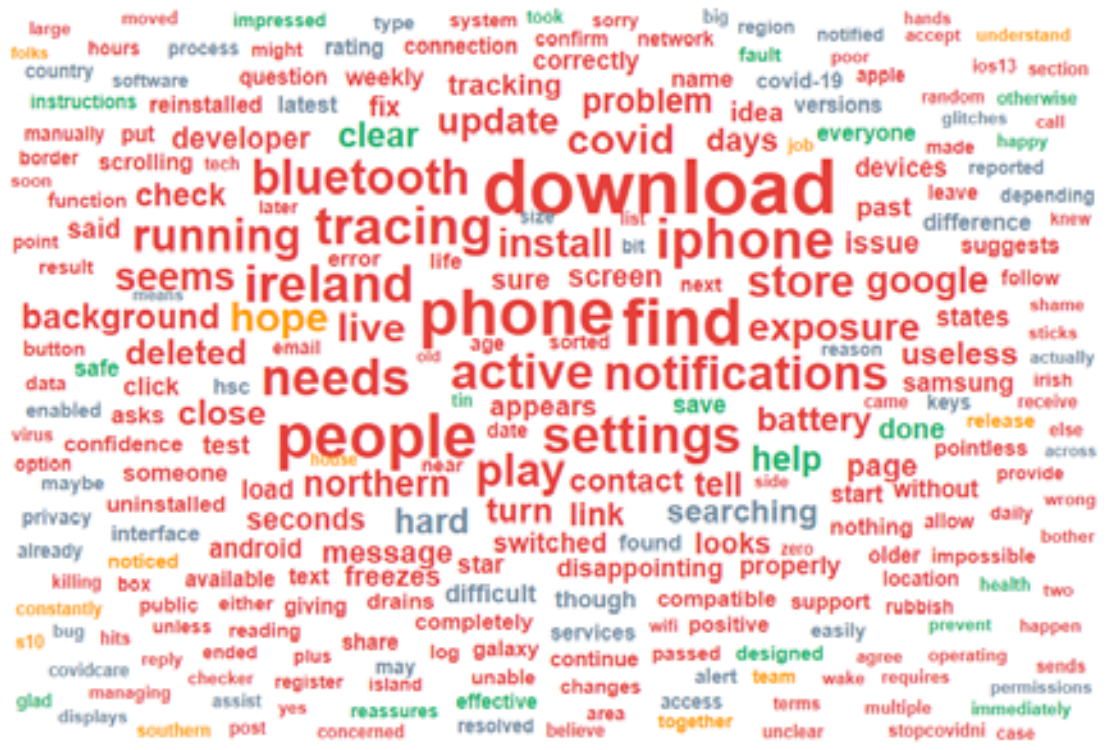

StopCOVID NI

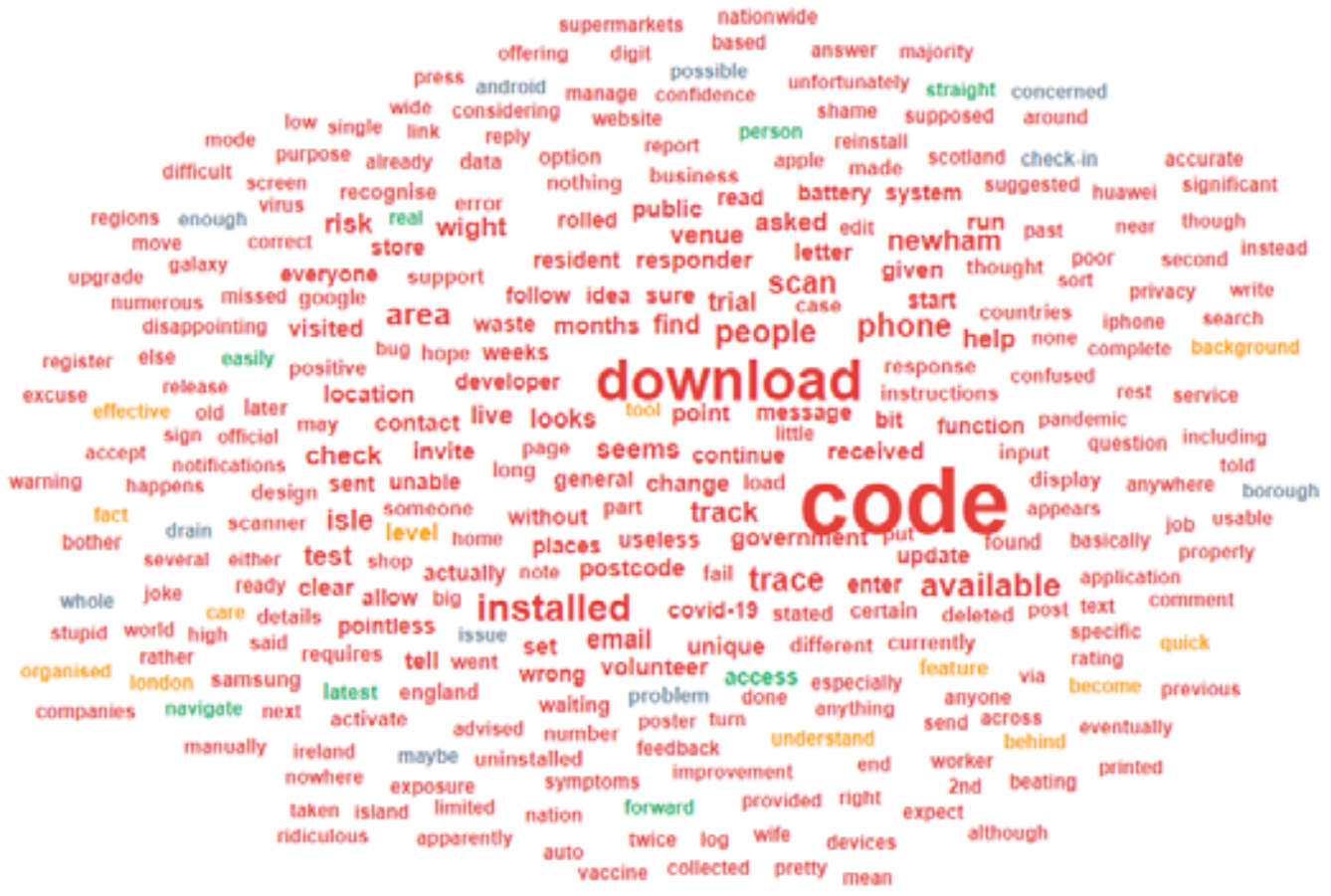

England NHS COVID

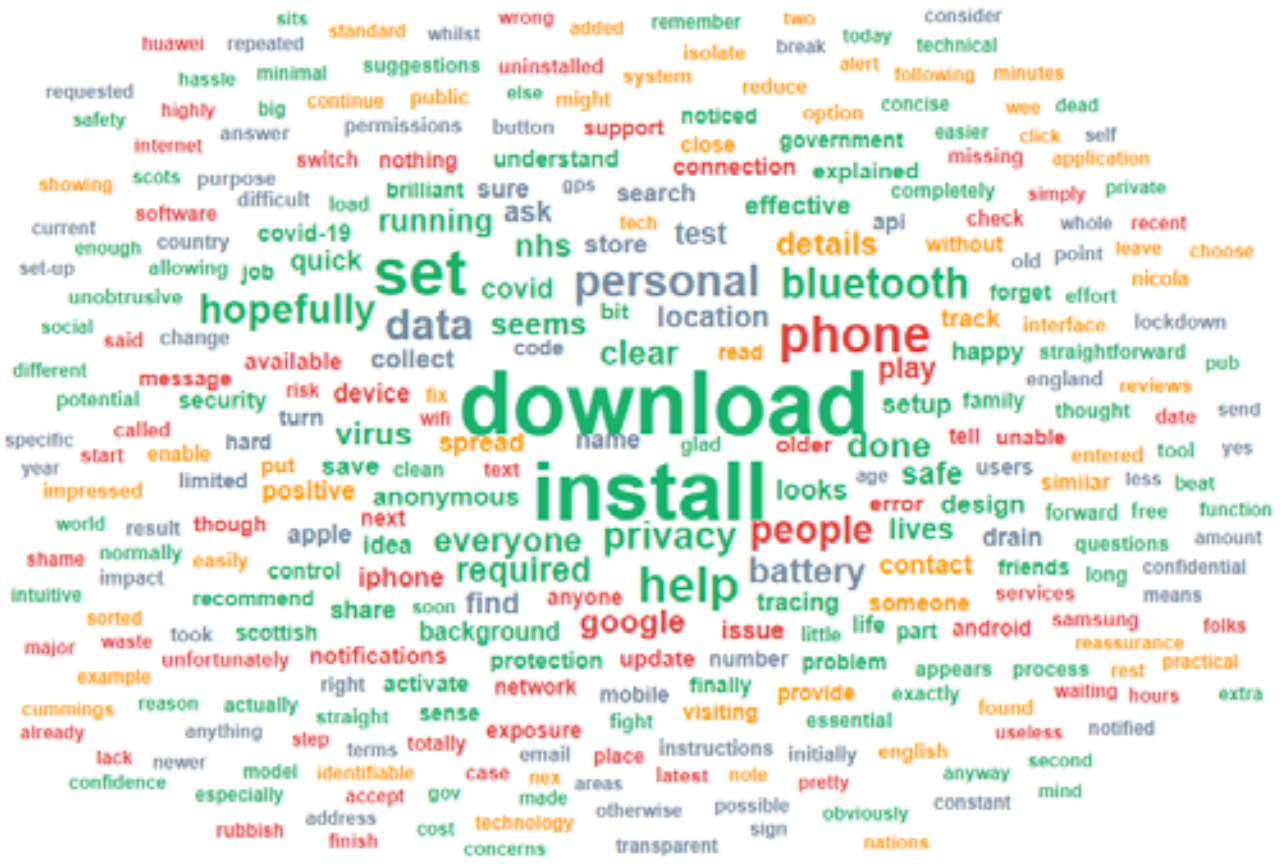

Protect Scotland

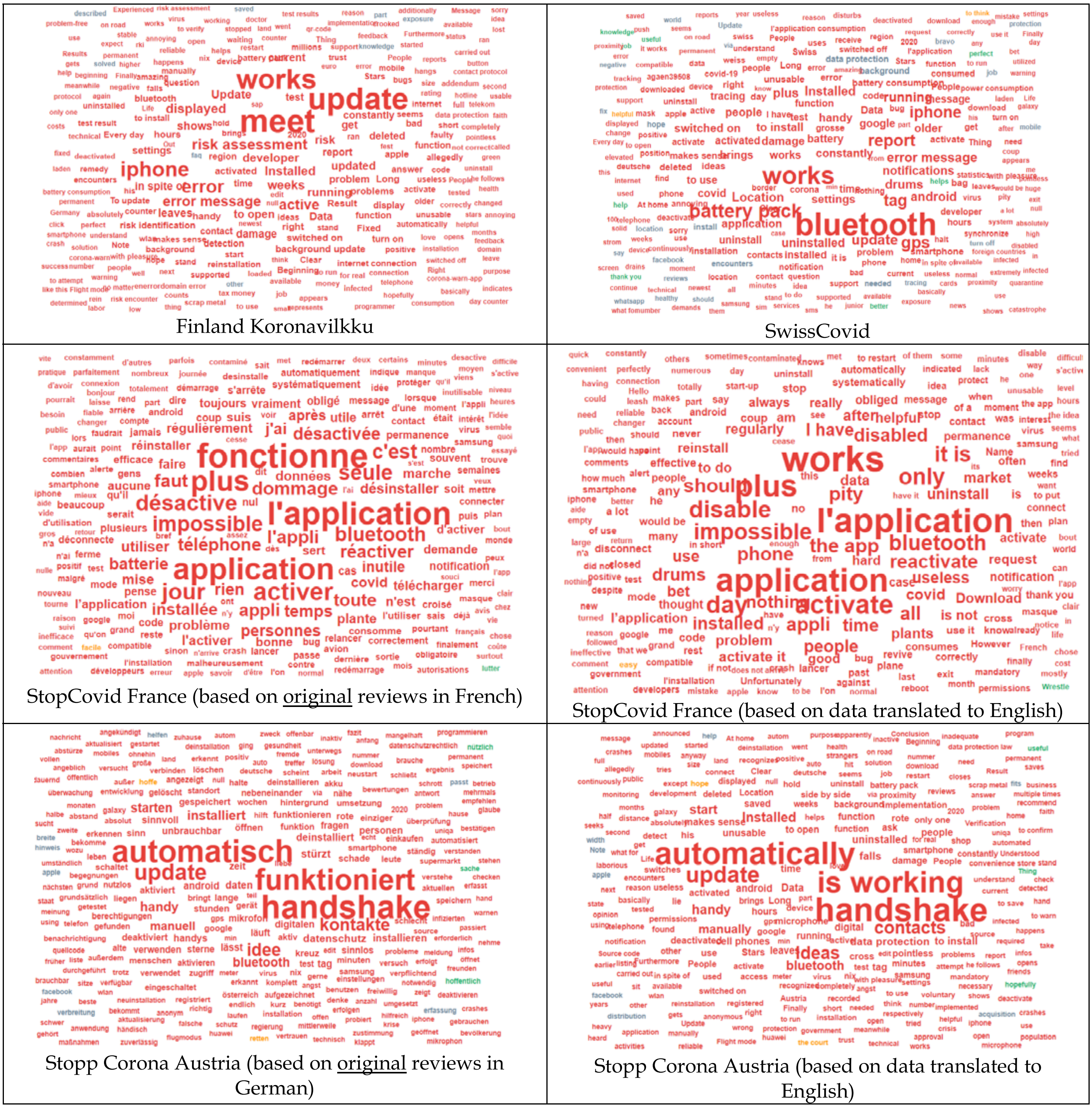

**Figure 9- Word-clouds for the app reviews of all nine apps in which phrases are color codes based on sentiment analysis: positive (green), negative (red), neutral (grey), and mixed (orange),**

One word-cloud in Figure 9 (the second one from the top) belongs to all data: when all the reviews of all apps are analyzed (n=39,425 reviews). This word-cloud shows that the reviews have a negative sentiment towards the functioning ("is working" in the word-cloud) of the German app. Furthermore, there seem to be major issues with the Bluetooth handshake protocol.

In the next three sub-sections, we look at three examples apps (countries) and their specific problems, as reported in user reviews. We select the two apps with the highest number of reviews: the German app (27,744 reviews, combined for both OS apps) and the French app (2,638 reviews). The case of several UK apps is also interesting since UK is a nation with four regions, for which three different apps have been developed: "StopCOVID NI" for Northern Ireland, "NHS COVID-19" app for England and Wales, and "Protect Scotland" for Scotland. We also analyze the case of UK and its apps next.

### 4.3.1 Problems reported about the German app

As visualized in the word-cloud in Figure 9, one of the frequent words with negative sentiments for this app is “*funktioniert*" (German), meaning "works" (in English), which has appeared in 5,550 negative reviews.

As discussed in Section 4.3, there is a useful feature in AppBot: when we click on each word in the cloud, all the reviews containing that word are listed (as shown in Figure 10). To ensure reproducibility of our analysis and for the interested reader, we show in Figure 10 the steps for retrieving the "critical" (negative) reviews in which a certain keyword (“*funktioniert*" in this example) is mentioned, using the AppBot tool. As we can see in Figure 10, the term “*funktioniert*" (German), = "works" (English), has appeared in 5,550 reviews in the time window under study (April-September 2020).

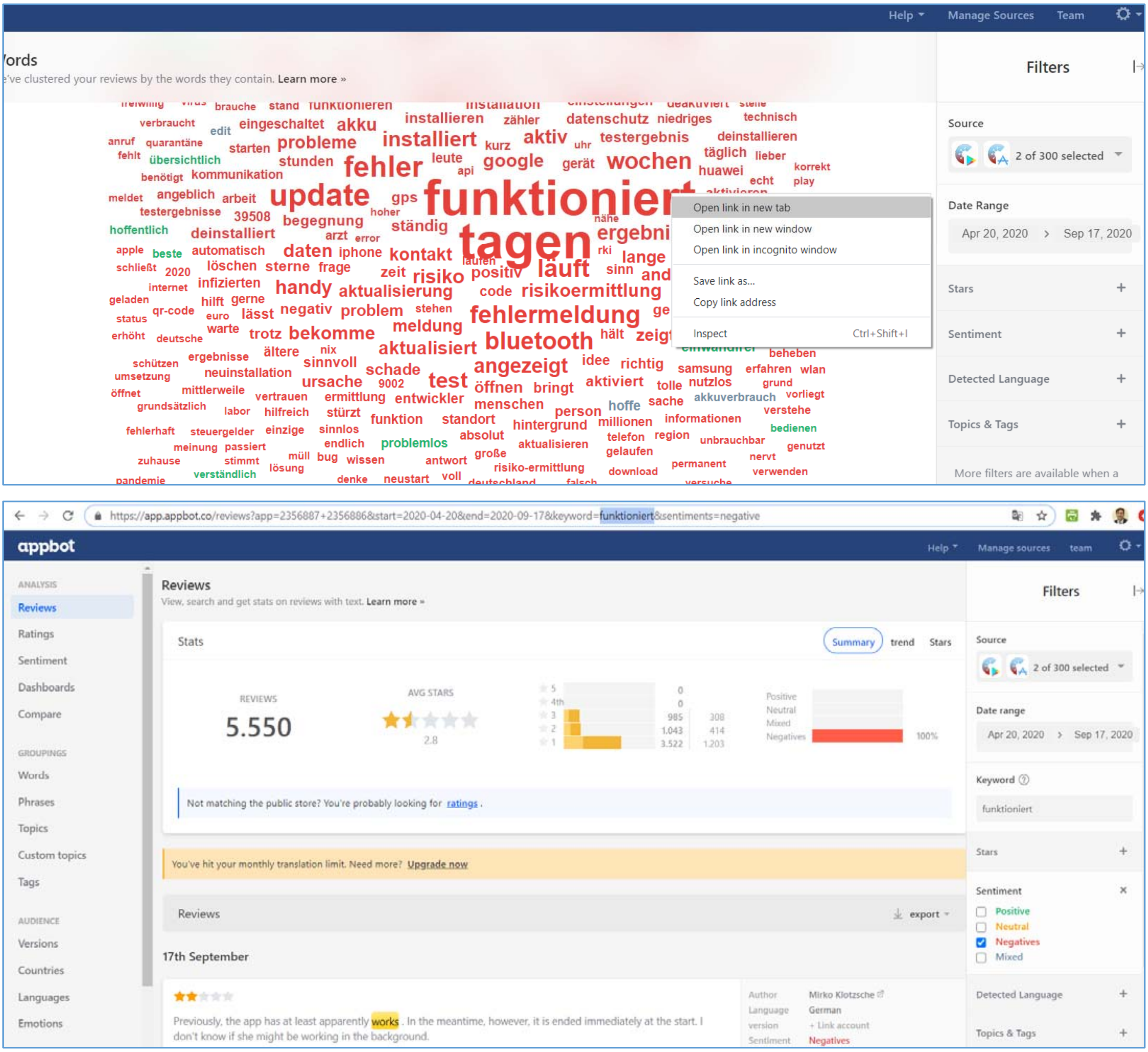

**Figure 10-Retrieving the "critical" (negative) reviews of the German app in which a certain keyword is mentioned, using the AppBot tool**

We looked at a random subset of that large review set (5,550 records) which contained the keyword "works". It turned out that most of the negative reviews with the keyword "works", were conveying the message that the app does not work and were actually a sort of bug reports (two more examples are shown below).

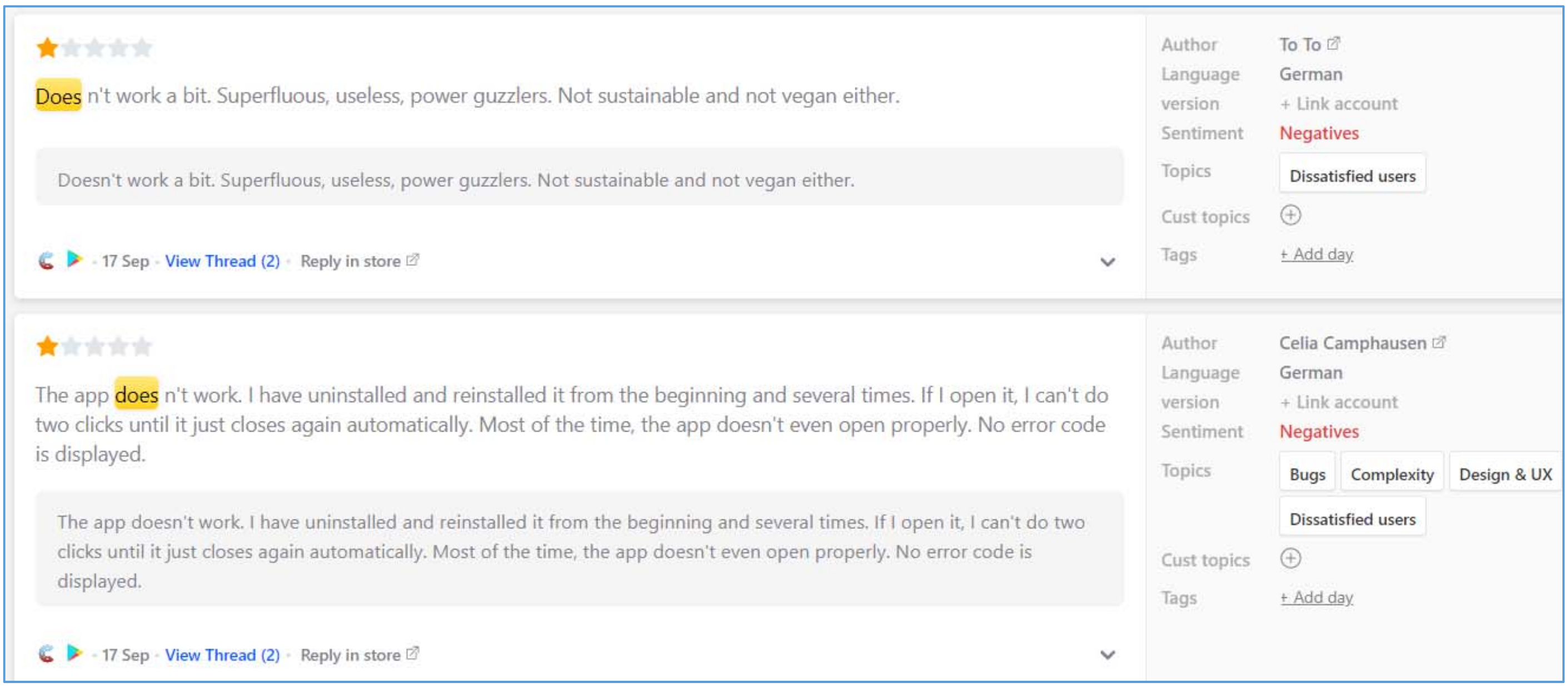

**Figure 11-Two critical (negative) reviews mentioning problems with running the German app**

**Lesson learned/recommendation**: For the German app, a substantial number of reviews are about the app not working, which can be seen as bug reports. But unfortunately, since most users are non-technical people, informative and important components of a bug report (e.g., phone model/version and steps to reproduce the defect) are not included in the review. Thus, it would be impossible for the app's software engineering team to utilize those reviews as bug reports. A recommendation could be that in the app itself (e.g., in its "tutorial" screens), explicit messages are given to the users, asking them that, if they wish to submit bug reports as reviews, they should include important components of a bug report (e.g., phone model/version and steps to reproduce).

We also noticed that many of the reported issues were about the app not working on certain mobile devices. One example was as follows (translated from German by AppBot):

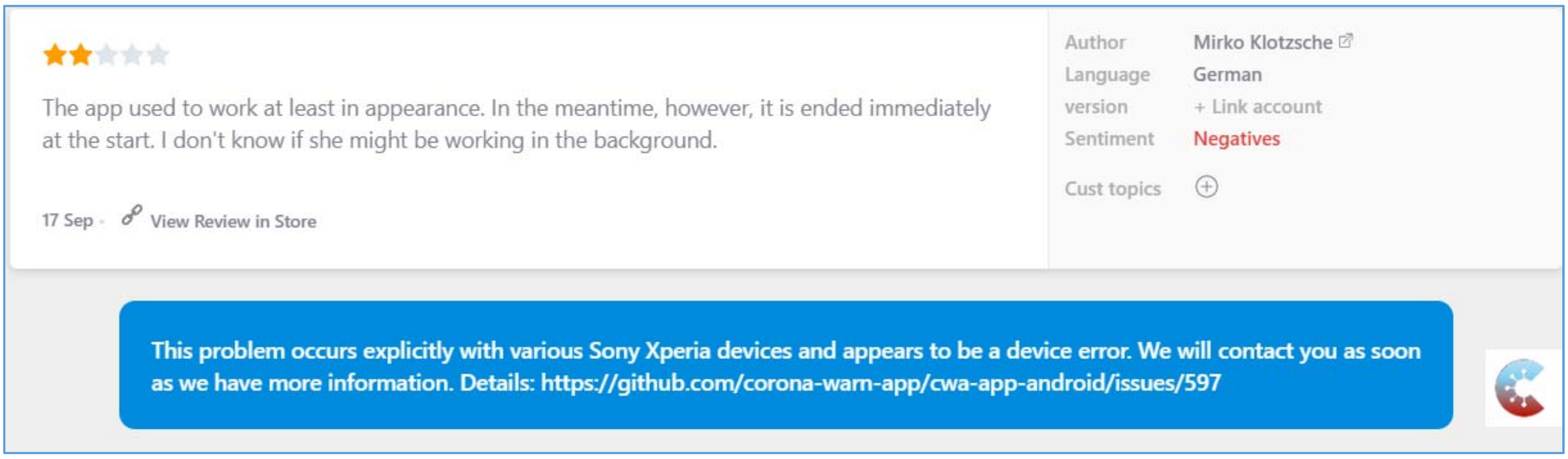

**Figure 12-A critical (negative) review mentioning sporadic (intermittent) app crashes**

This example review, and many other reviews that we looked at, implied occurrence of sporadic (intermittent) app crashes for specific mobile device models. Such challenges are quite common in industry and have been studied in software engineering, e.g., in [43]. We see above that the development team has replied to this review, mentioning that they will contact the user when they have more information, but there is no newer follow-up reply about the solution. It quite be possible that the development team has fixed some of those issues in the upcoming updated versions (patches).

**Lesson learned/recommendation**: A large number of cross-(mobile) device issues have been reported for the German and other apps too. This denotes inadequate cross-device testing of the apps, possibly due to the rush to release the apps to the public. Given the nature of the apps, and since the apps could be installed on any mobile device model/version by any citizen, the development and testing teams should have taken extra care in cross-device development and testing of the apps. There are many sources both in the academic literature [44, 45] and also grey literature[1] on this issue, which the development and testing teams can benefit from.

As visualized in the word-cloud of Figure 9, another frequent keyword within negative sentiments for the German app is "Tagen" (German), which has been translate to "meet" by AppBot (it uses Google Translate), but the correct translation should actually be "days", when we looked at the full sentences in the reviews. This keyword has appeared in 4,465 negative reviews, and three examples are shown in Figure 13. Two of the example reviews indicate that a specific functionality (called "Risk assessment") was not available over several days. The third example in Figure 13 is crash report.

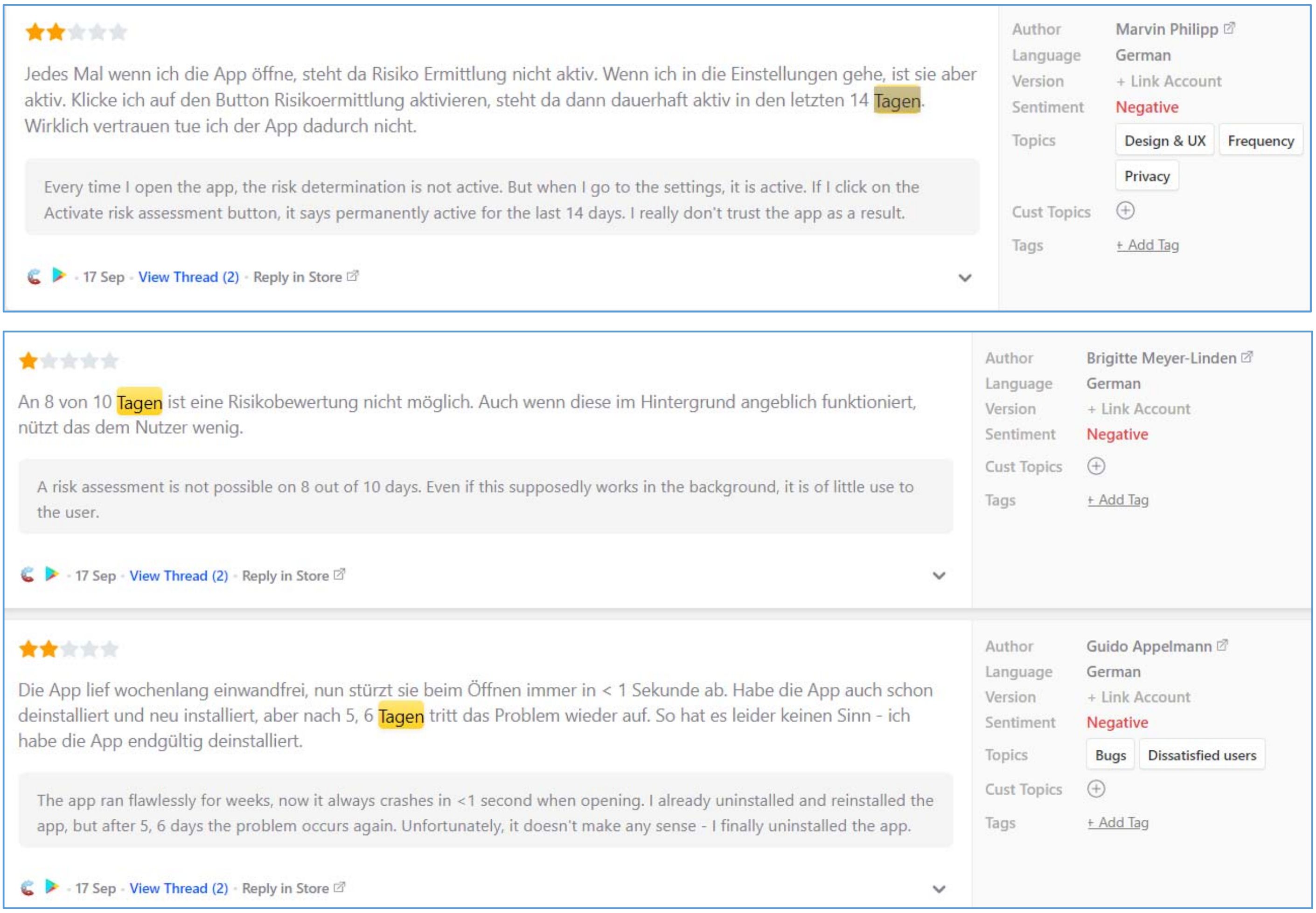

**Figure 13-Several critical (negative) reviews of the German app, mentioning problems with the keyword "days"**

**Lesson learned/recommendation**: We see that, for the German app, a specific functionality (called "Risk assessment") did not work for many users for several days. Such a malfunction usually gives a negative perception to users about an app, even if the other features of the app do work properly. It is thus logical to recommend that app developers should not include a feature in the app release if they predict or see from reviews that the feature does not work for certain users or on certain times/days.

As another issue type, 2,273 negative reviews mentioned the keyword "Bluetooth". Two example reviews from that large set are shown in Figure 14. Both these example reviews are bug reports, but again without important information (e.g., phone model/version and steps to reproduce the defect) to trace and fix the bug.

[1] www.google.com/search?q=movbile+app+%22cross+device%22+testing

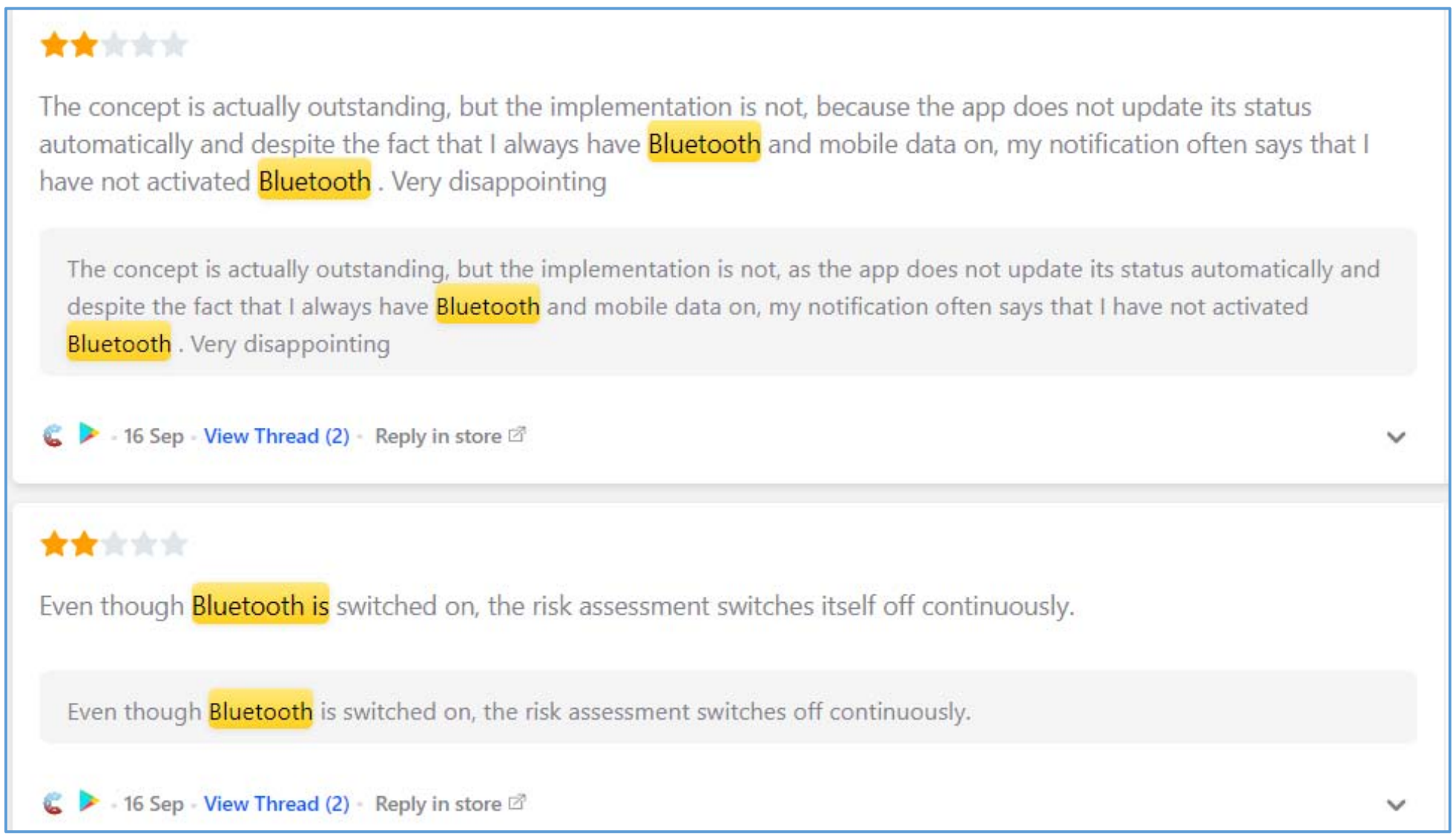

**Figure 14-Two critical (negative) reviews of the German app, mentioning problems with the keyword "Bluetooth"**

As another issue type, 1,264 negative reviews mentioned the keyword "battery" ("Akku" in German). Two example reviews from that large set are shown in Figure 15. Related to this issue, there have been a lot of discussions in the media (such as [1]) and also apps' support pages[2] about the high battery usage. Thus public (users) and media have complained about the issue. In response to this, the Android team has apparently made improvements[3] to the Apple-Google Exposure Notification API, i.e., "*In contrast to Classic Bluetooth, Bluetooth Low Energy (BLE) is designed to provide significantly lower power consumption*".

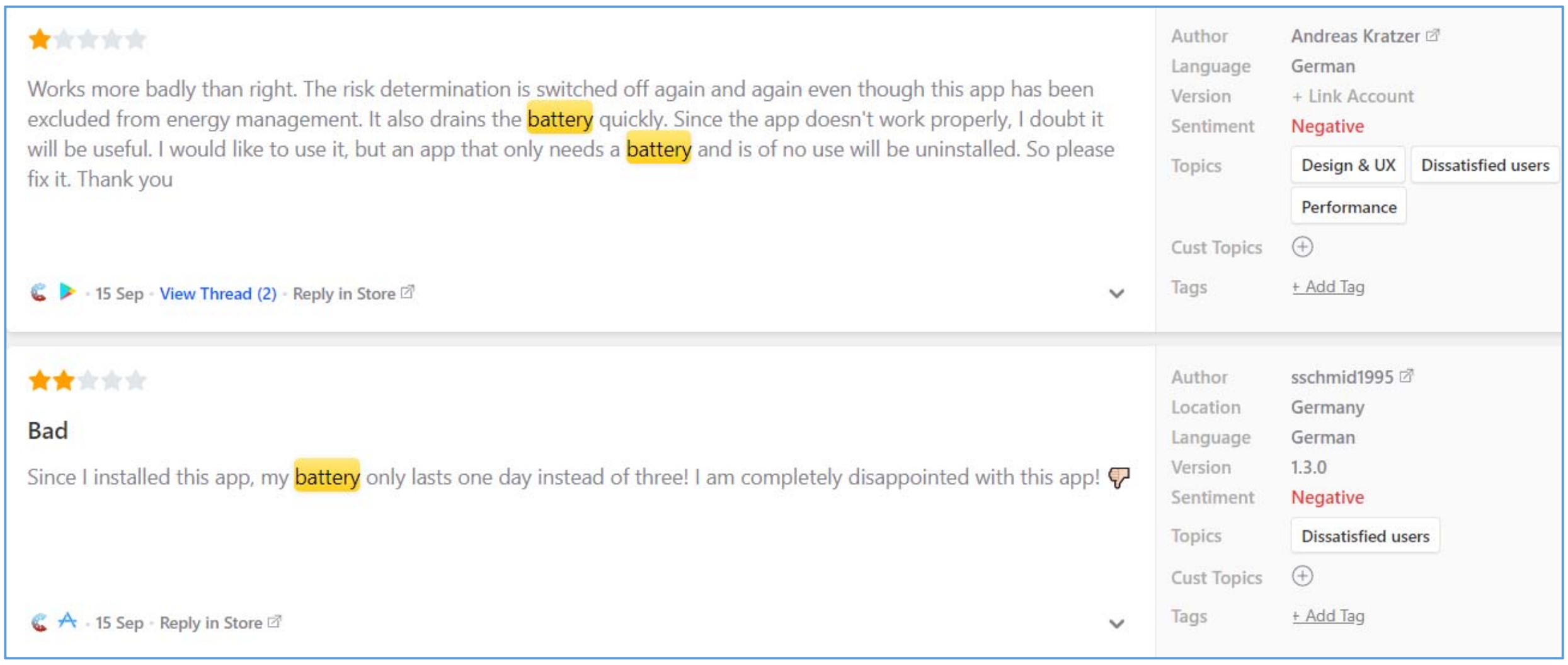

**Figure 15-Two critical (negative) reviews of the German app, mentioning problems with the keyword "battery"**

Many users have reported the battery usage of the apps on their phones, e.g., a YouTube video[4] shows the battery usage screenshot of an iPhone on which the Switzerland's SwissCovid is running. The video showed that, on a time period of 10 days, the app ("Exposure Logging" or "Exposure Notification" service on iPhone) had consumed only 4% of the battery (shown in Figure 16). Also, many bug reports have been filed in the German app's GitHub repository about its high battery usage, e.g.[2], in which screenshots of battery usage have been submitted. We show two of those screenshots in Figure 16 (they are in German, but the usage ratio is clearly understandable), along with a bug report in which the systematic bug report items have been provided, e.g., Describe the bug, Expected behavior, Steps to reproduce the issue.

[1] www.hitc.com/en-gb/2020/09/25/nhs-covid-19-app-battery-usage-explained-will-the-app-drain-your-battery
[2] github.com/corona-warn-app/cwa-app-ios/issues/671
[3] developer.android.com/guide/topics/connectivity/bluetooth-le
[4] www.youtube.com/watch?v=yEXiH4UIvSk

The first author of the paper also installed the StopCOVID NI app on his iPhone, and let it run for more than a week non-stop. We provide a screenshot from the battery usage screen of his phone in Figure 16, in which the "Exposure Notification" service has consumed only %4 of the battery, which we consider a reasonable power consumption (not high). But we should mention that he moved out of his home a few times only during that week, and he barely came close to anyone. Thus, the app did not have to exchange information with other uses who had the StopCOVID NI app on their phones.

We also found some discussions[1] in an online forum about the UK NHS app, in which one user mentioned: "*Our IT director reported massive battery drain after installing the app as in 70% to 15% on his journey home (three trains). I wonder if the drain comes from the amount of contacts you have with other people - if you are sat at home it has no contacts to ping but on trains etc. there might be hundreds*". Thus, just like any other mobile app, we can observe that the type of usage and movement of the user in different environments could indeed impact the battery usage of the app.

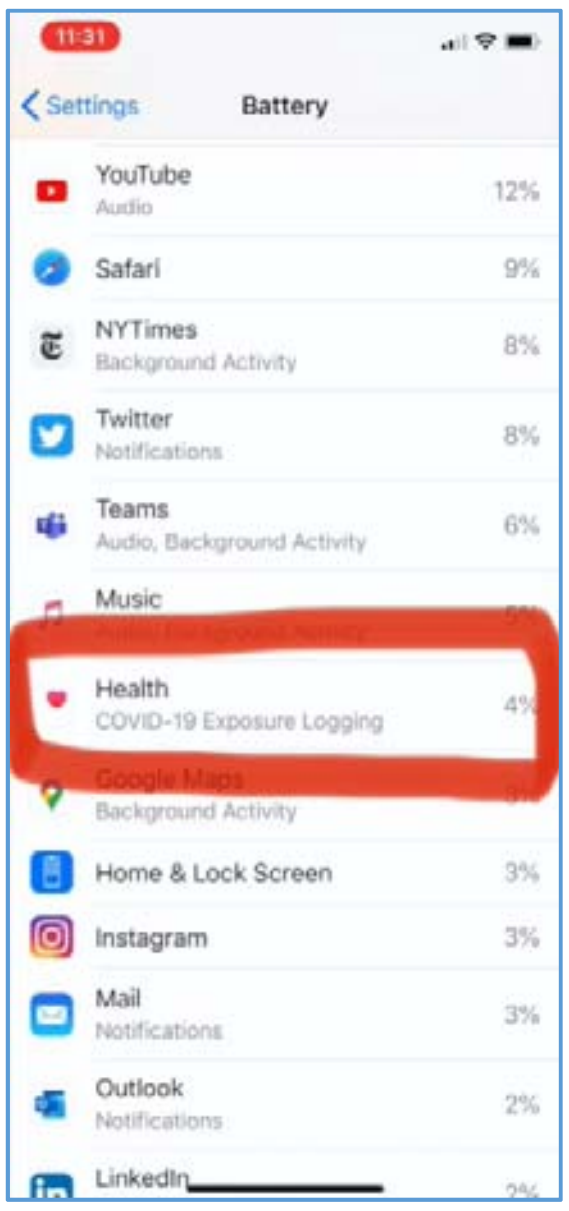

App: SwissCovid
[youtube.com/watch?v=yEXiH4UIvSk](youtube.com/watch?v=yEXiH4UIvSk)

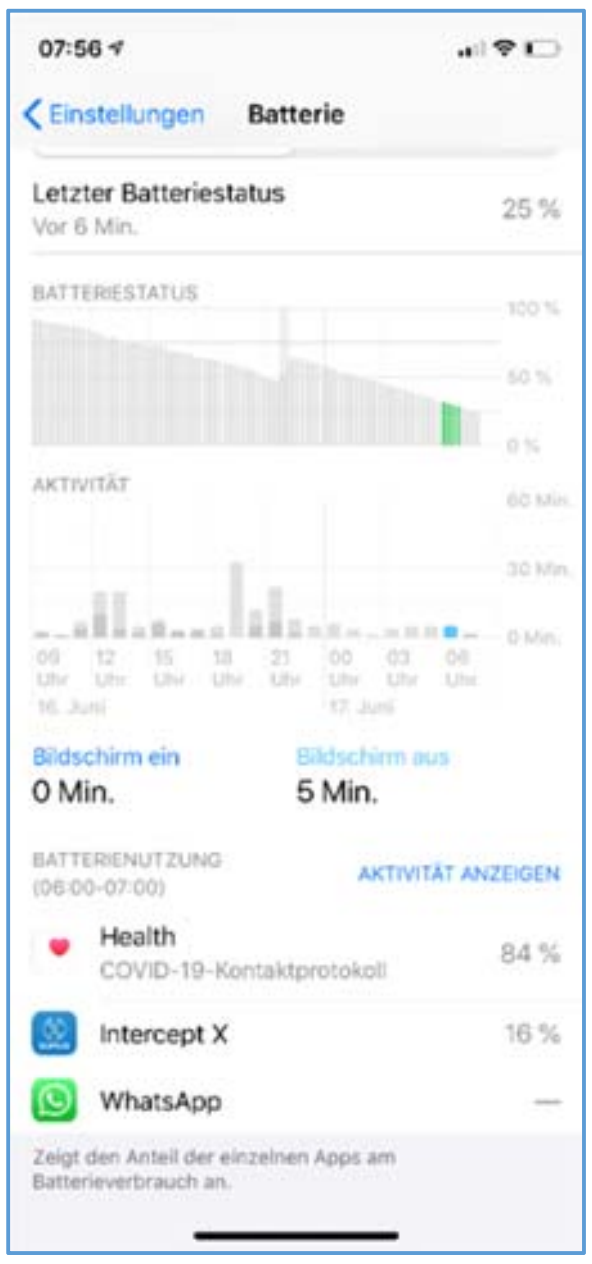

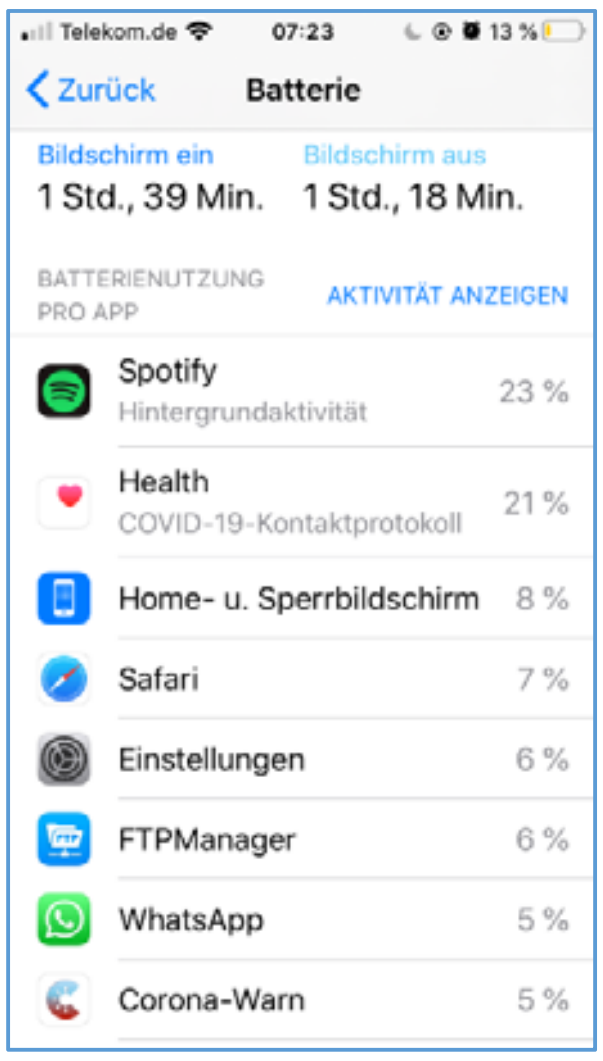

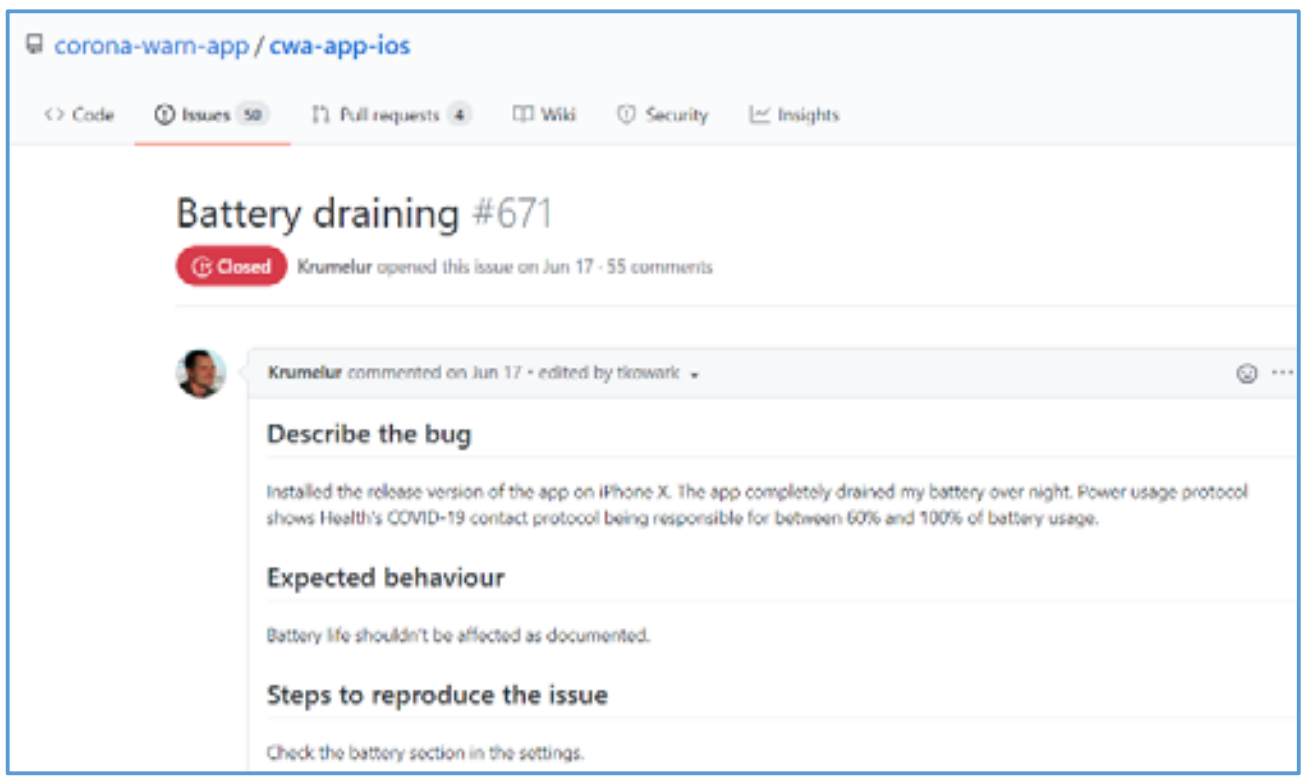

App: Corona-Warn Germany. Source: [github.com/corona-warn-app/cwa-app-ios/issues/671](github.com/corona-warn-app/cwa-app-ios/issues/671)

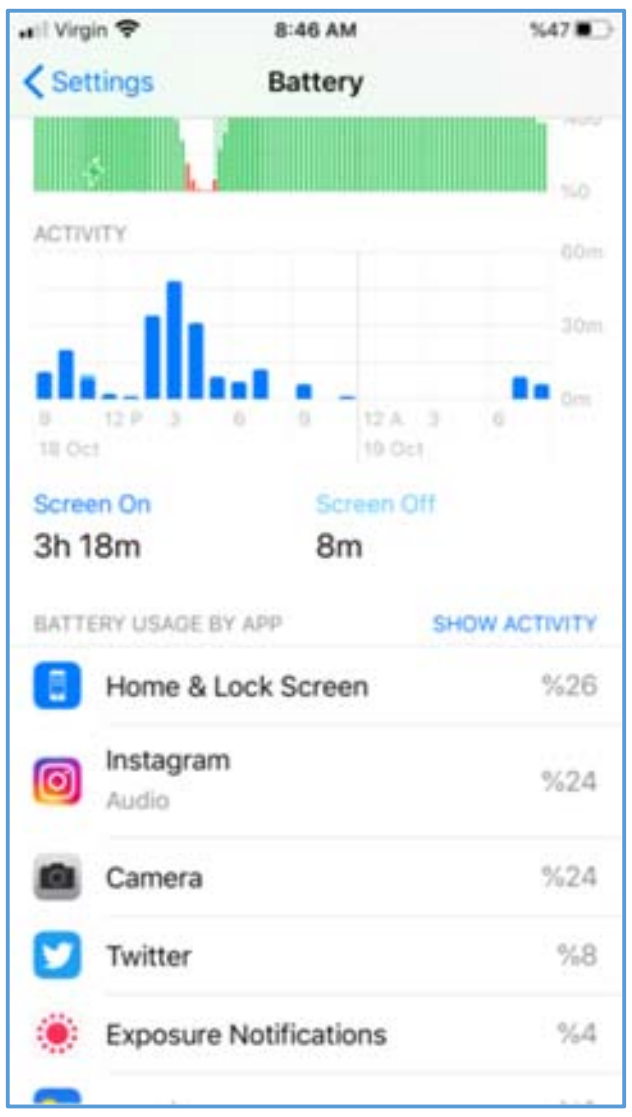

Screenshot from the first-author's iPhone, on which the StopCOVID NI app was running for more than a week non-stop

**Figure 16-Four screenshots and one bug report submitted by users about battery usage of the apps**

### 4.3.2 Problems reported about the French app

As visualized in the word-cloud of Figure 9, for the French app, two of the frequent words with negative sentiments for this app are “application” and “l'application”, referring to “app”, which are trivial terms. When translated to English, the other frequent terms are "activate”, “install”, and “Bluetooth”. We discuss a small randomly-chosen subset of those reviews next.

[1] [se23.life/t/nhs-track-and-trace-mobile-app/15699/24](se23.life/t/nhs-track-and-trace-mobile-app/15699/24)

237 reviews critical reported problems with the keyword "activate" for the French app. Two example reviews from that set are shown in Figure 17. We also include the permanent links to the reviews, for traceability. These users have reported serious problems with activating the app, which is unfortunate.

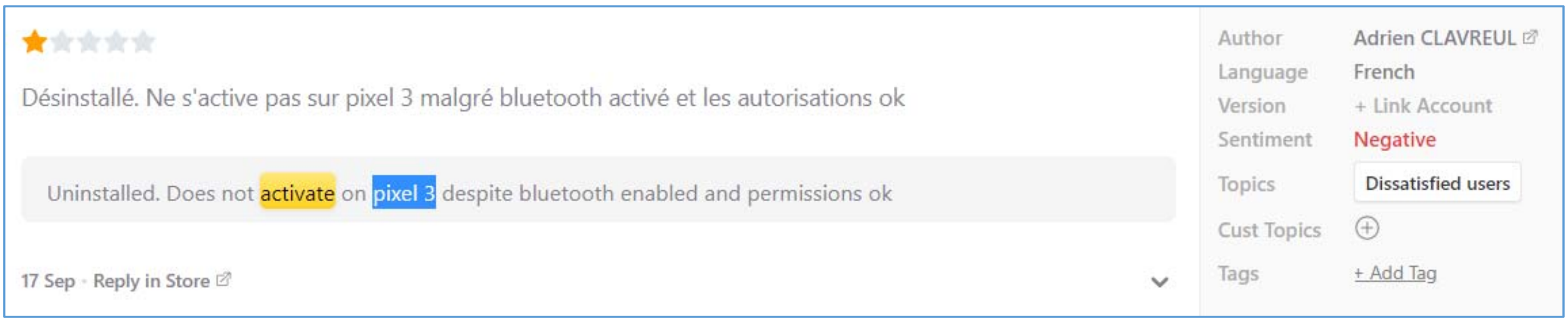

appbot.co/apps/2370437-tousanticovid/reviews/1950620865

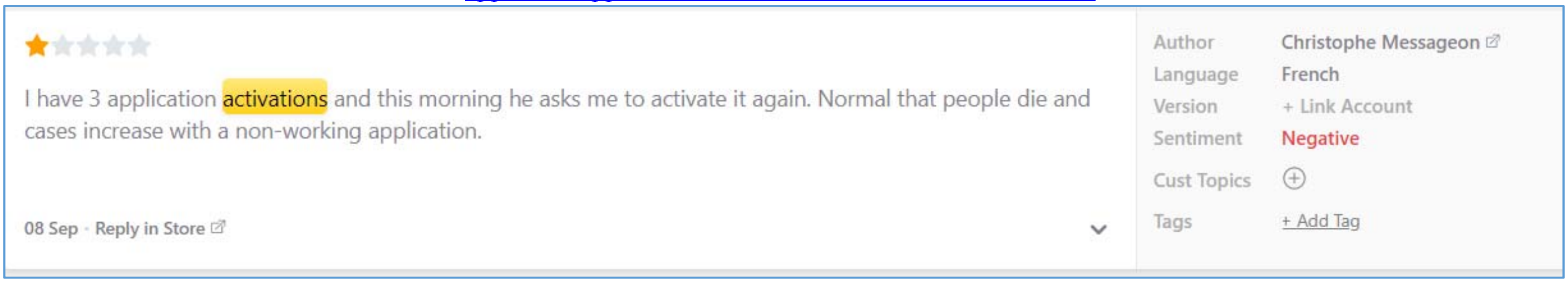

appbot.co/apps/2370437-tousanticovid/reviews/1937553799

**Figure 17- Two critical reviews of the French app, mentioning problems with the keyword "activate"**

**Lesson learned/recommendation**: We are seeing rather trivial issues in the apps, i.e., users have to “activate” multiple times, instead of just once. We would have hoped that the test teams of the apps had detected and fixed those trivial issues before release.

130 reviews critical reported problems with the keyword "install" for the French app. Two example reviews from that set are shown in Figure 18.

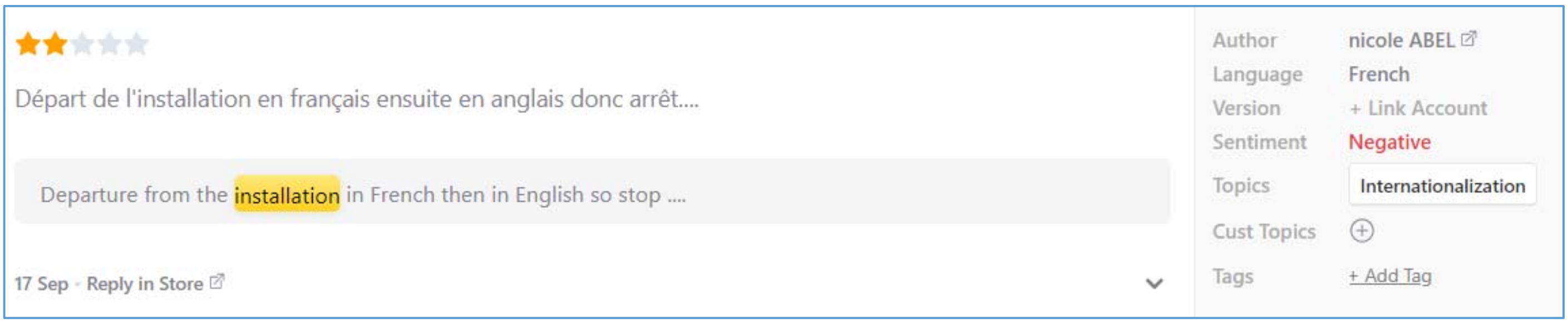

appbot.co/apps/2370437-tousanticovid/reviews/1950621276

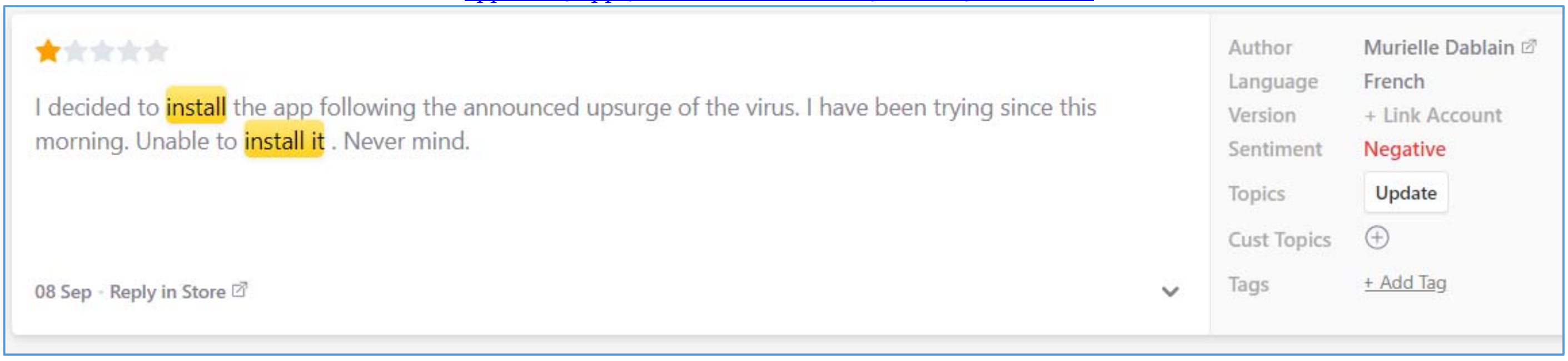

appbot.co/apps/2370437-tousanticovid/reviews/1937553584

**Figure 18- Two critical reviews of the French app, mentioning problems with the keyword "installation"**

**Lesson learned/recommendation**: The first example review of Figure 18 denotes issues w.r.t. internationalization (language settings). It is important that a given app automatically switches to the home country’s language since some non-English users will feel odd if they see a sudden switch from their native language to English in the app’s GUI.

Exactly 300 critical reviews reported problems with the keyword "Bluetooth" for the French app. Two example reviews from that set are shown in Figure 19. The first example review is about high battery drainage of Bluetooth, like all other apps in the pool.

The second example review below is about incompatibility of the app on old phones. The second example review also raised an important issue: a large ratio of elderly are known to not have latest smartphones or even how to install and use apps like these on their phones. In fact, a paper has been published on this very subject, entitled: "*COVID-19 contact tracing apps: the 'elderly paradox'*" [14].

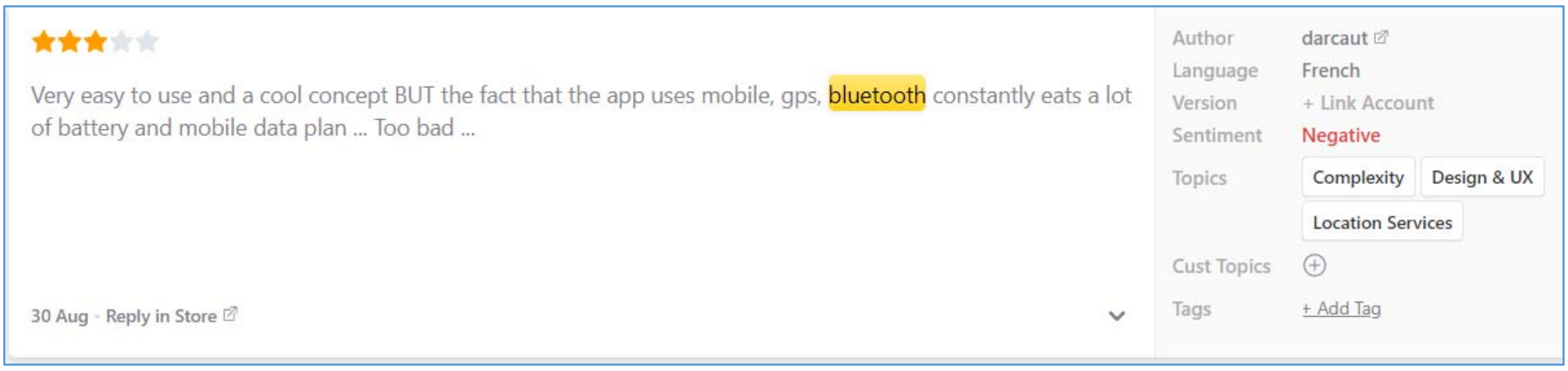

appbot.co/apps/2370437-tousanticovid/reviews/1923403135

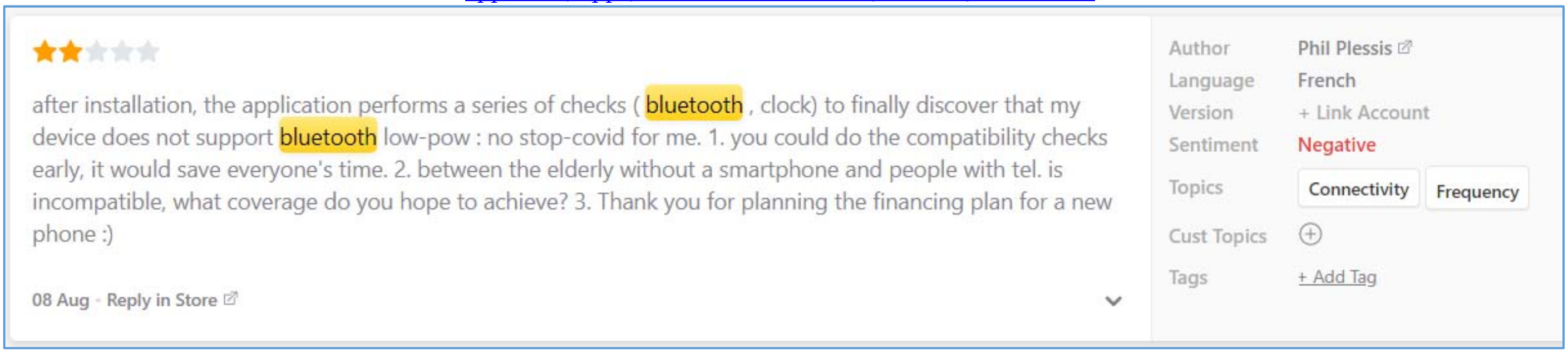

appbot.co/apps/2370437-tousanticovid/reviews/1896028609

**Figure 19- Two critical reviews of the French app, mentioning problems with the keyword "Bluetooth"**

**Lesson learned/recommendation**: High battery drainage of Bluetooth has also been reported for the French app.

### 4.3.3 Problems reported about the three apps in the UK

For the four regions of the UK, three apps have been developed: *NHS COVID-19* for England and Wales, *StopCOVID NI* for Northern Ireland, and *Protect Scotland* for Scotland. We review next a subset of the common problems reported for all three, and then review a subset if issues reported for each of them.

Common problems reported for all three apps:

One major issue reported by users is the lack of "interoperability" between the apps, i.e., if a user from one region, using that region's app, visits another part of the UK, the app will not record the contact IDs in the new region and in case of entering a positive COVD result, the app will not notify those contacts. This issue has been reported in a large number of reviews, e.g.:

- *"Complete and utter waste of space. Only works if I come into contact with someone else using the same backstreet application, who has managed to get tested without being turned away, and inputs a code into their app. If I bump into someone from England, Wales, Ireland, or anywhere else for that matter with COVID-19 then this app does diddly squat - What's the point??"*[1]
- *"it's not linked to apps used in other parts on the UK, again a missing feature."*[2]

[1] appbot.co/apps/2437310-protect-scotland/reviews/1957885880/ or directly in Google Play store: play.google.com/store/apps/details?id=gov.scot.covidtracker&hl=en_GB&reviewId=gp%3AAOqpTOHVDLw1vCASIapQQmiyem1xyhXCw4SQBcOPdRXy0v1YPz95_hcZ5CUz7kWFe8004v5TbARTEjYyuEJoGw

[2] appbot.co/apps/2437310-protect-scotland/reviews/1961498492

- *"Live in Scotland and work in England. Only one app will work at a time. Do I choose the NHS Covid or Protect Scotland version!!"*[1]

Also a number of users, understandably, compared the features of the three apps, and complained about the case of a given app, not having the feature provided by another UK-based app. An example review:

- *"Looks great, easy to use but oh how they missed out some useful features such as a [NHS] Covid-19 [app's] alert state notifier, scanning business QR Codes, etc. so user's data needn't be handed over in pubs, etc."*[2]

Many users reported having problems installing the apps, e.g., 10 reviews of the 573 *Protect Scotland Android* app. Some of those installation problems were due to having older phone models, but we still saw several reviews reporting newer (phone models not being able to install the apps, e.g., *"Waste of time have tried to install numerous times got the very latest Samsung S20 and it doesn't install on my phone"*[3]. This raises the issue of the development team not doing adequate installation testing of the app using latest phone models. There are indeed many advanced commercial testing tools on the market, e.g., Testinium (testinium.com) to conduct that testing efficiently. As discussed in Section 2.3, the first author of the current paper served as a consultant to the development team of the StopCOVIDNI app and conducted an inspection of test plans and test cases of the app. One of the comments that he had made was indeed installation testing of the app on multiple phone models using such test tools.

Another common issue that we noticed for the UK apps was the lack of response by apps' development teams to almost all reviews in the app store (only the Google Play store allows replies to reviews). This was in contrast to the case of some other apps, e.g., the German app, whose development team has been proactive in replying and communicating with users directly via the review threads.

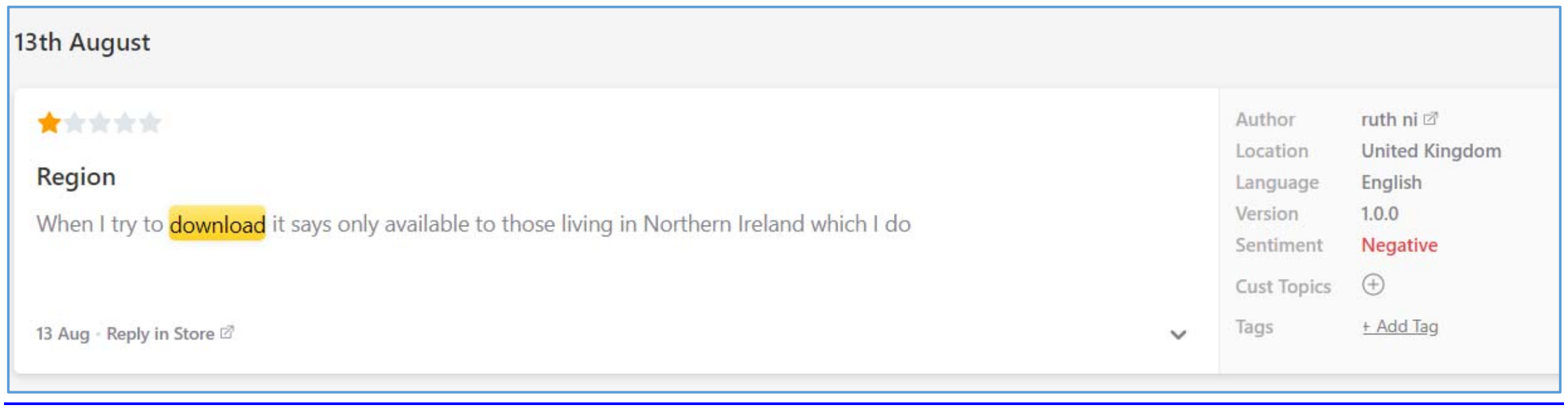

[1] appbot.co/apps/2437310-protect-scotland/reviews/1960556725
[2] appbot.co/apps/2437310-protect-scotland/reviews/1961498492
[3] appbot.co/apps/2437310-protect-scotland/reviews/1960556689

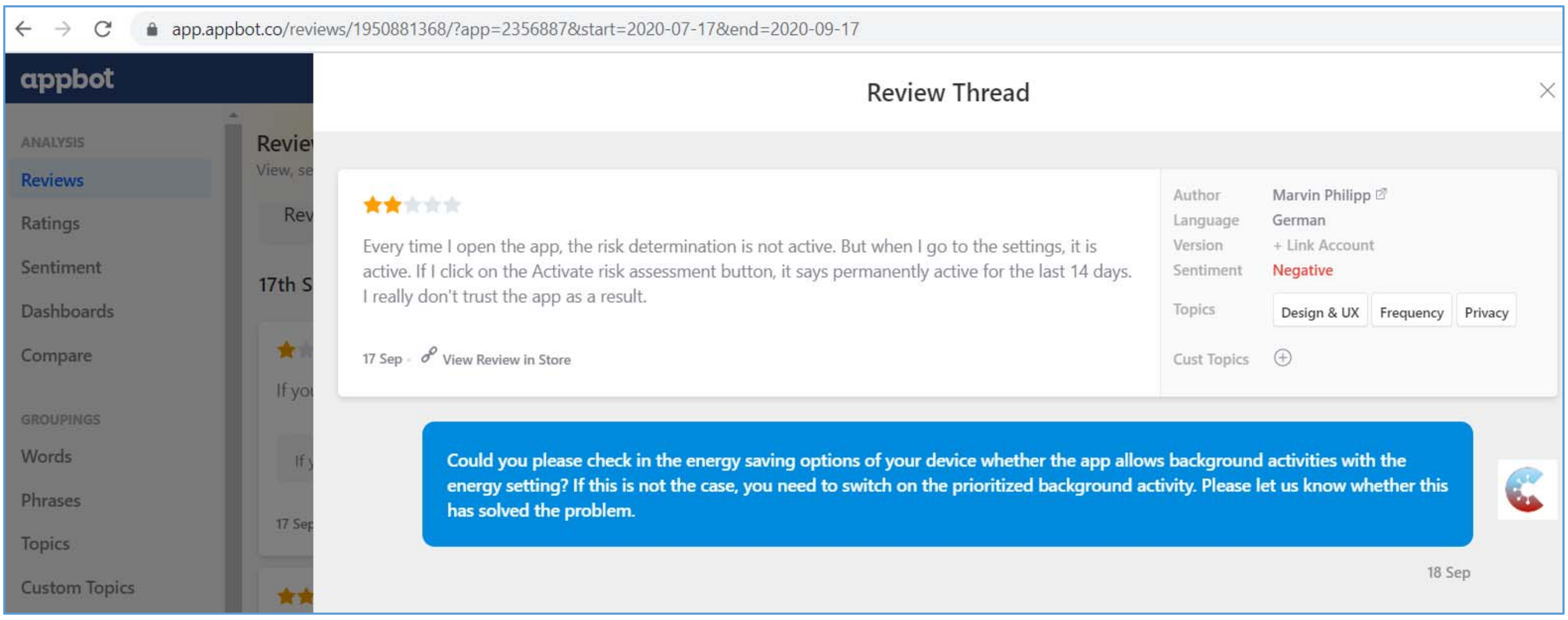

**Figure 20-Comparing a review[1] without any reply from the development team for the StopCOVID NI app and a review[2] with a reply from the development team for the German app**

**Lesson learned/recommendations**: The development team of all apps should be proactive in replying to user reviews, and filtering informative reviews and getting more information (e.g., steps to reproduce the defects/problems) from them, e.g., by direct replies to the reviews in app stores.

*StopCOVID NI* app:

As visualized in the word-cloud in Figure 9, one of the frequent words with negative sentiments for this app is "notifications" which has appeared in 12 negative reviews, e.g.:

- *"Want to get people to uninstall it? Don't produce audible notifications you haven't been exposed this week at 6am on a Fri morning, waking people up"*[3]
- *"I am getting a warning that exposure notifications may not work for the area I am in. As this is Northern Ireland I am unclear why it is saying this. The exposure log does not appear to have made any checks since early August. This does not give confidence that the app is working properly. I do hope the designers are reading these reviews as this appears to be a recurring issue"*[4]
- *"I was keen to install and safe. I have iphone7 with latest update. And like so many others, I get the exposure notification error. Can't select to turn on exposure notifications. Useless. Very disappointing"*[5]
- *"Hopefully the app does what it says on the tin - but I get another error message that says "Exposure Notifications Region Changed", followed by "COVID-19 Exposure Notifications may not be supported by "StopCOVID NI" in this region. You should confirm which app you are using in Settings." I am using an iPhone 11 Pro Max running iOS 13.5.1. ... so I have no confidence that the app is working properly at present."*[6]

**Lesson learned/recommendations**: There seems to be rather trivial usability issues with some of the apps (e.g., the case of exposure notification errors in the NI app). This raises the question of the inadequate usability testing of the apps and possibility of releasing them on a "rush".

Another frequent word with negative sentiments for this app is "download", which appeared in 16 negative reviews, e.g.:

- *"When I try to download it says only available to those living in Northern Ireland which I do"*[7]
- *"The app just tells me I need to live in NI to use it. I do. I deleted and downloaded again. Same problem"*[8]

---

[1] appbot.co/apps/2392818-stopcovid-ni/reviews/1950428670
[2] appbot.co/apps/2356887-corona-warn-app/reviews/1950881368
[3] appbot.co/apps/2392818-stopcovid-ni/reviews/1950428641
[4] appbot.co/apps/2392818-stopcovid-ni/reviews/1950428665
[5] appbot.co/apps/2392818-stopcovid-ni/reviews/1950428700
[6] appbot.co/apps/2392818-stopcovid-ni/reviews/1950428741
[7] appbot.co/apps/2392818-stopcovid-ni/reviews/1950428670
[8] appbot.co/apps/2392818-stopcovid-ni/reviews/1950428675

- *"I just downloaded this and used the 'share this app' function to all my contacts in N Ireland and the link doesn't work!! Not a good start for the app and doesn't build my confidence that any other part of the app works! I am now getting multiple messages from people asking what is the link for. Very disappointing"*[1]

Some randomly-sampled negative review were:

- *"Tried to download on an elderly relative's Samsung phone but the app isn't compatible. Nowhere can I find a list of compatible devices or Android versions. Sadly the app won't help the most vulnerable"*[2]
- *"What is the point of urging people to install this app to stop the spread of covid 19 yet when the app is not working for some people the developers don't even bother to fix or reply to the email that they ask people to send if there is a problem. Google tried their best to resolve the matter immediately and also notified the developer yet 5 days past and nothing"*[3]
- *"The app does not seem to work correctly unless automatic battery optimisation is switched to manually allow app to run in background. Settings -> Battery -> App launch -> StopCOVID NI. Might also be under Applications -> StopCOVID NI -> Battery Optimisation depending on version. Once I switched this I went from 5 checks over 10 days to 8 checks in a single day"*[4]
- *"As others have said, the app does not properly run in the background as intended - the app needs to be open and the phone unlocked. Good idea in theory, however poor execution, going forward this app will be useless without correction."*[5]

<u>*NHS COVID-19* app:</u>

As visualized in Figure 9, one of the frequent words with negative sentiment for this app is "code" which has appeared in 153 of the 341 negative-sentiment reviews for this app. This phrase does not refer to source code, but to a QR code which is used in the app (see the real photo example of a restaurant with the QR code in Figure 21). There were a great number of criticisms, and the followings are only some examples:

- *Well, as a business we are directed to register for track and trace. Having registered for a QR code and subsequently printed said code. I thought, in good naval tradition, 'Lets give it a test before we put the poster up'. So download the app from Play Store. Scanned the code and a message pops up 'There is no app that can use this code'. Next move, open the application. What do we find!! Currently only for NHS Volunteer Responders, Isle of White and Newham residents'. What is the point of publicizing this if it does not have basic functionality? Measure twice cut once.... MrX Also there should be an option for no Star as it appropriate for this application!*[6] → Poor alignment of publicity timing
- *QR location doesn't seem to work for me. Used a standard QR reader on my phone and it took me straight to venue but the QR reader in the app said QR code not recognized.*[7] → Poor testing of that module software
- *It is not working. My daughter has been working within 2 meters of someone all week who tested positive yesterday. She's had no notification and they both have the app. The other person has put the code in she received at 5pm yesterday and daughters had no alert.*[8] → Raises serious concerns about efficacy and quality of the software
- *After months in the planning, this app has been such a let down. At first I was receiving notifications, but when I clicked to open and read them, they disappeared. Last week I developed symptoms and updated the app. My husband has the app and, 6 days later he still hasn't received a notification that he's been near someone with symptoms. Yesterday I updated this app with my positive test result code, and still - nothing on my husband's phone. I have no confidence in this app at all!*[9] → Raise serious concerns about efficacy and quality of the software
- *"I move about a lot with my job. I can't change the post code to the area I'm in unless I uninstall the app"*[10] → The need for better software requirements engineering
- *Keeps asking for post code but doesn't give any where to put it in*[11] -> Poor usability and UX/UI design

---

[1] appbot.co/apps/2392818-stopcovid-ni/reviews/1950428858
[2] appbot.co/apps/2436851-stopcovid-ni/reviews/1950429102
[3] app.appbot.co/apps/2436851-stopcovid-ni/reviews/1950429005/
[4] appbot.co/apps/2436851-stopcovid-ni/reviews/1950429052
[5] appbot.co/apps/2436851-stopcovid-ni/reviews/1950429071
[6] appbot.co/apps/2411517-nhs-covid-19/reviews/1951839432
[7] appbot.co/apps/2411517-nhs-covid-19/reviews/1994520039
[8] appbot.co/apps/2411517-nhs-covid-19/reviews/1993224335
[9] appbot.co/apps/2411517-nhs-covid-19/reviews/1993228781
[10] appbot.co/apps/2411517-nhs-covid-19/reviews/1993233064
[11] appbot.co/apps/2411517-nhs-covid-19/reviews/1991681430

- *I booked a test through the app. Test is negative. SMS and Email with negative result. No code in SMS or email. App still counting down isolation. I don't want to go out with phone and potentially give false positives to loads of people.*[1] -> Poor integration of the software in the healthcare (business) processes
- *This app requires Bluetooth and location to be on all the time, goodbye battery life. Some qr codes won't scan and no manual input option as a fall back. The alerts are useless as well, it tells you that someone in my area has been confirmed to have Covid, but doesn't tell you if you had been to a place they checked in, so is a pointless notification*[2] -> Poor quality

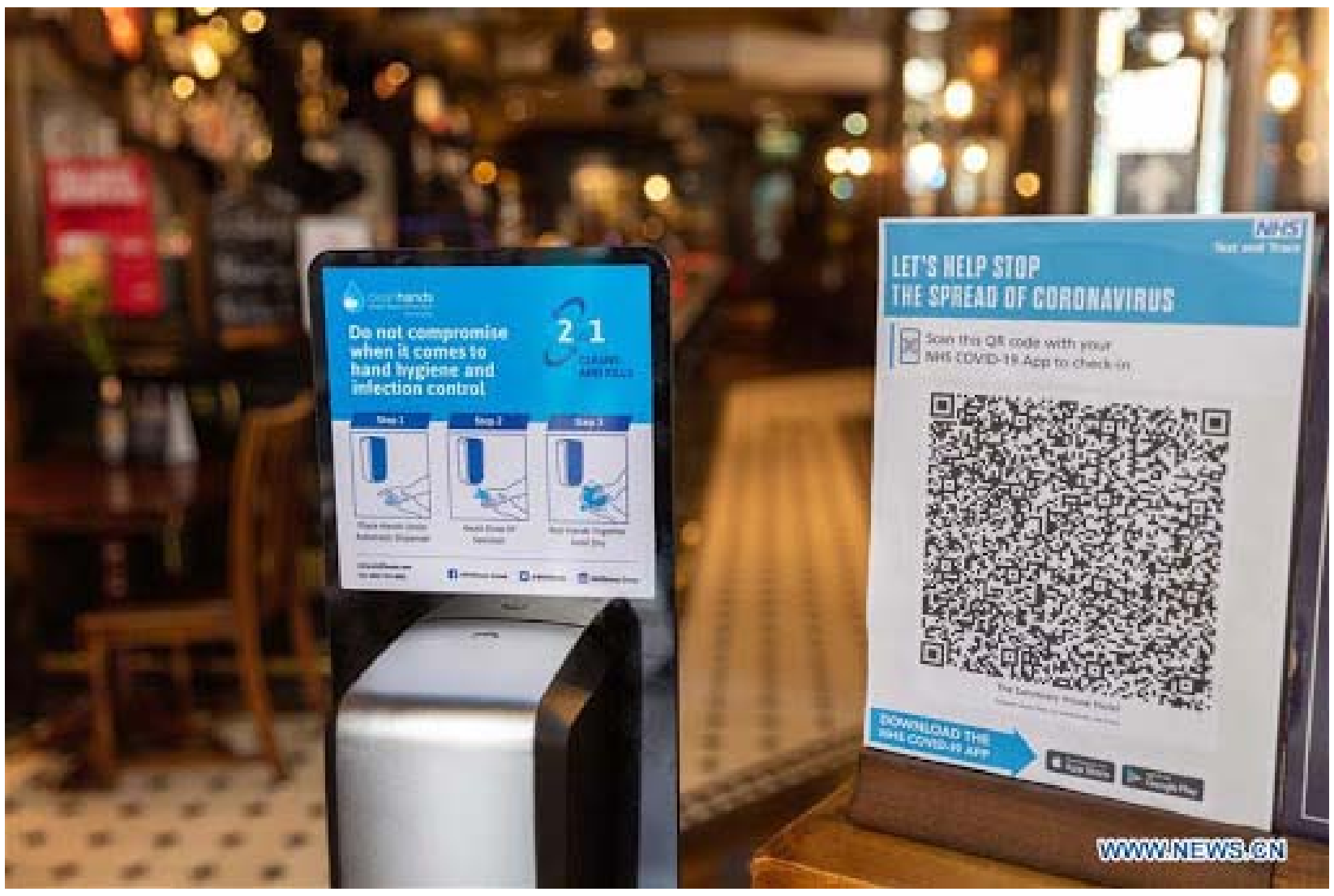

**Figure 21- A QR code to be scanned with the NHS COVID-19 app before entering a venue in London, UK. Source:** [3]

Confusion about the QR code aspect of the app has also been covered in many news articles, e.g.[4], which stated that: "*Some residents [of Newham, UK] reported that the QR code throws up an error message in the app or simply takes too long to scan, causing queues to enter a shop — hardly ideal in these times of social distancing*". An app reviewer on Google Play said[5]: "*Although the app looks good, if I can't use the QR scanner, it defeats the object of the app's purpose*".

Decoding the software engineering issues from the above issues could give us some insights: "*QR code throws up an error message in the app*" denotes that not enough testing has been done on all possible QR codes; and "*simply takes too long to scan*" denotes that not enough performance testing has been done.

**Lesson learned/recommendations**: Some of the reviews provide insights on software engineering issues of the apps, e.g., not enough testing has been done on all possible types of QR codes, and not enough performance (load) testing has been done**.**

<u>*Protect Scotland* app:</u>

As visualized in the word-cloud in Figure 9, one of the words with negative sentiments for this app is "people". By inspection of the comments by the AppBot, we saw 18 comments having this term, e.g.:

- *"Think about this... What's the point unless 100% of people have this app? I could be in a supermarket with 100 people. One person has Covid-19 in said Supermarket, but is the only one who does not have the app. That person inflects several people, but they won't know where they caught it - because that one person didn't have the app"*[6]: the user stresses the need for wide adoption of the app, which is a valid issue.

Another frequent term with negative sentiment was "Google", which was related to the confusion about the updates to the Google Play Services and Google Exposure Notification API. Most (layman/nonprofessional) users, understandably, could not figure out how to update and check their updates. Two example reviews were:

[1] appbot.co/apps/2404650-nhs-covid-19/reviews/1993813691
[2] appbot.co/apps/2411517-nhs-covid-19/reviews/1989859049
[3] www.xinhuanet.com/english/2020-09/25/c_139396283_6.htm
[4] www.wired.co.uk/article/nhs-covid-app-trial-newham
[5] app.appbot.co/apps/2411517-nhs-covid-19/reviews/1951842716/
[6] appbot.co/apps/2437310-protect-scotland/reviews/1951729752

- *"Keeps telling me I need to update Google Play Services for the app to work - even though it is fully up to date. So app not usable"*[1].
- *"I can't get it to work, getting message 'google exposure notification api not available on this device' even though it's running Android 11"*[2].

**Lesson learned/recommendation**: Especially for Android phones, the update mechanism of the OS and its components (e.g., APIs) should be "seamless" (automatic) since we cannot expect all users to have the "technical" skills to do such tasks properly.

Just like for other apps, there were also multiple reviews about high battery usage and other issues related to when the phone's Bluetooth is on, e.g.:

- *"This app requires Bluetooth to be permanently on. A real battery killer, I also get bombarded by nearby devices that see my Bluetooth is on. Google location is accurate to around 1m, why is this not enough? Uninstalled until something better comes up"*[3].

### 4.4 RQ4: A GLIMPSE INTO POSITIVE REVIEWS AND WHAT USERS LIKE ABOUT THE APPS

While the focus of our analysis so far has been mostly on negative comments, it is important to realize that the reviews of the apps are all not negative, as many positive reviews have also been reported. When looking, as we are, at multiple apps in different countries, the positive reviews can be very important to determine what has been done well on one app that could be generalized into others. One way to see a bird's-eye view of positive reviews is to look at review sentiment categories, as shown in Figure 6-(b). The ratios of positive reviews, among all reviews of an app, based on sentiment analysis, range between 5% (for the StopCovid France-iOS app) to 56% (for the *Protect-Scotland* Android app).

Using the AppBot tool, it is possible to drill down into a specific app and gain an understanding of sentiment both over time and overall through an easy-to-use user-interface (UI) as shown in Figure 22 (the case of the *Protect-Scotland* Android app). While it is true that the overall sentiment for this app is highly positive, the time-based chart shows the bulk of reviews were made on the 10th and 11th of September immediately after the 10th September launch. Three days after launch, the number of reviews per day had fallen below 50 and, by day four, below 25, a pattern which has continued. Such a large number of positive reviews being made immediately post-release will skew any overall mean or median measurement of overall sentiment. It is also interesting to consider how accurate these reviews will be given the short period of use the users must have had before posting them in the app store, for example how would battery usage or alert efficacy be measured in the 48-hours immediately post launch? How does this fairly measure any future updates?

**Lesson learned/recommendations**: When considering review sentiments, time-boxing should be used to look at specific updates or recent opinions as well as all-time data, to allow mitigation of any lasting effects of large numbers of reviews in a short period such as launch.

The second pane in Figure 22 shows the ability of AppBot to produce the textual data filtered as needed, shown is the Protect Scotland-Android App reviews filtered for positive sentiment only in the date range September 9th to September 17th 2020. Use of this feature allows the quick gathering of overall positive statements and can avoid the biases mentioned above with date filtering.

[1] appbot.co/apps/2437310-protect-scotland/reviews/1951721703
[2] appbot.co/apps/2437310-protect-scotland/reviews/1951721414
[3] appbot.co/apps/2437310-protect-scotland/reviews/1951722760

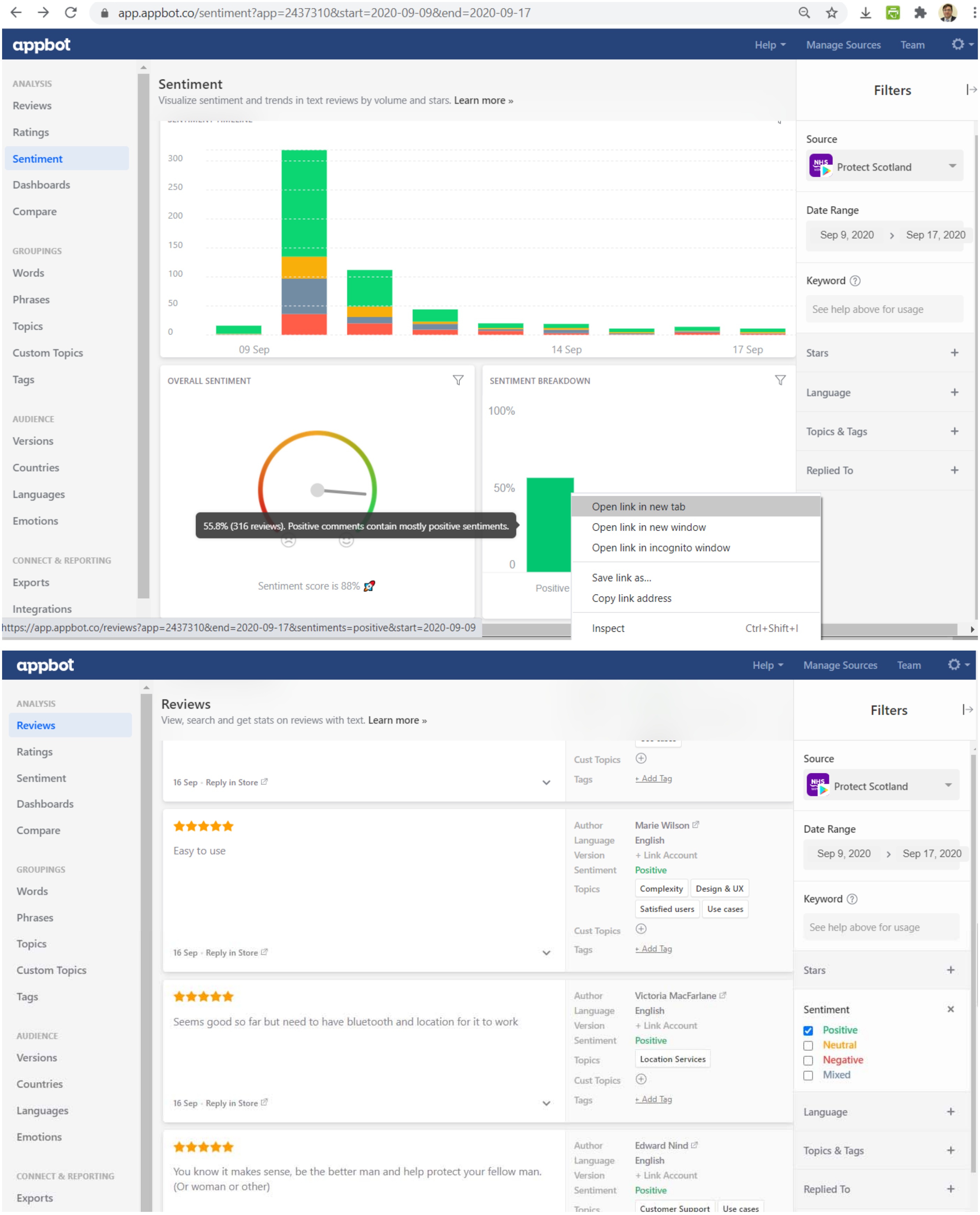

**Figure 22- A simple way to view the "positive" reviews of the *Protect-Scotland* Android app in the AppBot tool**

In the sections below, we sample a few of positive reviews of several apps below, and interpret the findings from those samples.

#### 4.4.1 Sampling positive reviews of the *Protect Scotland* app

As expected, a large majority of reviews are entered by regular (layman) citizens, in which users have expressed their non-technical feedback on the apps, e.g., (each of the following items is exactly one review comment for the Protect Scotland-Android app):

- *"Useful and anonymous*
- *Super easy and not needing internet/data just Bluetooth no details needed brilliant*
- *Very easy to set up, happy to play my part in hopefully returning to our normal way of life soon, if this app helps that in anyway then it's fine by me*
- *Simple and easy to use.*
- *Excellent privacy and will help Scotland. Easy to use.*
- *The more people that use this App the more effective it is. It is ANONYMOUS tracing and will help reduce spread and lockdown.*
- *Plain and simple, and uses the Apple/Google model. Well done Scotland.*
- *Easy to download and activate*
- *Very simple to download, just remember to keep your Bluetooth on when going out*
- *Seamless and invisible. Well designed app we all need.*
- *Super easy layout and quick to set up.*
- *I love that it is confidential and keeps you informed of any exposure.*
- *Just what is needed*
- *Great idea, hopefully everyone who can download it does soon!*
- *Easy to install*
- *Well done Scot.Gov*
- *Great app and can helpful to get info.. recommend to download this app*
- *Finally...*
- *Really good idea! Hopefully more people will download it to make it more worth while!*
- *Good idea*
- *This app is visually appealing, easy to use and gives peace of mind*
- *Very simple and transparent. Get it downloaded*
- *Impressed with the app and also the clear messaging about how it works and what it does.*
- *Fingers crossed I never have to use it.*
- *Really simple to use and the more people that use it the more effective it becomes.*
- *Very easy to set up and gives users genuine privacy. Just install it!*
- *Takes seconds to set up, simple and easy to understand.*
- *So easy to install*
- *Very easy to set up*
- *Easy to download and use*
- *Simple and easy to use. Impressed by the privacy and security controls. Encourage everyone to download it."*

As we can see in the above randomly-sampled subset, some users have liked the good usability of the Protect Scotland app, while some others have commended the seemingly well-designed privacy and security controls of the app. Many users have been "optimistic" in their reviews by mentioning terms such "hope" and "hopefully" in their reviews, hoping that more people will use the app. The term "hopefully" appears three times in the small subset above, and the string "hop" appeared in 45 of 573 reviews (7.8%) for the Protect Scotland-Android app.

We found that almost all positive comments were not "technical", but rather were in "surface" level, i.e., users just praised the ease of installation and use of the app. But given the complex nature and function of the app (sending notifications to recent contacts in the case of COVID test being positive), very few users could actually see whether the app is really doing what it is supposed to; as discussed in the quality, efficacy and accuracy aspects of these apps (Section 2.4). To clarify our point, let us take for instance a regular (non-health) mobile app such as a typical "Notes" mobile app. It is much easier for a user (layman) to figure out if such an app is working or not, e.g., by creating some example notes (to do's) and then checking in another day in the app, if those notes are showing up properly.

Although a contact–tracing app can be easy to install and easy to use, and even have a nice UI, there is no guarantee for its "core" feature to work properly. This reminds us of the French news article[1] that we reviewed in Section 1, as it had reported that, as of mid-August 2020, "*StopCovid [the French app] had over 2.3 million downloads [of a population of 67 million people] and only 72 notifications were sent [by the app]*". As we can imagine and also as reported in the grey literature[2]: "*A faulty proximity tracing app could lead to false positives, false negatives*", which stresses the fact that these apps are safety-critical. Some reviews actually made explicit notes about this very important issue: "*Obviously I can't know how well the tracing works, but everything else about the app is great. It clearly explains all you need to know about how it works (especially in relation to privacy) and the interface itself is simple and effective. A very well made app*".

Returning back to the type of positive (and mostly optimistic) statements mentioned in positive reviews, some reviews were mentioning using this app to save lives, e.g., "*We've all got to do this guys! Save your granny's life!*". Some other positive reviews of the Scottish app praised its possible integrability with other European apps: "*Glad it's up and running using the template adopted by other European nations, allowing potential international use in the future*".

### 4.4.2 Sampling positive reviews of the *StopCovid France* app

As discussed above, the ratio of positive reviews, among all reviews of the French app, based on sentiment analysis, was the lowest (5%) among all the nine apps in the dataset.

Note that almost all reviews for the French app were in French, and thus we used the Google Translate tool to use them in this paper. Some of the positive reviews in the French reviews dataset were insightful, e.g., "*Assuming it's working it's a great app, you don't have to do anything. Just have it running. However some way to know its working would be nice. Even just a counter of how many other apps it's discovered in that day, just to know its doing something*". We can treat this review as a feature request by users, a topic that has been studied in focus in research papers [6, 39]. The review also refers to the unclear efficacy of the app: "*Assuming it's working*". Many other users also expressed their uncertainty on whether the app actually does its core feature, e.g., "*Apparently it works. And it's transparent. Nothing to say ... It would be good from an epidemiological point of view if it was used more as an app!*", which also referred to the important epidemiological aspects [46] of the app and the need for its wide usage.

Another feature request was: "*Works very well. An improvement would be to know the number of phones using the cross app in the week, month, day. It will help everyone to perceive the usefulness of their action by installing this app*".

Similar to the other apps in the study, some users reported in their review their satisfaction with how the app is preserving their privacy: "*Perfect compliance with the GDPR [EU's General Data Protection Regulation] so nothing to fear. Let's all activate this app to help fight the virus and prevent our businesses from being closed!*". Some other comments were happy of how usage of this app could have made their lies easier in lockdowns, e.g., "*Thanks to this application, I no longer have to close high schools*" (Note that we did not find any online news article linking usage of contact tracing apps to ease of school closure in France).

Another positive and detailed comment that we came through was: "*The choice of activating or not is free and Bluetooth is activated at the same time. I find that the battery holds up very well. Data protection is fully explained. The only problem is that there aren't enough people using it. It should be publicized, and then it to be really effective*", which again commented on the wider public usage epidemiological aspects [46].

Furthermore, there were reviews which were somewhat thank-you feedbacks to the development team, e.g., "*Thanks to the developers for removing the permanent notification!*", which referred to an apparently-annoying notification mechanism which was in the app in its previous versions, and apparently has been fixed based on earlier user feedbacks.

Another review reported an interesting feature of the app: "*Good application example when you are at the bakery, there is a message that says you have to put on a mask and stay 1 meter [apart]*".

Many bug reports were informally mentioned in some of the reviews: "*Like a lot of people, I put my phone in airplane mode at night, but you have to reactivate the app every day. It's annoying and above all we forget ... Otherwise the application is very useful*".

### 4.4.3 Sampling positive reviews of the *StopCOVID NI* app

46 of the 195 reviews (23.5%) of the *StopCOVID NI Android* app had positive sentiments, according to the AppBot tool. We discuss a few insightful examples of those comments below.

---

[1] www.lefigaro.fr/secteur/high-tech/stopcovid-2-3-millions-de-telechargements-et-seulement-72-notifications-envoyees-20200819

[2] www.eff.org/deeplinks/2020/04/challenge-proximity-apps-covid-19-contact-tracing

- Positive feedback about the app and its features:
  - *"the app seems to be working well in the background and notifies me on a weekly basis for the searches it has done. "*
- Some reviews expressed that the app is not easy to be found in app stores:
  - *"Excellent App, giving it 4 stars as it was difficult to find on play store, ended up looking it up on Google"*
  - *"Very easy to setup, but hard to find in app store."*
- Some feature requests:
  - *"Great app, glad it's out now. Would like to be able to find out just how many have downloaded it though."*
  - *"… only fault I find is it would be good if you could make it so you can put your post code in and find out how many is in your area that would be a big plus"*
  - *"App for southern Ireland gives a few overall general statistics, this gives no information at all."*

**Lesson learned/recommendation**: The apps must be clearly identifiable and searchable in app stores to maximize the number of users downloading it.

**Lesson learned/recommendation**: Where possible, some feedback (such as statistics about COVID cases in the region and also number of close-by phone IDs recorded in the past) should be provided as a feature of the app, to encourage users that the app is working to emphasize the pro-social and individual benefit it is having.

### 4.5 RQ5: Feature requests submitted by users

Each submitted review can have different messages in it, e.g., error report, feature requests, user just mentioning her/his satisfaction or dissatisfaction with the app. If a review's text contains some form of suggestions for new features, it can be used as a 'feature request' by the app's development team in further improving it. Mining feature requests from app reviews and also from Twitter data have been recognized as a form of "requirements engineering" (elicitation) in the software engineering community, as several papers have been published in this area, e.g., [6, 47-53]. The topic has also been referred to as "crowd-based" requirements engineering [54].

Given the large number of reviews in general, and also in our study (our dataset has 39,425 records), however, pinpointing and extracting only the reviews which contain 'feature requests' manually is not an option. As discussed in Section 3.3, AppBot has a useful feature of to filter reviews to show only those with "Feature requests" submitted by users for an app. We show an example of using this feature for the case of the German *Corona-Warn* app in Figure 23. This filtering feature is listed under a "Topic and tags" dropdown list, in which the items have been generated by applying the "Topic modeling" technique [55], which is an NLP technique, by the AppBot tool.

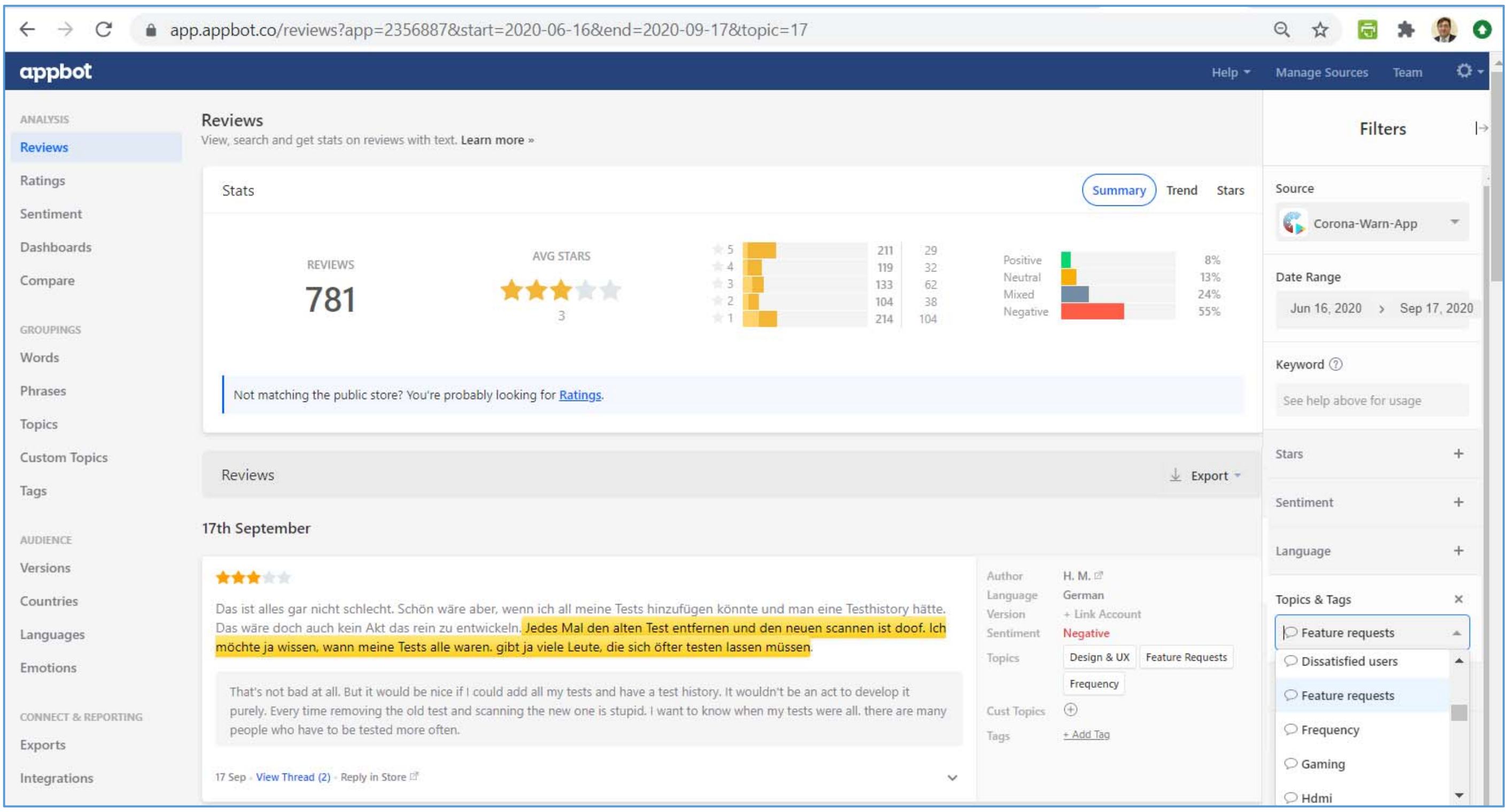

**Figure 23- The useful feature of AppBot to filter reviews to only see the "Feature requests" submitted by users for an app (example in this screenshot: the German app)**

As representative examples from our set of nine apps, we look at feature requests for the cases of the German *Corona-Warn* app and the *COVID Tracker Ireland* app, next.

#### 4.5.1 The case of German *Corona-Warn* app

The suggestion for a new feature, in the example review shown in Figure 23, is the following phrase: "*... it would be nice if I could add all my [COVID] tests [in the app] and have a test history*" (even the original English text is highlighted in yellow by the AppBot tool for easy finding). The app's development is quite active and actively replies to almost all comments when needed, e.g., see the "*View Thread (2)*" link just below the example review shown in Figure 23. In this case, the app's development replied by saying that (English translation from German): "*We have the wish-list repository on Github, in which comparable suggestions for expanding the app have already been created, see github.com/corona-warn-app/cwa-wishlist*".

As we can see in Figure 23, when we filter all the reviews in the time window under search (from the app's first lease until Sept. 17, 2020), 781 of all 20,972 reviews (or 3.7%) have been identified as those containing suggestions for new features.

Once we had the 781 feature-request-reporting reviews, we were interested in grouping them to actually see a refined list of suggested features by users. Our chosen tool (AppBot) did not have such a feature. In fact, we saw the need for some form of thematic analysis (or qualitative coding) to group and combine that subset of reviews. We looked into the literature but did not see a ready-to-use technique for this purpose, and of course, we did not have the time resources to do "manual" qualitative coding of the reviews. We thus raise the need for such a technique and analysis to future works.

Nevertheless, we think it is worthwhile to mention in the following several insightful feature requests given by users for the German app. For full traceability, we also provide the "permanent" links to each review in the AppBot's database:

- *Would be great if you can see how many people you exchanged the keys with.* appbot.co/apps/2356887-corona-warn-app/reviews/1900405383
- *I think the app is generally well done and as a computer scientist, I would also like to praise the public interaction (including OpenSource)! The only thing I would still like would be if I could see how many relevant encounters have taken place in the last 14 days. So it might be a little more transparent that the app is actually doing something in the background.* appbot.co/apps/2356887-corona-warn-app/reviews/1896344794
- *It would be great if you could get more information such as infection numbers and spread at district level.* appbot.co/apps/2356887-corona-warn-app/reviews/1822553545
- *Improvements are always possible, e.g., how many encounters there were with other app users and the like.* appbot.co/apps/2356887-corona-warn-app/reviews/1818108607
- *More info would be nice. The risk assessment is all well and good, but it would be better if you could see how many other app users you had contact with* appbot.co/apps/2356887-corona-warn-app/reviews/1809389199
- *It would be great if this app were also available for tablets. We are always on the road with an iPad on business. This way, a possible infection could also be tracked after meetings.* appbot.co/apps/2356887-corona-warn-app/reviews/1805884538
- *This is a great app, but the need for internet is annoying because I don't have a lot of mobile data and it would be better if you could use the app without internet.* appbot.co/apps/2356887-corona-warn-app/reviews/1804643241
- *Kudos to the activists and developers who made it that way. Every little helps in the fight against Covid19. Download it and activate the anonymised decentralised logging of physical proximity to other users of this app! One suggestion: the nerd in me wants to see a list of the beacons detected on my device.* appbot.co/apps/2356887-corona-warn-app/reviews/1803093210
- *Good, but I would have liked optional GPS tracking. If, for example, Corona breaks out on the train and you are not directly next to a person, you can still be informed based on the location if you want* appbot.co/apps/2356887-corona-warn-app/reviews/1798228592
- *Works great, but it would be awesome if I could check how many people I've met that also had the app installed (not only those who are/were infected)* appbot.co/apps/2356887-corona-warn-app/reviews/1798010216

Even with this small sample set of reviews (above), we can see that the new feature of seeing how many encounters there were with other app users, is a common feature request as mentioned by users, at least for the German app.

On the other hand, we were curious to see the precision of the tool in identifying reviews with feature requests, as we observed several reviews being incorrectly-classified as feature requests. To do this, we looked at a random subset (more than 50) of those 781 reviews, and manually identified whether they did not have obvious feature requests in them. We list several of those cases below, along with our hypothesis of why we think the AppBot tool has incorrectly classified them as feature requests. We should note that this observation raises questions on the precision of the AppBot's feature to filter reviews by those containing feature request.

1. *The app has been telling me in the weekly overview for several weeks that I had 223 checks within the last 14 days. How can it be that I am presented with the same number 223 week after week? I am out and about every day so there must be fluctuations. So something doesn't work there and that makes the whole thing very questionable.* appbot.co/apps/2356887-corona-warn-app/reviews/1946942728
2. *The app has only caused one problem for me so far: I have never removed my test result, now it has probably been removed from the databases and the app tries forever to retrieve it, but I cannot add a new test. Please fix.* appbot.co/apps/2356887-corona-warn-app/reviews/1941257952
3. *Would be even better if more people participated!* appbot.co/apps/2356887-corona-warn-app/reviews/1859275190
4. *Uncomplicated and (for me) calming. Since I use public transport a lot for work, I would like to be warned, as well as others, if you had contact* appbot.co/apps/2356887-corona-warn-app/reviews/1799464996

Reviews #1 and #2 above should have been classified as bug reports. Review #2 even has clear words in it to hint to the classifier tool that it is a bug report, e.g., "*Please fix*" and "*The app has only caused one problem for me*". One possible reason on why we think AppBot has incorrectly classified review #3 as a feature request is the phrase "*Would be even better*" in it. For review #4, it could be the phrase "*I would like to*".

**Lesson learned/recommendation**: A variety of insightful feature requests have been provided by users, e.g., by user of the German app: How many encounters there were with other app users (how many people you exchanged the keys with); infection numbers and spread at district level; can the app be used without internet? As a form of "iterative" requirements engineering" (elicitation) [6, 47-53] or "crowd-based" requirements engineering [54], app's software engineering teams are encouraged to review those feature requests and select a subset to be implemented.

**Lesson learned/recommendation**: While AppBot's feature to filter reviews to see feature requests only is a useful feature, we found many example reviews which AppBot incorrectly classified them as feature requests. We realize that an NLP/AI-based algorithm has been used to do that classification and such an algorithm will have limited precision, but still there is need to improve such algorithms by developers (vendors) of App review analytics tools, such as AppBot.

#### 4.5.2 The case of *COVID Tracker Ireland* app

Using the "Topic modeling" feature [55] of AppBot (just like what we did in Figure 23 for the German app), we used the same tool feature to review and analyze the feature requests submitted by users for the *COVID Tracker Ireland* app, as shown in Figure 24.

Out of the 1,737 reviews submitted for both OS versions of the *COVID Tracker Ireland* app, 169 (9.6%) were automatically tagged by AppBot as feature requests. While we did not check all of them to ensure they are indeed feature requests, the three examples in Figure 24 are indeed so.

There are various feature requests (inside reviews) such as the first one in Figure 24: "*Would be better if we just saw trends instead of daily figures*". The app's software engineering teams could review those comments and consider implementing those feature if there are good reasons to do so. For this particular comment, the team does not have to "replace" the old feature ("*daily figures*" [of COVID cases]) by the new feature ("*trends*"), but could add the new feature and users can choose to use them alternative as options, inside the app UI.

The last review shown in Figure 24 implies that many users find this particular app not that useful and such reasons lead them to uninstall the app. The user also rightly mentions that the app "*gives you information you can get anywhere*", e.g., media, news sites, etc.

**Lesson learned/recommendation**: Many users have casted doubts on the usefulness of the apps, i.e., they do not provide most of the "right" and much-needed features that many users are looking for. Thus, using "crowd-based" requirements engineering [54] techniques for these apps are critically needed.

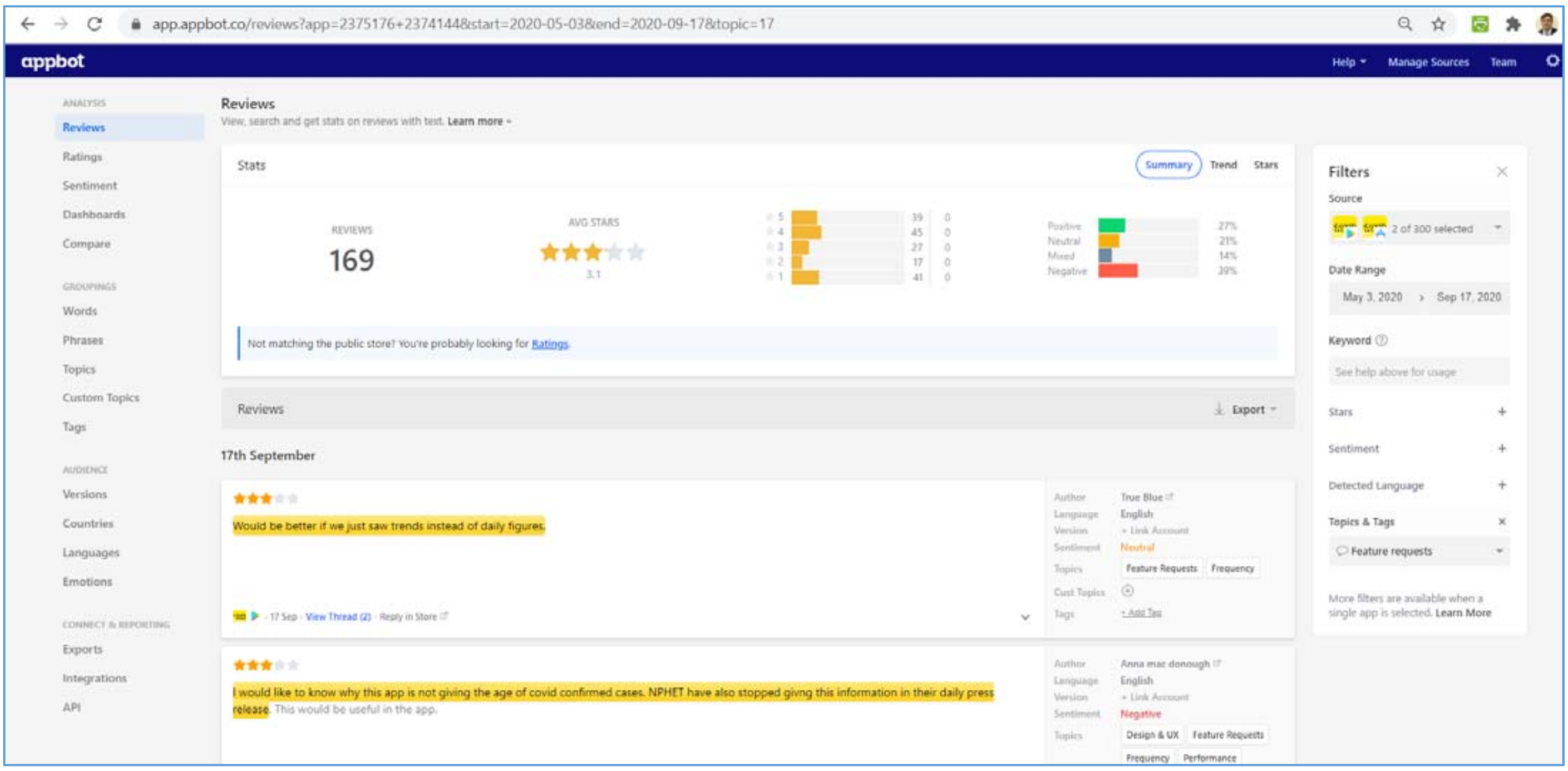

appbot.co/apps/2375176-covid-tracker-ireland/reviews/1950639714
appbot.co/apps/2375176-covid-tracker-ireland/reviews/1950411937

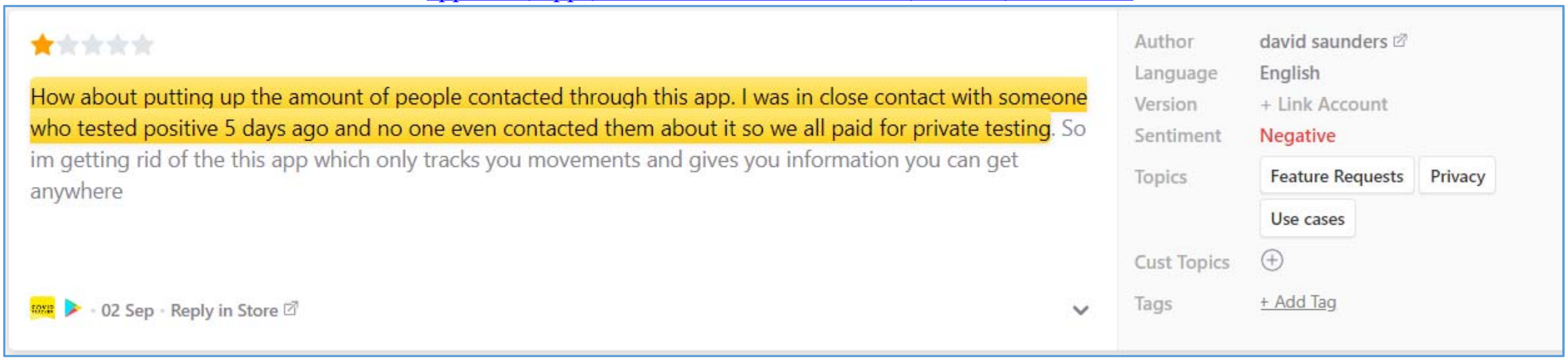

appbot.co/apps/2375176-covid-tracker-ireland/reviews/1927510885

**Figure 24- Reviews a subset of the "Feature requests" submitted by users for the *COVID Tracker Ireland* app**

## 4.6 RQ6: Comparing the reviews of Android versus the iOS versions of the apps: Similarities and differences

Given the wealth of the information in the dataset, we found it a good opportunity and area of interest to compare the reviews of Android versus the iOS versions of the apps, and to observe the similarities and differences (if any). For the sake of space in this paper, we selected three of the most insightful aspects, which we report next:

- Popularity of each OS app version as measured by the average number of 'stars'
- Sentiment of reviews for the Android versions of the apps versus their iOS versions
- Problems reported for each OS version

### 4.6.1 Popularity of each OS app version as measured by the average number of 'stars'

We take the average number of 'stars' (as reported by app stores) for each of the two OS app versions of all the nine apps under study, and visualize the values as a scatter (XY) plot in Figure 25.

While a clear correlation can be seen here, most apps do show a slight disparity between Android and iOS ratings, as seen more broadly with other apps in the literature [28], though in no cases is this more than 0.5 stars. For most apps, star ratings of the iOS version are slightly higher than the Android version; except the England & Wales and the Republic of Ireland apps, in which star ratings are slightly highly on iOS than on Android.

Also as seen in Figure 25, in both platforms, the ranking order of the apps is much the same, with only some mid-table positions being different. The highest and lowest-ranked apps, Protect Scotland and England & Wales, respectively, are clearly the highest and lowest regardless of platform.

One caveat to this analysis is that the average star rating, reported by app stores, take no account of the volume of users which may be worth considering in future work, especially in cases where there is a noteable disparity between the OS versions. Outside of the scope of this work also is a consideration of how the apps were developed for different platforms, specifically to what extent did they share a codebase or interface. While the correlation implies that there is a great degree of similarities between country apps on different platforms, it would be interesting to further examine the differences at a technical level and see how this may relate to platform disparity.

**Lesson learned and recommendation for future work**: It would be interesting to examine the differences among the apps and also their two OS versions, at a technical level, e.g, their code-base, software architecture.

Of interest to note here is that the relatively close nature of the rankings on both platforms, and with some crossover between which has a better score for individual apps, implies that the underlying decentralized services provided by Google for Android and Apple for iOS, which all the decentralized apps will use, make little or no difference to user perception.

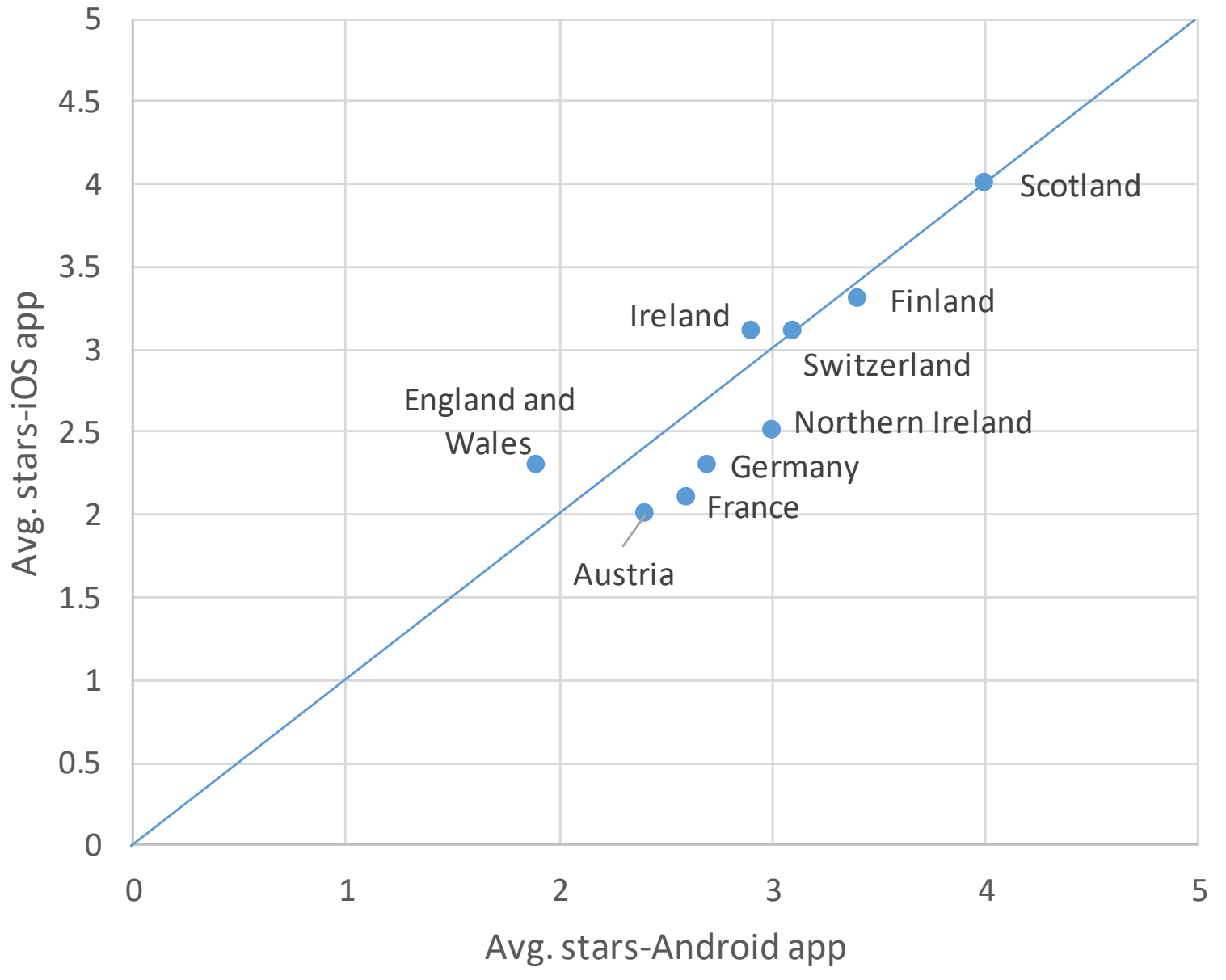

**Figure 25- Popularity of Android versions of the apps versus their iOS versions, for each country, as measured by the average 'stars'**

**Lesson learned**: There is a clear correlation between user (star) ratings on different platforms, which shows that the underlying Google or Apple decentralized technology makes little difference in user perception compared with the frontend/OS implementation. Also, since most probably, the Android and iOS versions of each app have the same features (we did not check/compare those in detail), users' satisfaction (stars) of either versions of a given app are quite similar.

### 4.6.2 The sentiment of reviews for the Android versions of the apps versus their iOS versions

As we discussed in Section 3.2, it was interesting to observe that, in the case of all nine apps, the Android apps have received more reviews compared to iOS apps. This seems to align with the general trend in the app industry, as reported in the grey literature: "*Android users tend to participate more in reviewing their apps*"[1] and "*Android apps get way more reviews than iOS apps*"[2].

Apart from the "volume" of reviews received for each OS version, we were curious to compare the "sentiment" of reviews between the two OS versions. As discussed in Section 4.1, AppBot calculates and provides a sentiment score of each review (a value between 0-100%). The higher this value, more positive the review sentiment, meaning that the review text has a positive tone in its message. AppBot also provides a single aggregated value from all reviews of an app, which we gathered and visualize them as a scatter plot in Figure 26.

[1] medium.com/@takuma.kakehi/we-need-app-reviews-but-we-need-to-ask-at-the-right-time-e2916b126c8e
[2] medium.com/@chiragpinjar/why-android-apps-get-way-more-reviews-than-ios-apps-30c5b9e7ee71

Comparing the sentiment analysis in Figure 26 with the equivalent comparison of star ratings in Figure 25 shows a clear difference in how sentiment analysis has detected comments compared with simple ratings. While again, as in Section 4.6.1, most of the apps show a correlation between platforms, there are some interesting differences here. For example, while the apps of both France and Austria have a very low (~ 6%) number of positive reviews on iOS, they have still low but many times higher levels of positive reviews on Android (~ 22%). Based on the sentiment analysis, the Scottish app remains the most highly regarded app by some degree with 60-80% positive reviews depending on platform and no other app having >40% positive reviews. Compared with star ratings, the sentiment analysis does show a difference however in the most negative end, with the German app garnering the lowest proportion of positive reviews on iOS (<5%). The app for England & Wales is still the worst reviewed on Android (~ 12% positive) with Germany just ahead (~ 17%).

Seeing again a disparity in general between iOS and Android reviews, we searched in the grey literature (online sources) to see if there were any discussions or reported evidence on why Android app reviews are slightly more "positive" than iOS app reviews. However, we could not find any. We think it is worthwhile to investigate this issue in future studies. It is also clear from sampling reviews that those especially towards the longer end are more nuanced and will discuss both positive and negative aspects of the app, perhaps making an overall judgement hard to make especially using automated sentiment analysis.

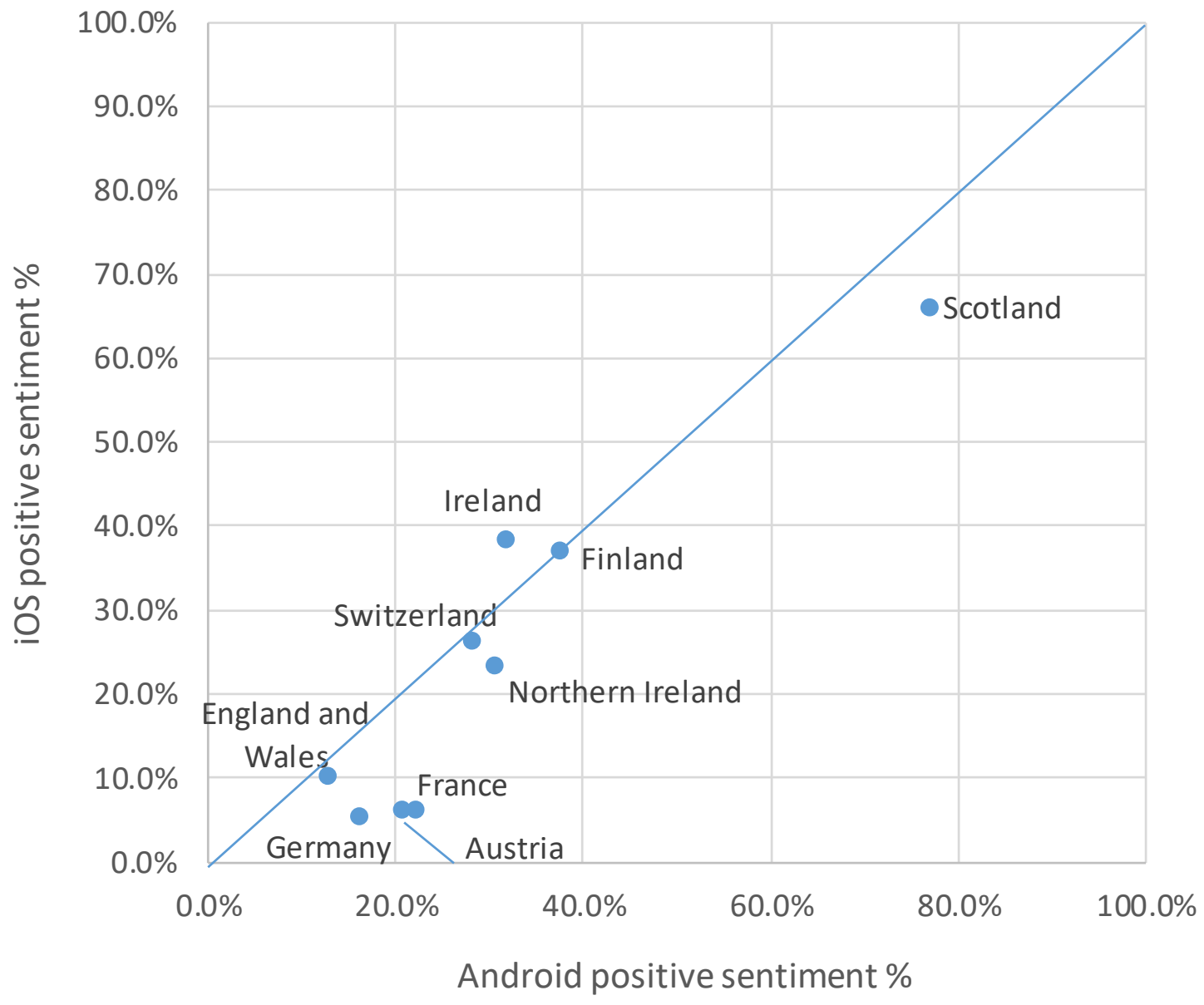

**Figure 26- Sentiment of reviews for the Android versions of the apps versus their iOS versions, for each country**

**Lesson learned/recommendation**: The sentiment analysis of apps can provide more complex granular output compared to just the "star rating", but there seems to be an inherent negative bias especially on Android which should be further investigated in future studies to better understand the phenomenon. A possible future Research Question (RQ) would be: Why is there an inherent negative bias in Android versions of an app, compared to the iOS version?

### 4.6.3 Problems reported for each OS version

In our dataset, we separated the reviews of each OS version and then fed them into the AppBot's sentiment analysis and word-clouds visualization, similar to what we had done in Section 4.2. As representative examples, we selected five of the nine apps. We mainly selected the apps with most reviews, to ensure that the sentiment analysis will have enough data to provide reliable and meaningful results.

We show in Table 3 the word-cloud visualization, enriched with sentiment results (by the color of texts), for each OS version of five example apps.

Visually, it is possible to see some differences and similarities, e.g., for the case of the *Protect Scotland* app, the term "install" seems to occur many times in the "positive" sentiment in the reviews of its Android version compared to the iOS version.

By a closer inspection, i.e., clicking on each term in the word-cloud in AppBot's UI and reviewing a sample of the corresponding reviews, we found that many users have said something along the lines of "*Easy to install*".

**Recommendation for a future work**: It would be worth investigating in future what differentiates the Android app users from the iOS app users to have different opinions about the app installation.

Again for the *Protect Scotland* app, there is red (negative) sentiment for the term "Google" in the Android version, and clearly that term does not occur in the iOS version reviews, since Android decides are strongly associated with Google and its services, e.g., Google Play app. One user mentioned[1] the review shown in Figure 27.

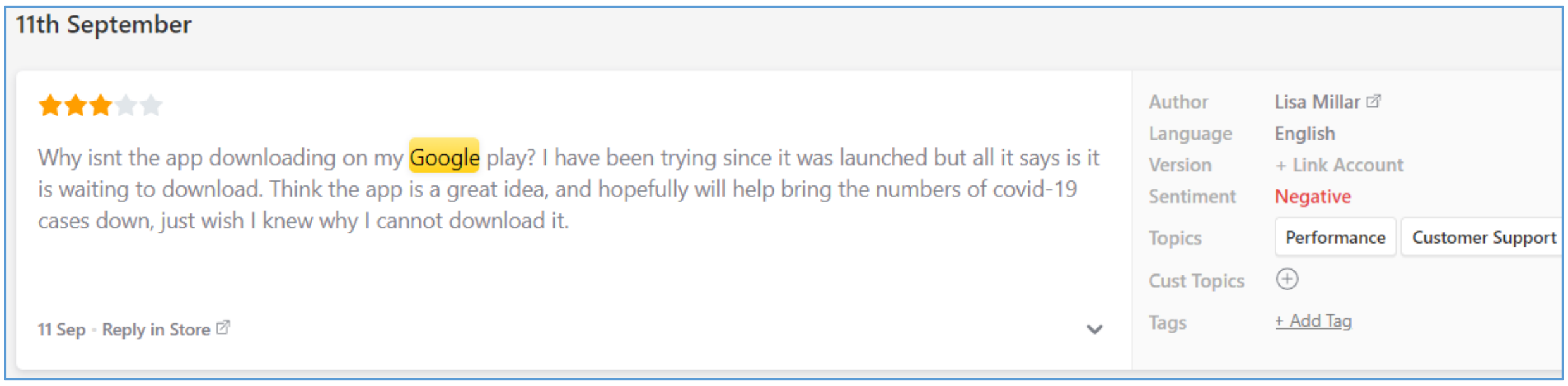

**Figure 27- A user feedback from an Android user of the *Protect Scotland* app**

Although visual analysis (manually) of the word-clouds in Table 3 could provide interesting insights, we were keen to find a way to more systematically and numerically analyze the similarities and differences of the OS versions of each app. We reviewed the text mining literature and found that there are indeed advanced methods to compare "semantic" similarities/differences of two large bodies of text, e.g., one widely used metric is *semantic overlap*, which is based on a concept called *semantic folding (fingerprinting)* [56].

There is an online tool[2], which provides an easy-to-use implementation of the semantic-overlap metric. We fed the entire reviews of each OS of each app to this tool, and the values are shown in the last column of Table 3. For the readers interested on how the cortical.io tool works, we show in Figure 28 a screenshot of the tool, comparing the semantic of the texts in the review dataset of the two OS versions of the *Stopp-Corona* Austria app.

The semantic-overlap measures, shown in Table 3, range between 45% (Stopp-Corona Austria) to 86% (COVID Tracker Ireland). A low semantic-overlap measure for an app could have various possible root causes, e.g., (1) the Android and the iOS versions of the app may have different features, thus leading to different user opinions; (2) there could be "platform" issues in either of the versions, which could cause negative or positive reactions (reviews) from users, e.g., in the example review of the *Protect Scotland* app shown in Figure 27, the user had problems downloading the app via Google Play.

We should mention that, as shown in Table 3, for the Corona-Warn German app, the cortical.io tool did not generate any output after letting it run for a long time (no response), possibly due to large dataset size of German app reviews.

**Table 3-Word-clouds of reviews for both OS versions of five example apps**

| Country / app | Android app | iOS app | Semantic overlap |
|---|---|---|---|
| Protect Scotland | 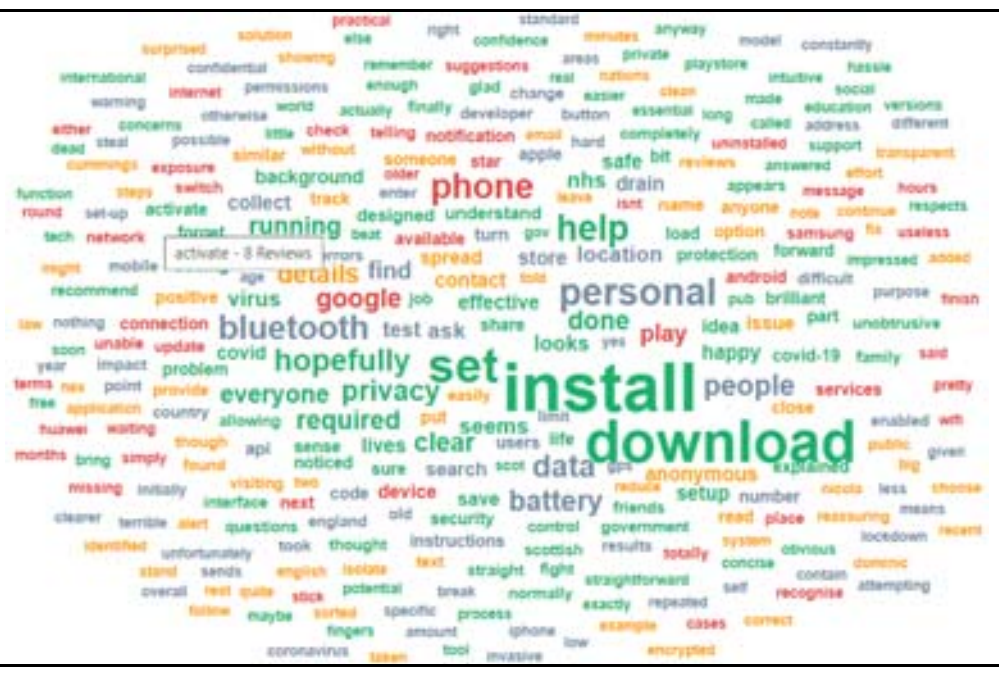 | 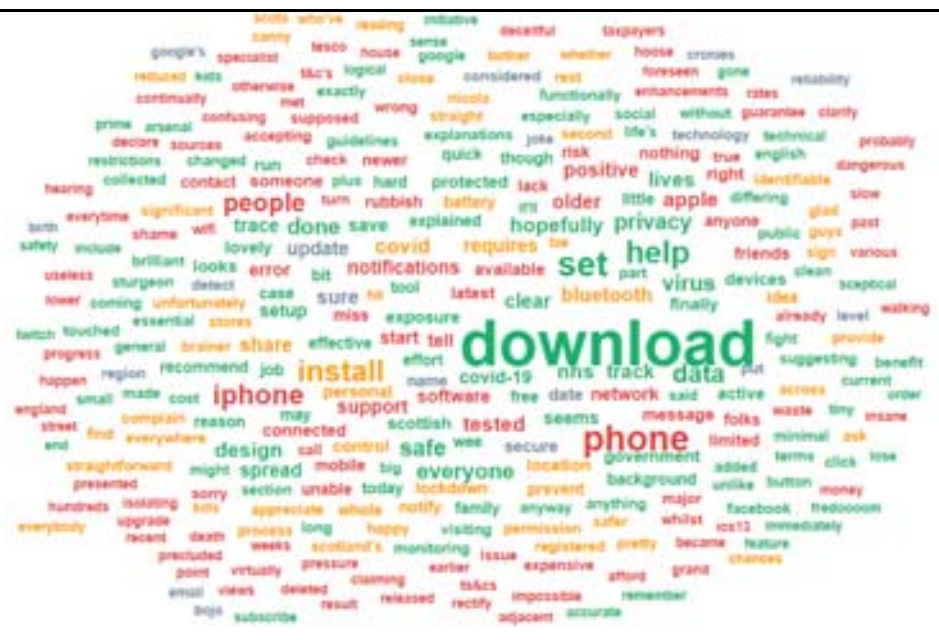 | 82% |

---

[1] appbot.co/apps/2437310-protect-scotland/reviews/1951719103
[2] www.cortical.io/freetools/compare-text

| | | | |
|---|---|---|---|
| StopCovid France | 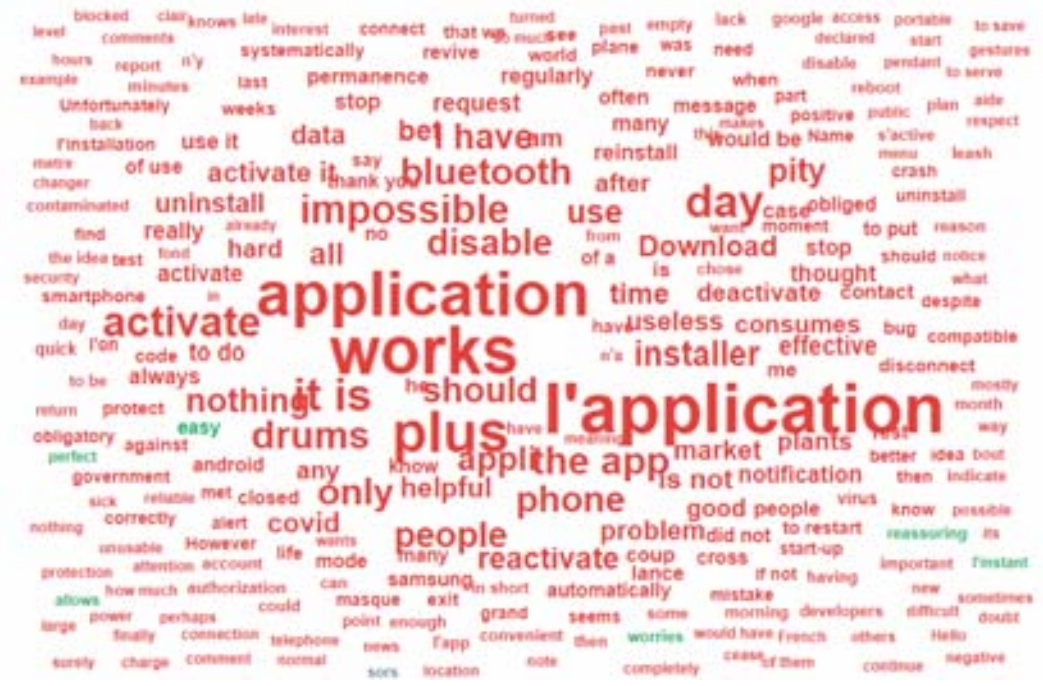 | 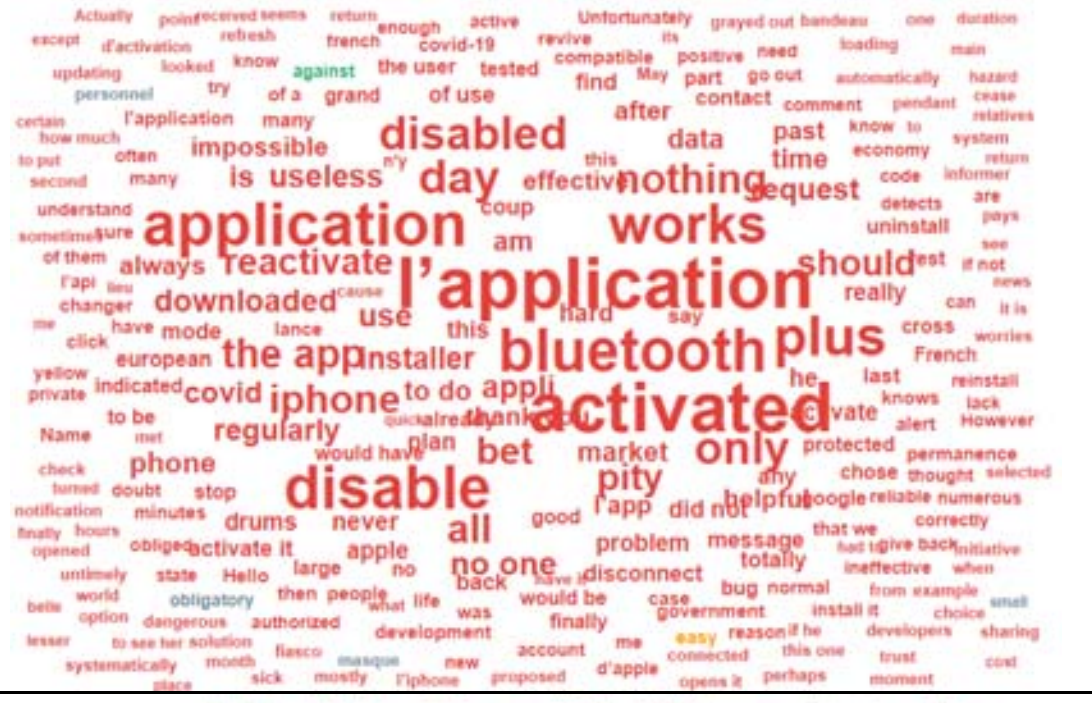 | 47% |
| Stopp-Corona Austria | 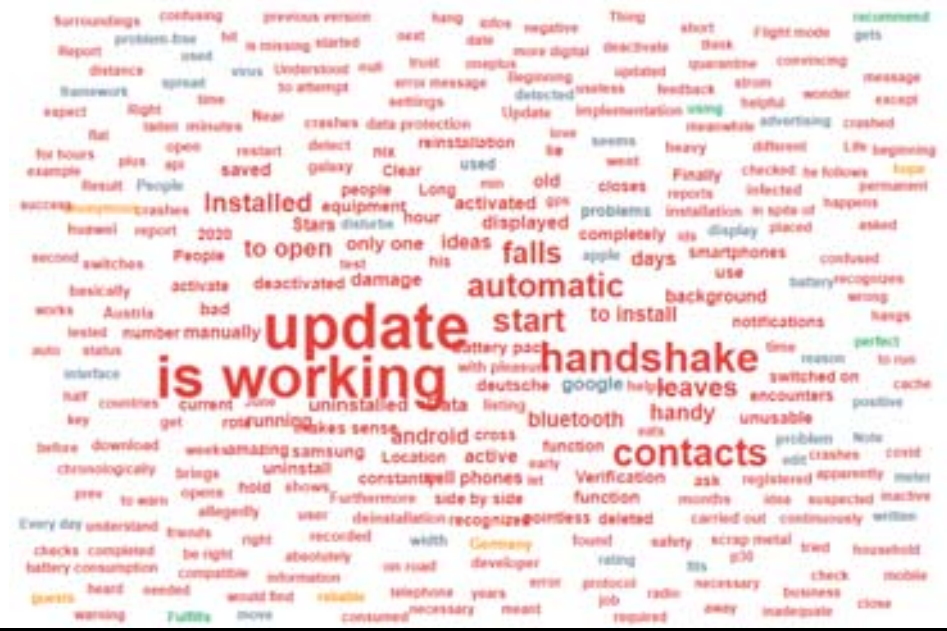 | 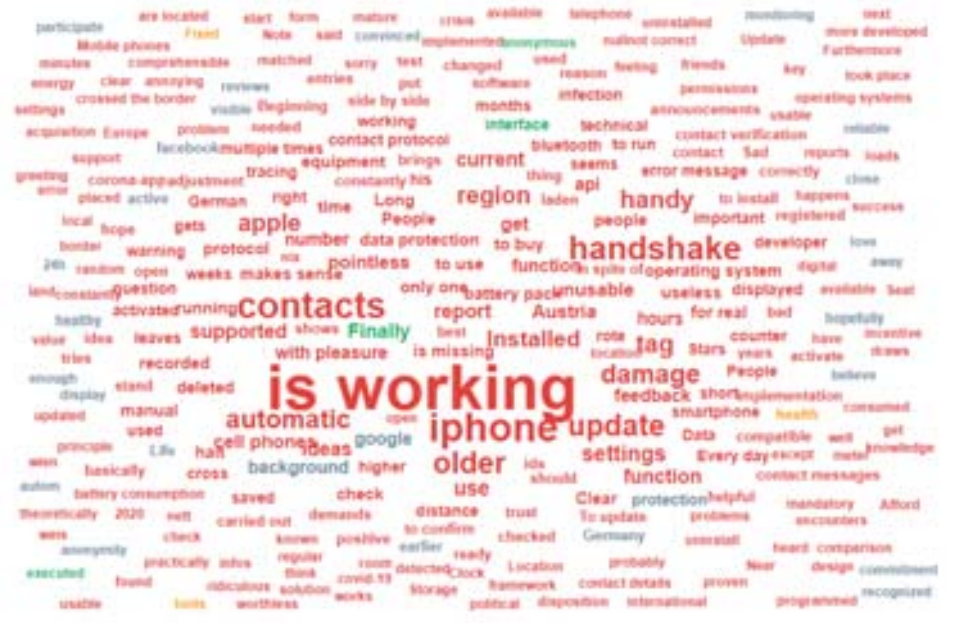 | 45% |
| COVID Tracker Ireland | 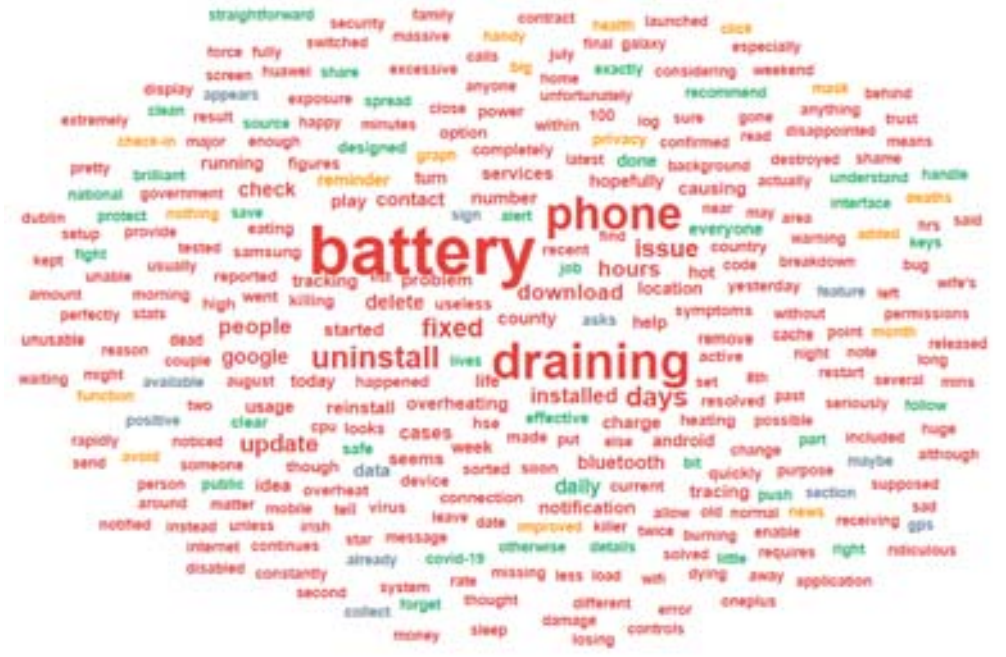 | 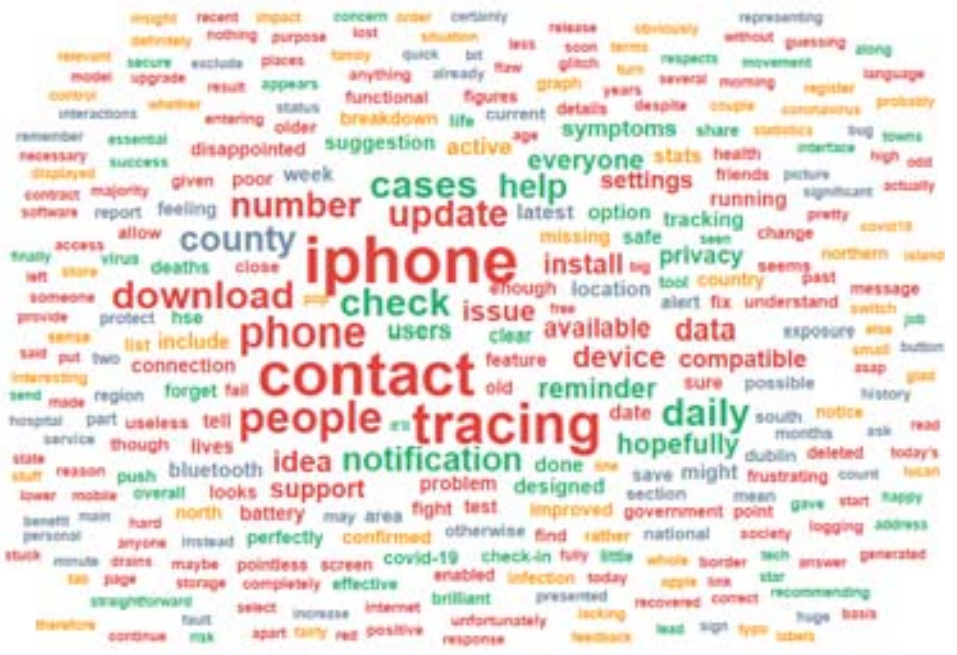 | 86% |
| Corona-Warn Germany | 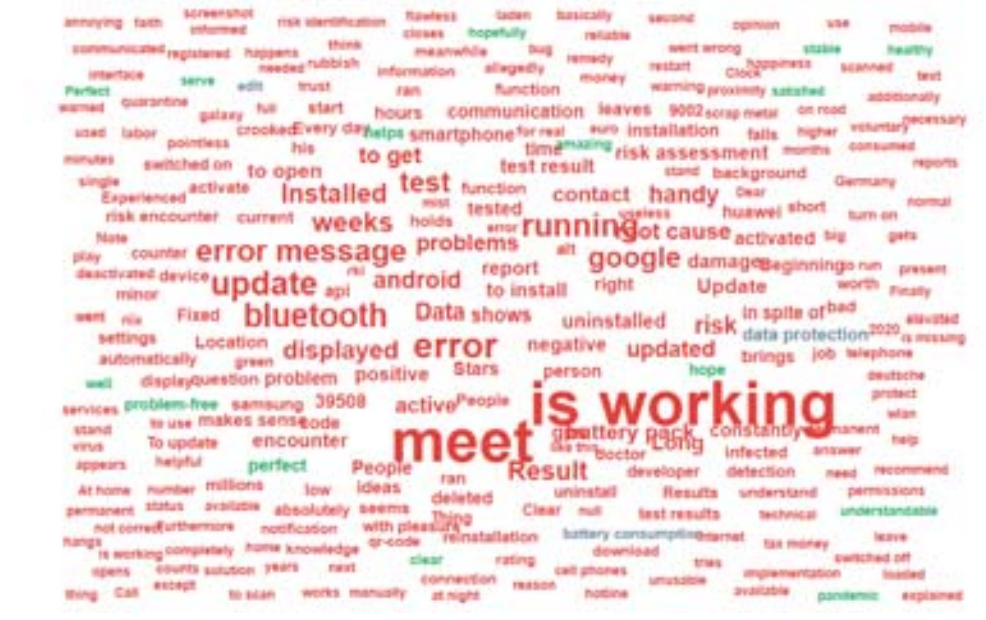 | 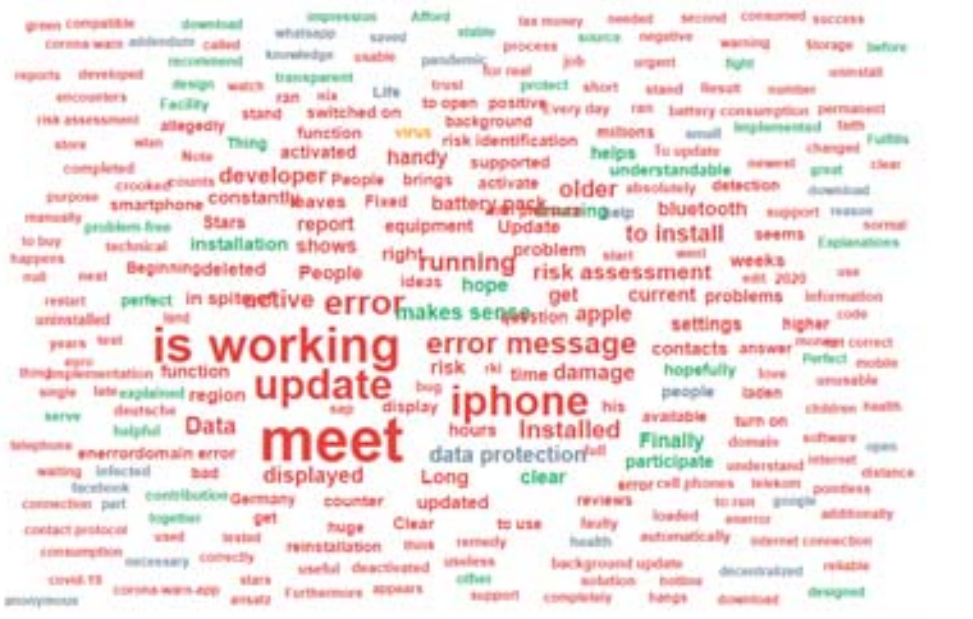 | The cortical.io tool did not respond, due to large data size |

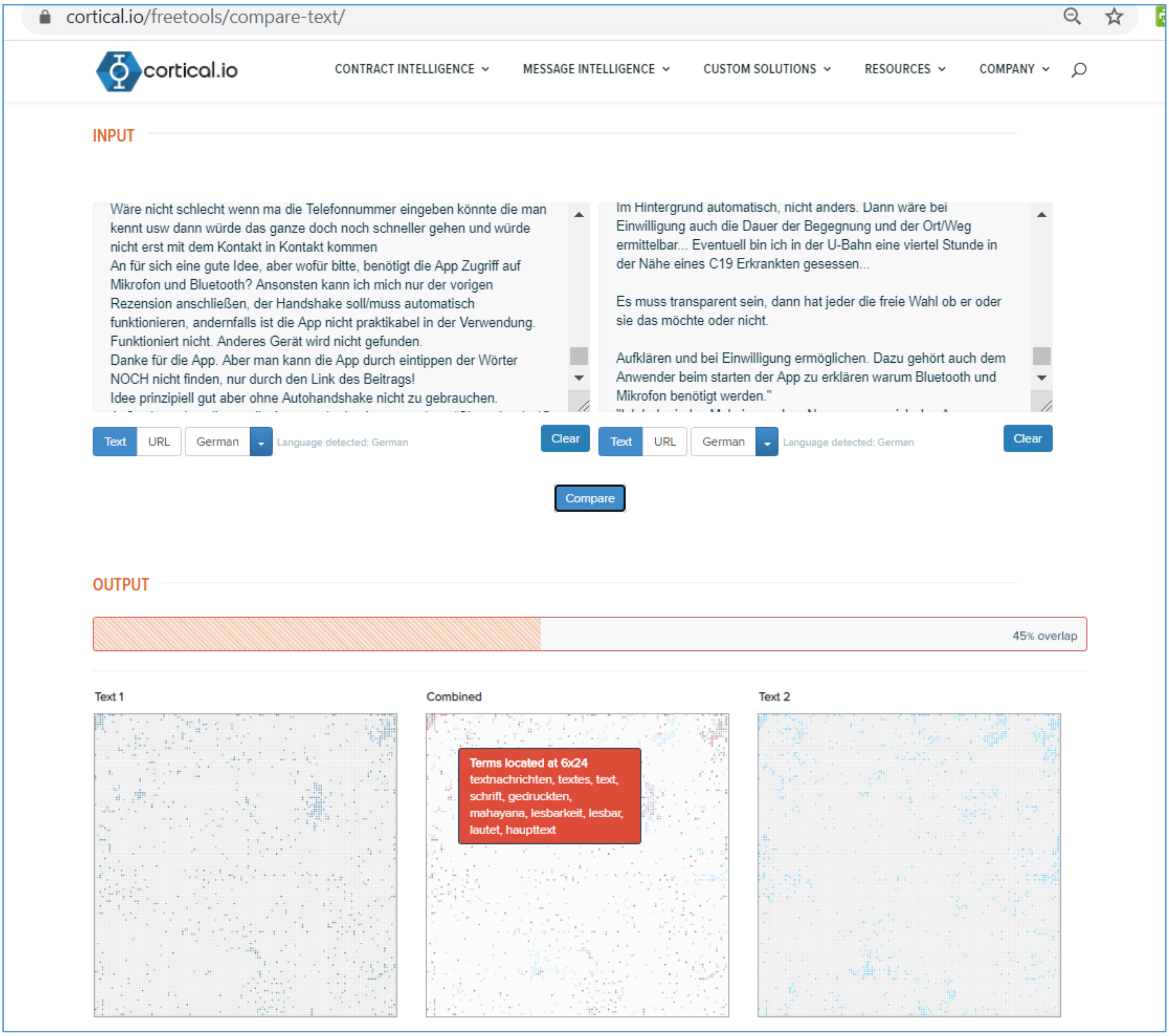

**Figure 28- A screenshot of the cortical.io tool, comparing the semantic of the texts in the review dataset of the two OS version of the *Stopp-Corona* Austria app**

**Lesson learned/recommendation**: The semantic-overlap measures between the two OS versions of the apps ranged between 45% and 86%. Possible root causes for low or high similarity should be studied in future works.

## 4.7 Correlation analysis

In this phase of our paper, we present a number of "exploratory" correlation analysis between various pairs of metrics in our dataset.

### 4.7.1 RQ7: Correlation of app downloads with country population

One would expect to see a correlation between the number of app downloads with each country's population size. Such an analysis would also provide us with a measure of usage (penetration) of the apps in each country.

We visualize the correlation of the number of downloads with country populations as a scatter-plot in Figure 29. We also show the linear trend lines in the chart. As discussed in Section 3.2, to get the number of downloads, we interpolated the number of downloads from Play Store's estimate, e.g., we averaged the Protect Scotland app's estimated download count of 100,000+ (meaning 100,001–500,000) to 300,000.

The Pearson correlation coefficient between the two metrics is 0.651. Thus, we can say that there is a reasonable correlation between the two metrics, i.e., for a country with a larger population, as one would expect, there are more downloads. However, the cases of Finland (FI) and Germany (DE) are special (in a positive way) since they are above the trend line.

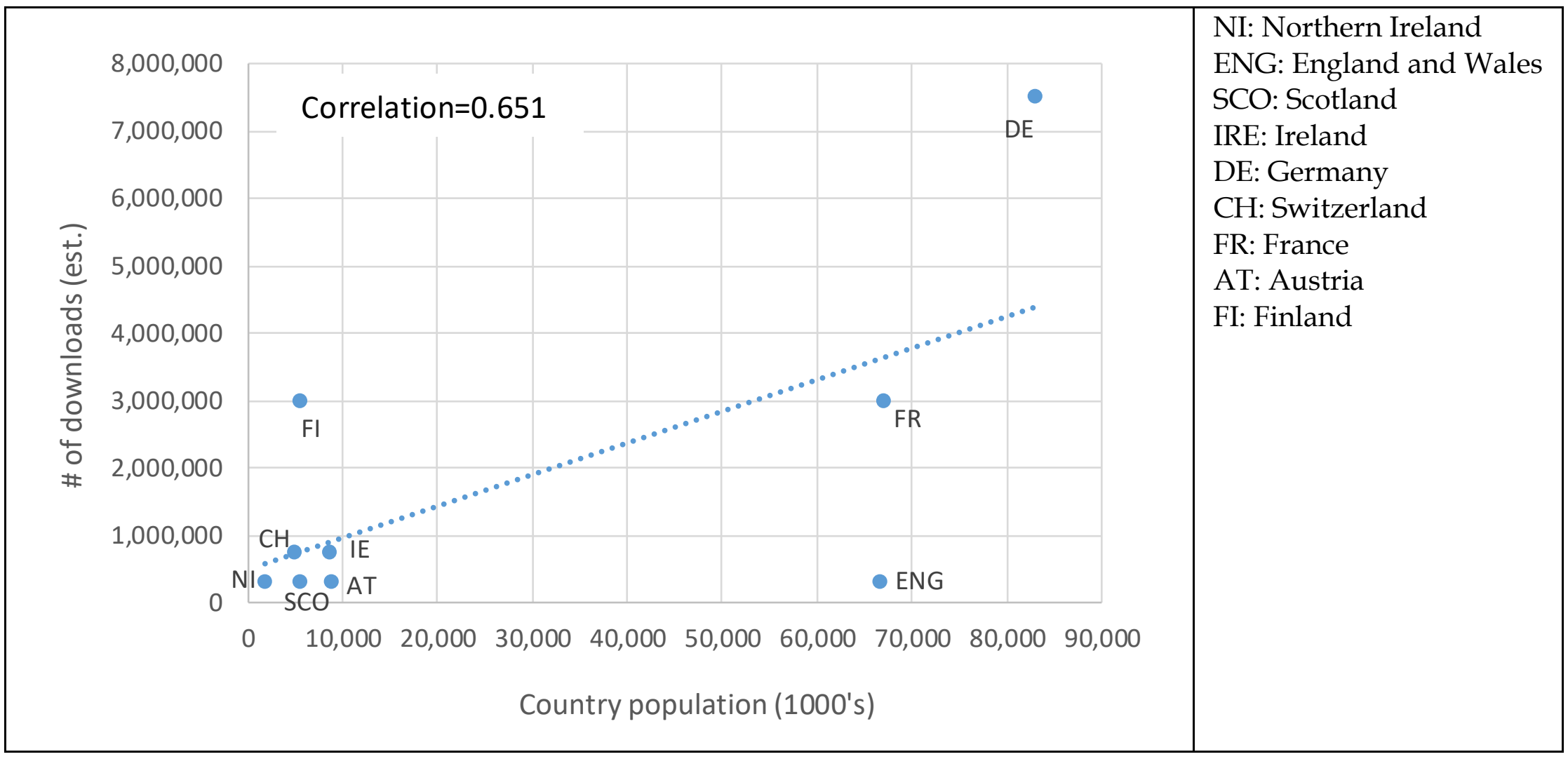

**Figure 29- Number of downloads versus country populations (1,000's) (estimated); including the linear regression lines**

On the other hand, countries such as England and France are below the trend line and raise some questions, e.g., one wonders why there have been relatively fewer downloads in those countries. One could analyze such questions based on the "social fabric" [57] of each country and also using different social metrics.

Although we are not social scientists, given our limited knowledge of how societies work and how citizens relate themselves to the societies that they live in, we believe that number of downloads, to some extent, portray, in a macro-scale in the context of a country, the level of its citizens' involvement (engagement) in society and the social responsibility. For these attributes, there have been many studies, advances and metrics (indices) in social sciences. We were able to find data for two such relevant metrics: (1) Trust In Public Institutions index (TIPI), from an OECD dataset[1]: a value between 0 (no trust at all, in public institutions) to 10 (complete trust); and (2) Civic Engagement Index (CEI): an indicators between 0-100, calculated mainly based on voter participation which some argue being "*the best existing means of measuring civic and political engagement*"[2].

We show the two correlations in Figure 30 as two scatter plots: correlations of the number of downloads (estimated), normalized by the population size, with two above social metrics (TIPI, and CEI). The Pearson correlation coefficients in the two correlation charts are 0.455 and -0.166, as also embedded in the charts. The former shows a moderate correlation, while the latter indicates a very weak negative correlation, which is somewhat surprising (discussed in detail in the following).

[1] ourworldindata.org/trust
[2] www.oecdbetterlifeindex.org/topics/civic-engagement/

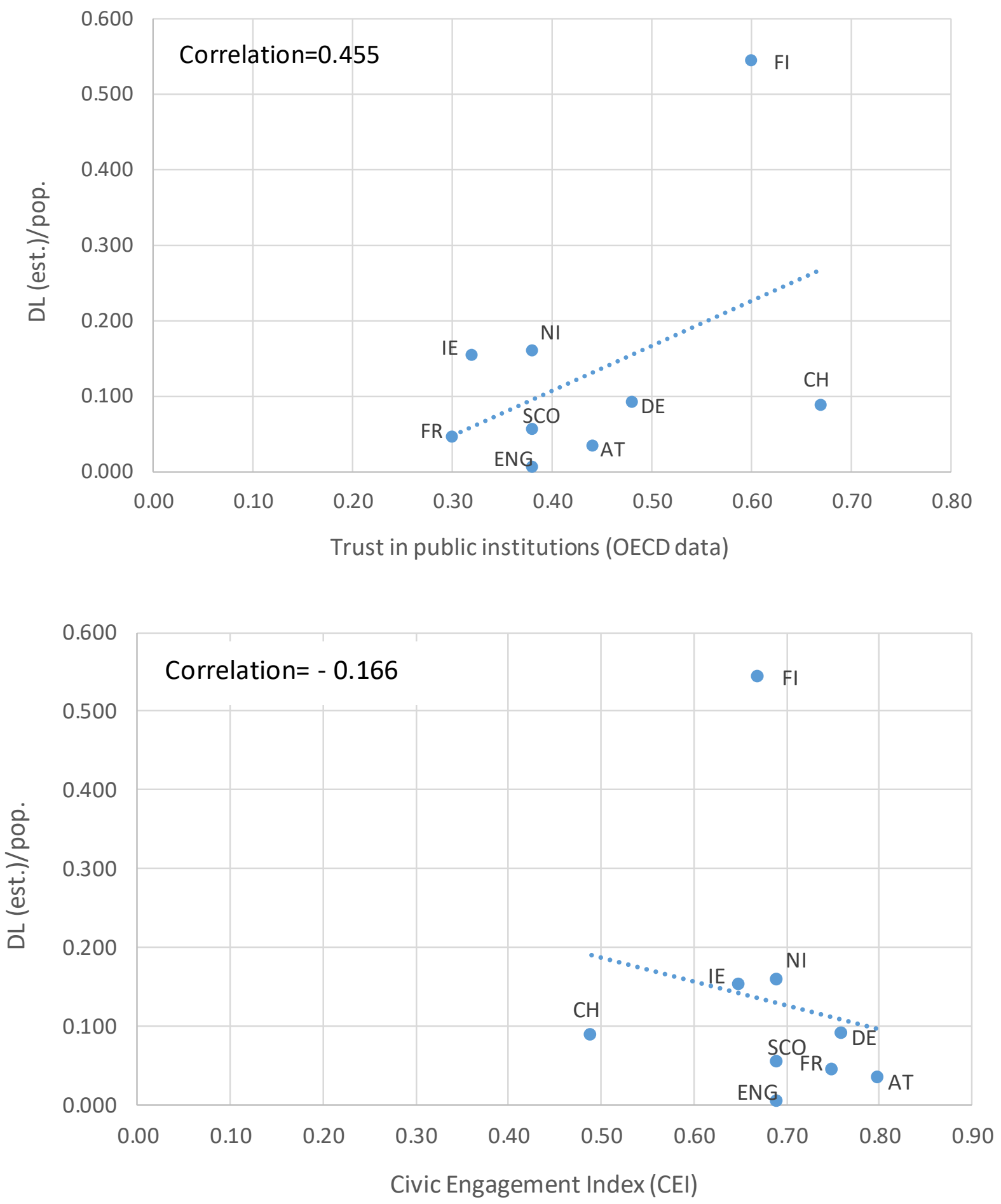

**Figure 30- Correlations of the number of downloads with two social metrics: Trust In Public Institutions index (TIPI), and Civic Engagement Index (CEI)**

We discuss the most important/interesting observations from the charts of Figure 29 and Figure 30 below:

- The Finnish app, German app, and English app ("ENG" for England in the charts) could be considered “exceptions” in the charts. The Finnish app is the "best" performing, while the English (covering England + Wales) app is the worst performing in all charts of Figure 29 to Figure 30.
- Finland ranks second in the TIPI index and first in the normalized measure of downloads to population ratio (DL/pop.). Among the nine data points, this value ranges between 0.005 (for NHS app) to 0.544 (for Finland Koronavilkku app), meaning that, for those two countries, respectively, about 5 of each 1,000 citizens and 1 of every two citizens have downloaded the app. This is such a diverse variation in usage (penetration) of the apps when comparing those countries. Finland’s “*good*” performance with such a high download ratio has also been covered in many news articles[1, 2].
- Higher relative download ratios of German and Finnish apps (Figure 29) could be due to a variety of reasons, e.g., those governments have taken more proactive measures to do more publicity for the apps in their countries or had "encouraged" their citizens to download and use the app. Again, going in-depth into these important issues is outside the scope of our paper since they related to the behavioral science, social aspects, and epidemiologic aspects of the apps and further work on these issues are needed, similar to the papers published already on these topics [24, 27] (as reviewed in Section 2.5).
- Switzerland (country code: CH) has the highest rank in the TIPI index, but the download to population ratio of the Swiss app is quite low. To find out what has possibly led to such a low download volume, we did a news search for the Swiss app and immediately found news articles that the legal actions and even a referendum has been

[1] qz.com/1898960/whats-behind-finlands-contact-tracing-app-success-user-privacy/
[2] uk.reuters.com/article/us-health-coronavirus-finland-app/one-in-four-finns-downloaded-covid-19-tracing-app-in-four-days-idUKKBN25U20H

"*launched against SwissCovid app*"[1,2]. It could be that such events have led the public to think again before installing the app.

Of course, we should interpret correlation data with caution since, as it is well known in the statistics and general scientific literature that: "*Correlation does not imply causation*" [58, 59], which refers to the inability to legitimately deduce a cause-and-effect relationship between two variables solely based on an observed association or correlation between them.

**Lesson learned/recommendation**: There is a moderate correlation between the number of downloads normalized by the population size and the *Trust In Public Institutions index* (TIPI). This seems to denote that the more trust a country's population, as a whole, has on their government, the higher the ratio of app downloads, and expectedly the higher the use.

#### 4.7.2 RQ8: Correlation of the number of reviews versus country population and also the number of downloads: What ratio of users have left reviews?

Similar to the previous RQ, one would expect to see a correlation between the number of app reviews with each country's population size. We visualize the correlation of the number of reviews with population data as a scatter-plot in Figure 31. The distribution of data points in Figure 31 is quite similar to the distribution of data points in Figure 29, implying that the number of downloads and reviews have had similar trends.

A higher number of reviews, relative to each country's population, could imply a variety of possible factors: (1) whether the population of one country tends to be more "outspoken" (or critical) of public (governmental) activities, in this case: contact-tracing apps, than other countries; (2) whether the apps of certain countries have really high or really low quality, as it seems that users tend to submit reviews in those extreme cases (according to comments mentioned in most reviews); and (3) Higher or more active engagement with national digital technologies (do citizens "care" to provide feedback to apps as a type of digital technologies?).

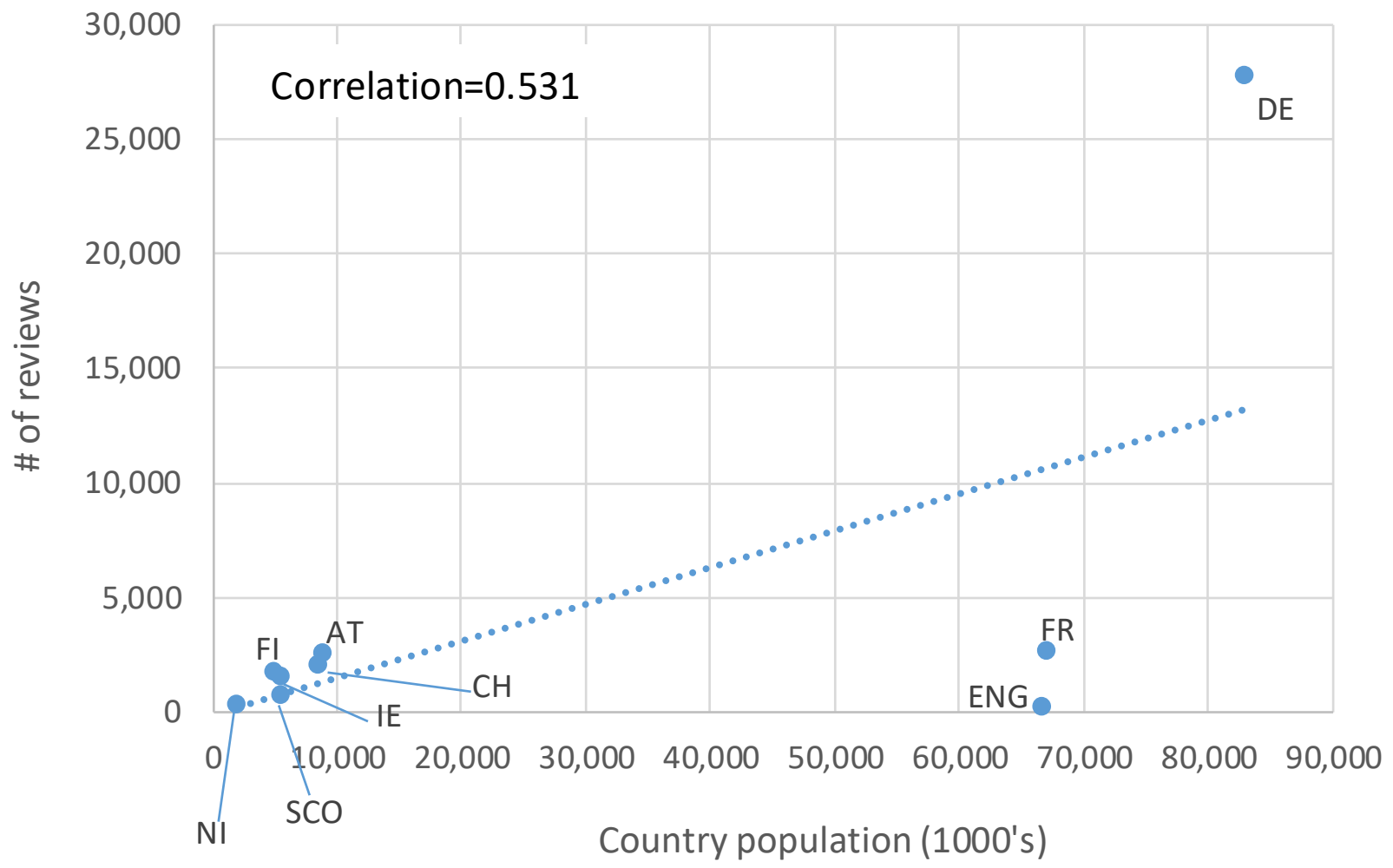

**Figure 31- Number of reviews versus country populations (1,000's); including the linear regression line**

As the last analysis in this section, we had the data to calculate the download-to-review ratios of the apps, i.e., to determine the ratio of users (people who have downloaded the apps) who have left reviews. We show the data as a scatter plot in Figure 32. We can see, for example, that the highest and lowest performers in this metric are: Austrian app (one review per 153 users), and the Finnish app (one review per 2,351 users), respectively. The German app is the second-highest, with one review per 358 users. Similar to the above discussions about similar metrics, root-cause analysis of this last indicator would also need to analyze the behavioral, social, and epidemiologic aspects of the apps, and we leave that to further works.

[1] lenews.ch/2020/07/24/referendum-launched-against-swisscovid-app/
[2] lenews.ch/2020/10/13/initiative-against-swiss-covid-app-fails/

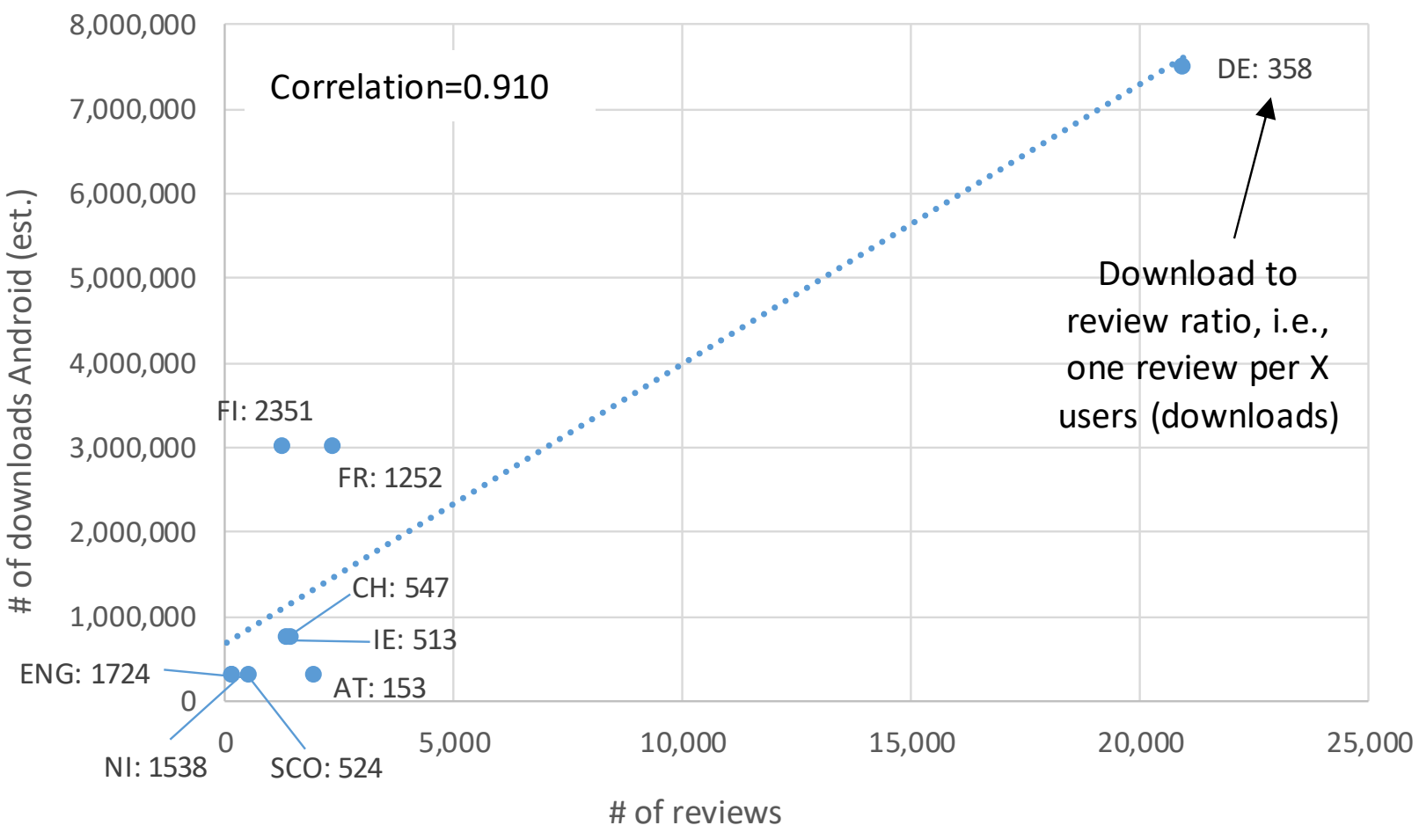

**Figure 32- Download to review ratios of the apps**

**Lesson learned**: The apps from the German-speaking countries Austria and Germany have the highest number of reviews per number of users.

### 4.8 RQ9: Trends of review volumes and their sentiments over time

Another insightful/interesting aspect of the review dataset that we found to be worth analyzing was the trends of review volumes and their sentiments over time. The AppBot tool again already has a feature to get such trend-charts easily and then we could do our interpretation / analysis on the charts.

We show those charts for all the nine apps, under study, in Figure 33. In the AppBot tool, there is even a way to select either or both of the OS versions of the app and have the chart generated separately. For the case of three apps (the Austrian, UK and Irish apps), chosen as examples, we provide both charts for the two OS versions (Android and iOS). For the sake of brevity and space, for the other apps, we provide only the charts for the iOS versions only.

One important piece of information in this timeline analysis is the release date of each version of a given app. The Apple App Store provides the version history (release date of each version) of a given app and AppBot uses that data to include those time instances in the generated charts (see the example of the Austrian app in Figure 33). The Google Play Store does not provide that information for its hosted apps.

As expected, the first release date of apps of different countries are different since each country's response to COVID and its decision for developing and releasing contact-tracing apps have been made independently. Back in Section 3.1, we included the first release dates in Table 2, and also discussed the number of releases after that date until the data-extraction date of our study (September 17, 2020).

| App | **Trends of review volumes and their sentiments over time** |
| --- | --- |
| Stopp-Corona Austria | iOS: 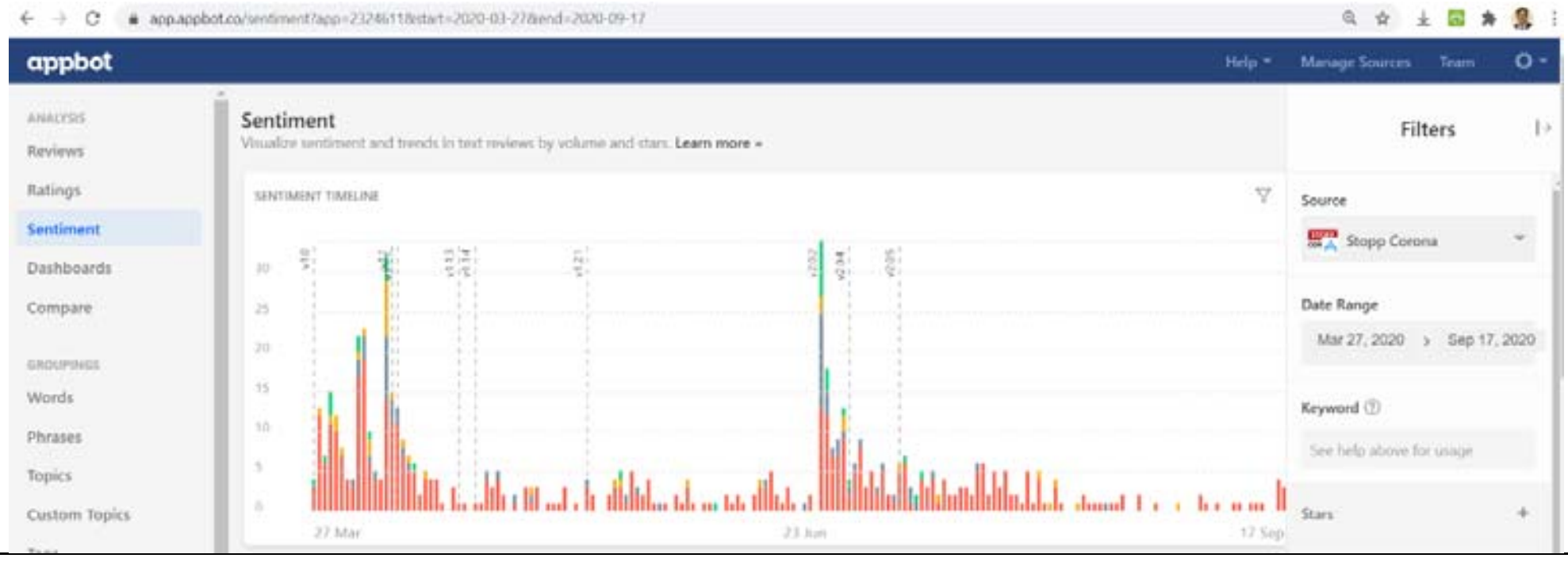  |
| | Android: |

<table>
<tr><td></td><td>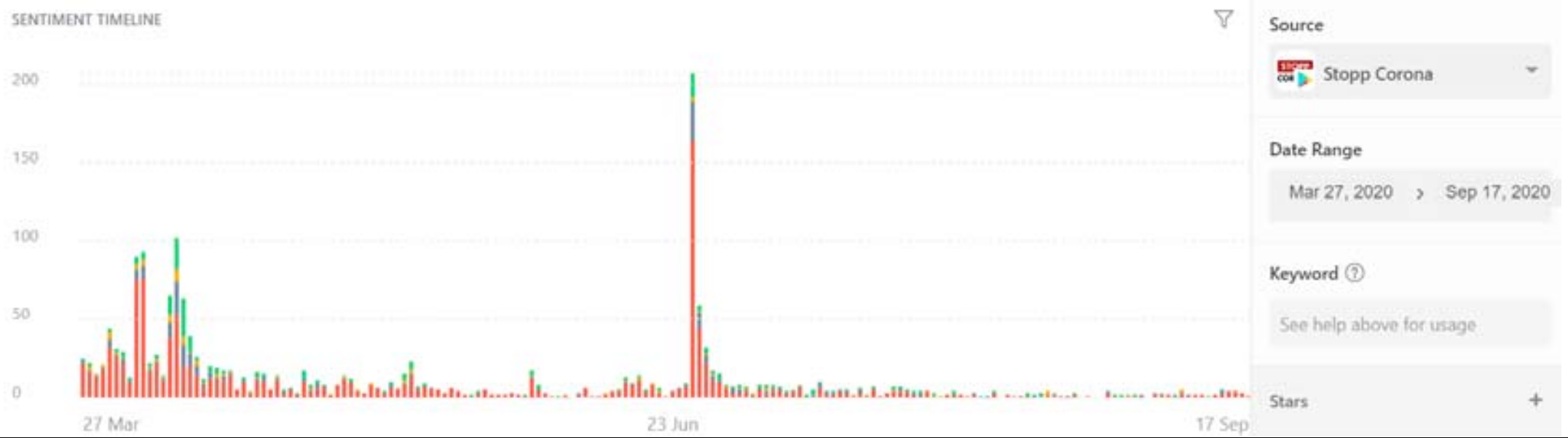
</td></tr>
<tr><td>COVID Tracker Ireland</td><td>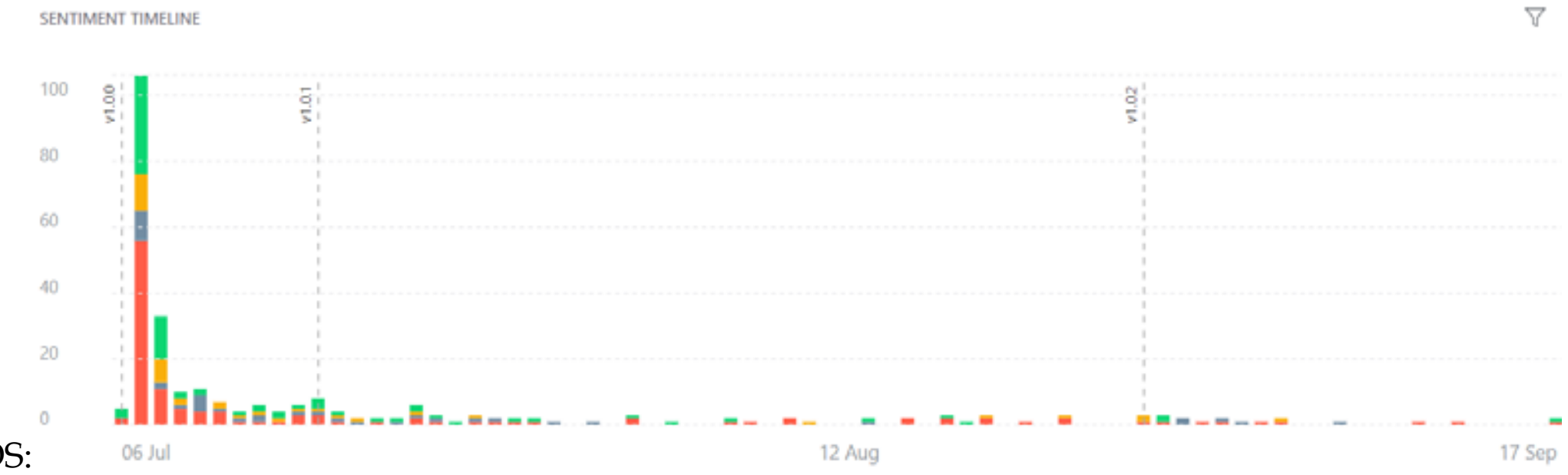

iOS:</td></tr>
<tr><td></td><td>Android: 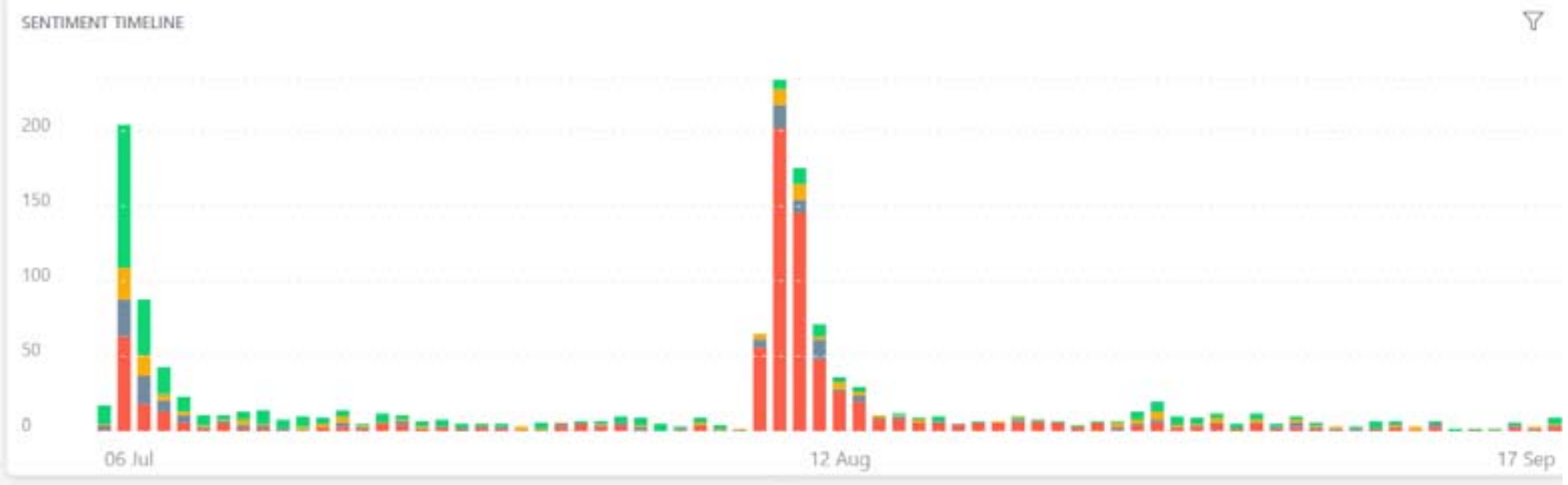
</td></tr>
<tr><td>StopCOVID NI</td><td>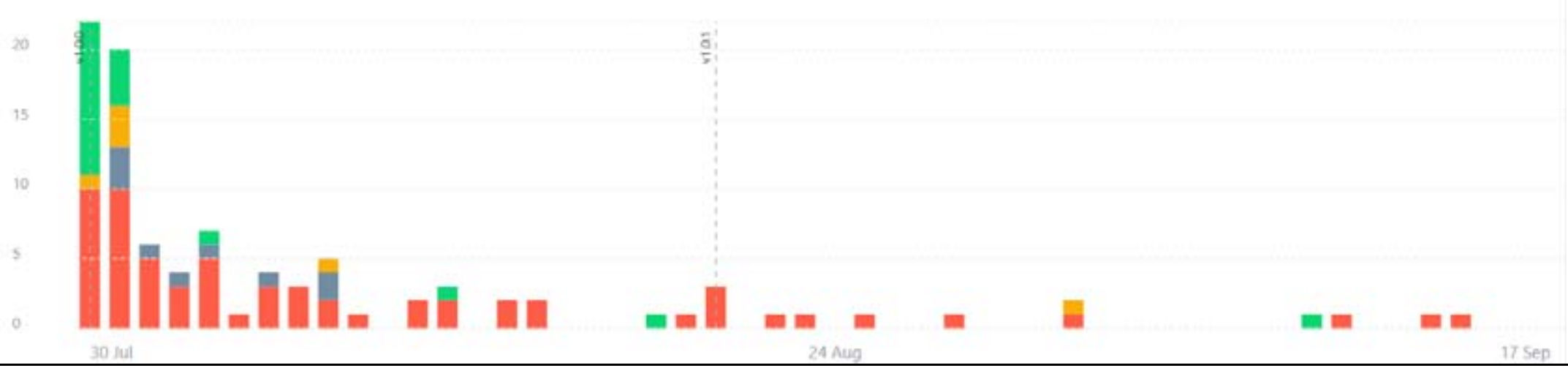
</td></tr>
<tr><td>NHS COVID-19</td><td>iOS:<br>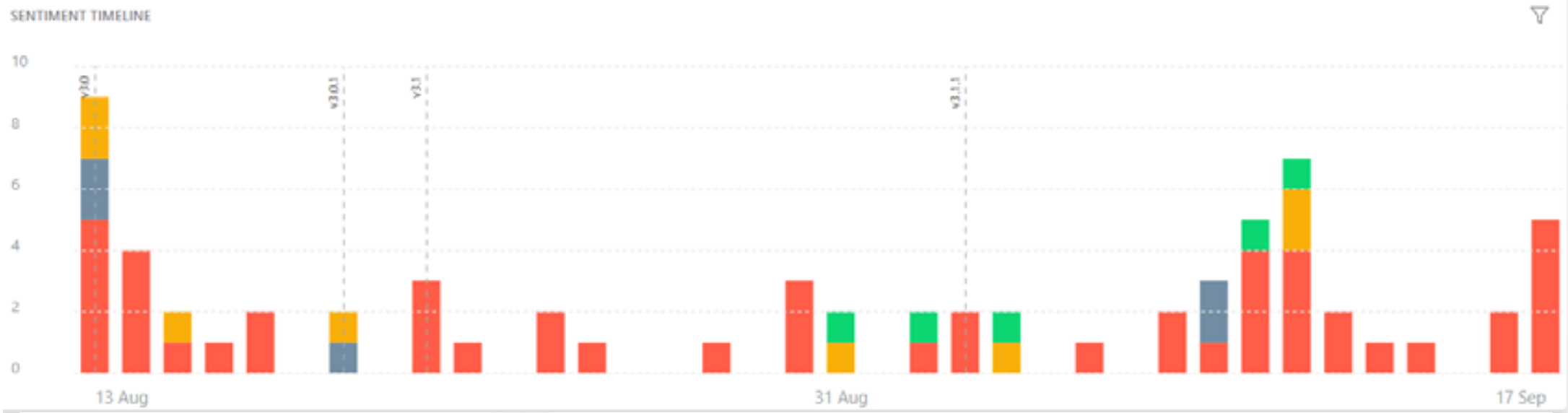

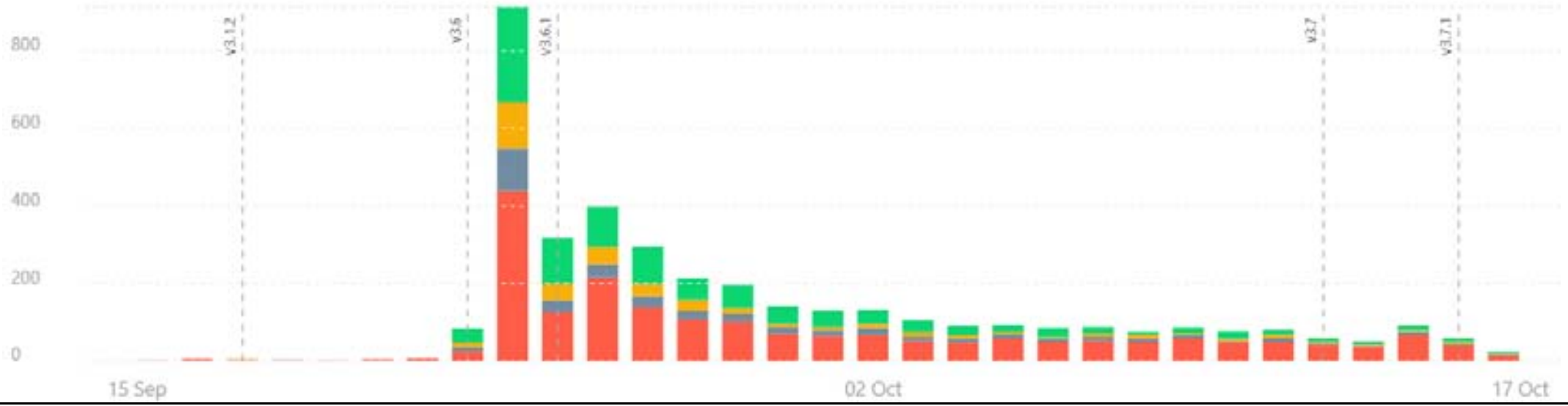
</td></tr>
</table>

| | |
|---|---|
| | 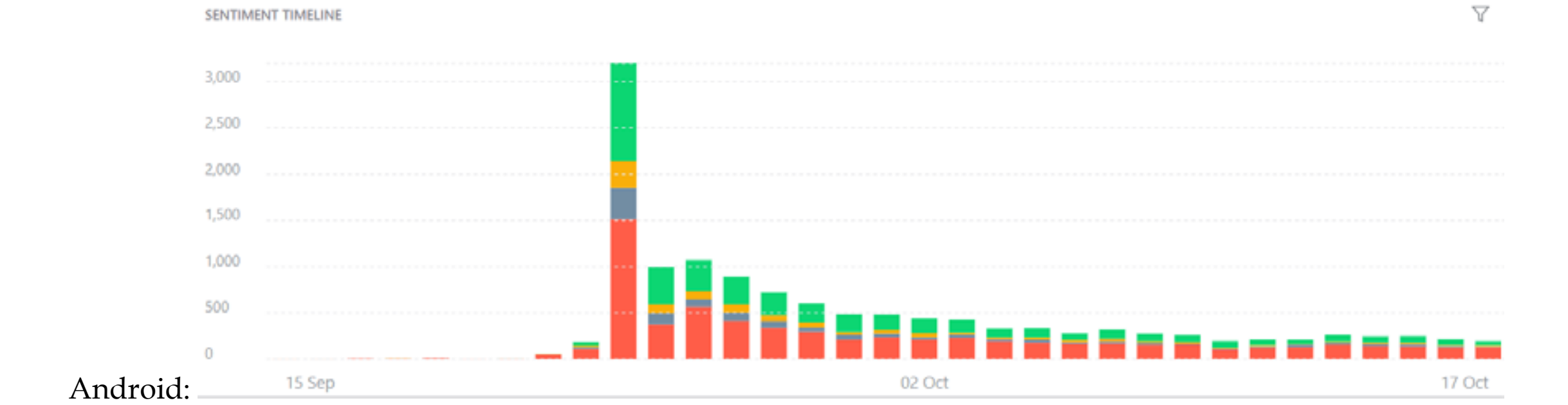  Android: |
| Protect Scotland | 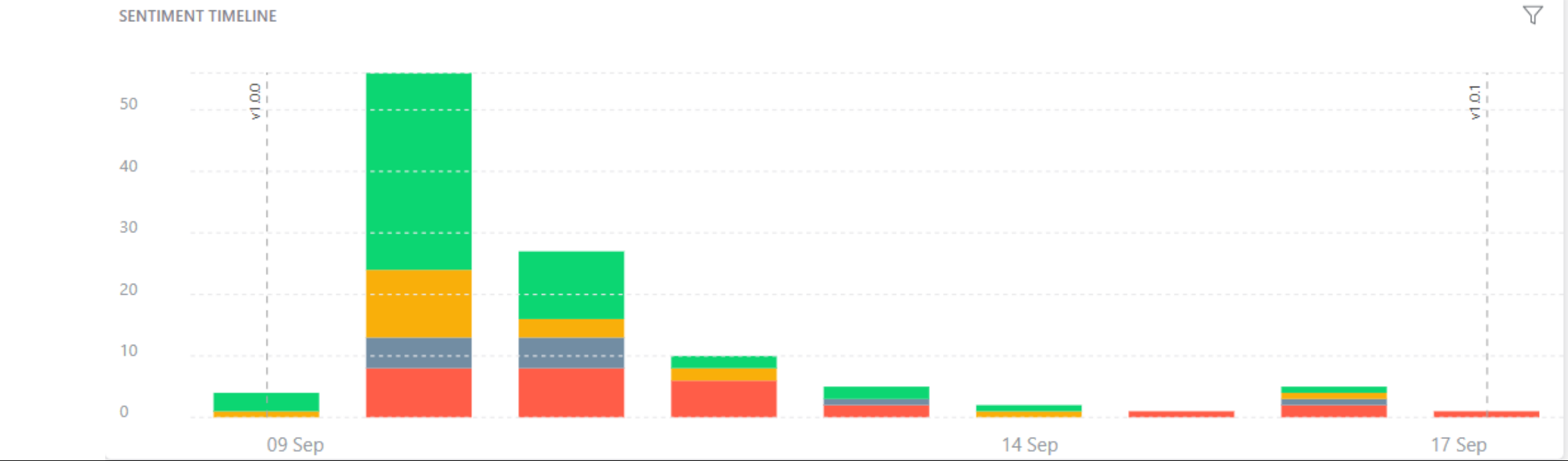  |
| Corona-Warn Germany | 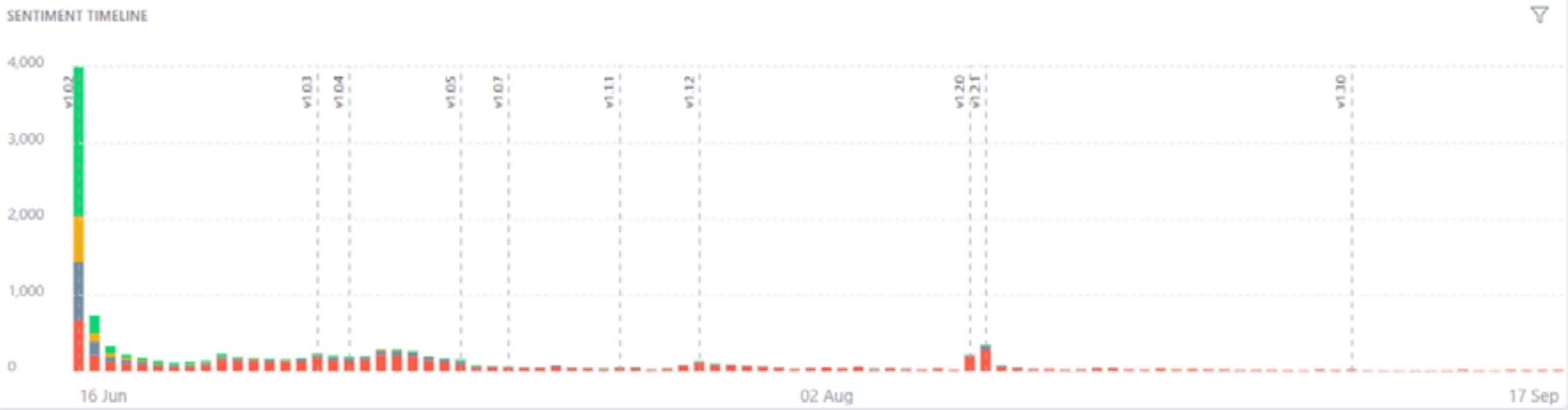  |
| SwissCovid | 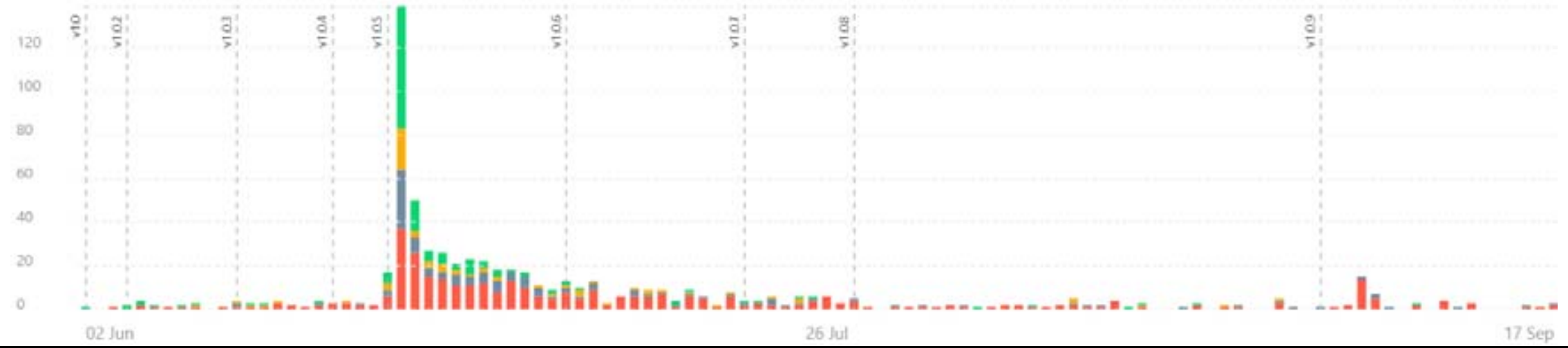  |
| StopCovid France | 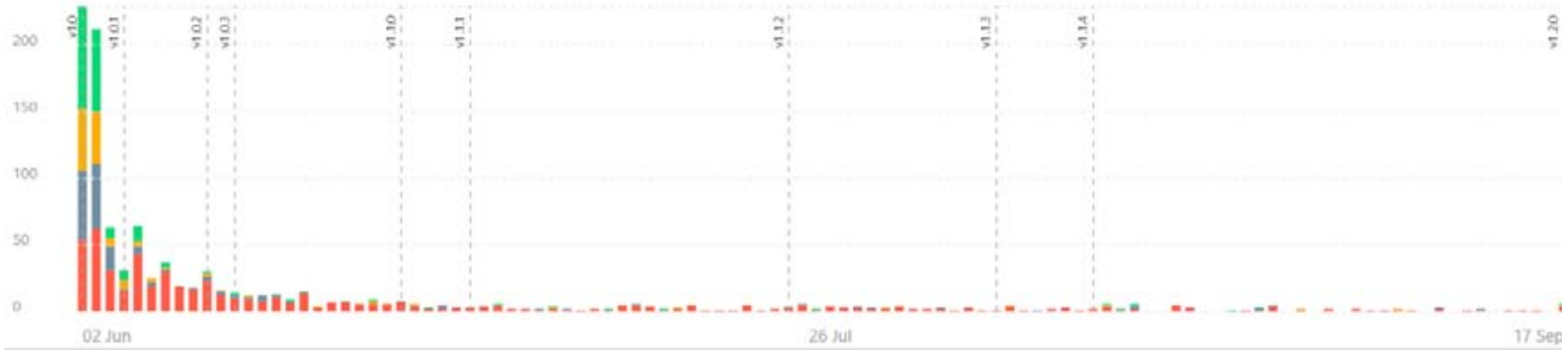  |
| Finland Koronavilkku | 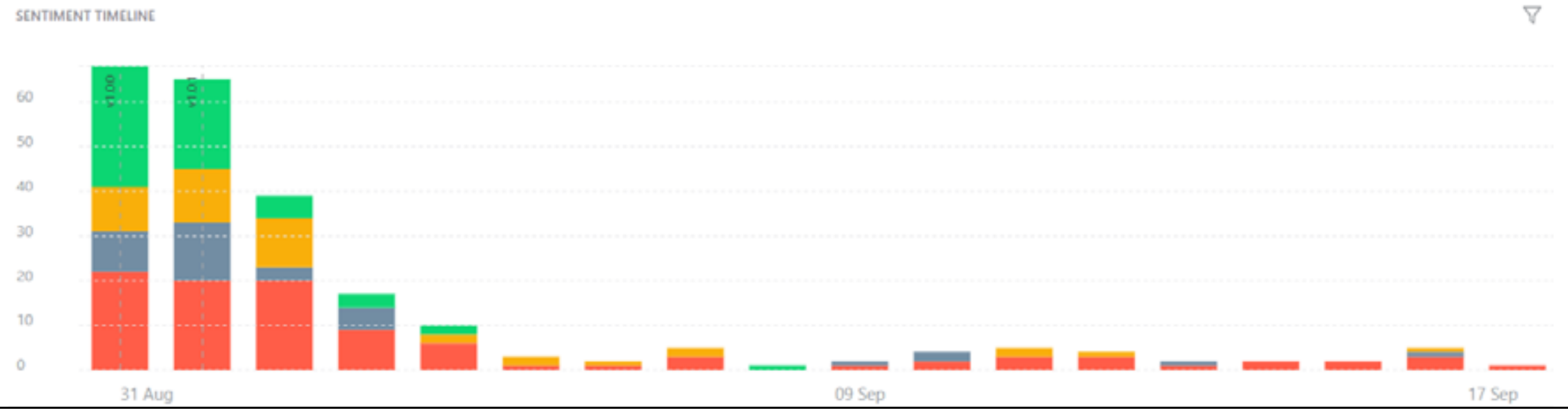  |

**Figure 33- Trends of review volumes and their sentiments over time for two of the apps**

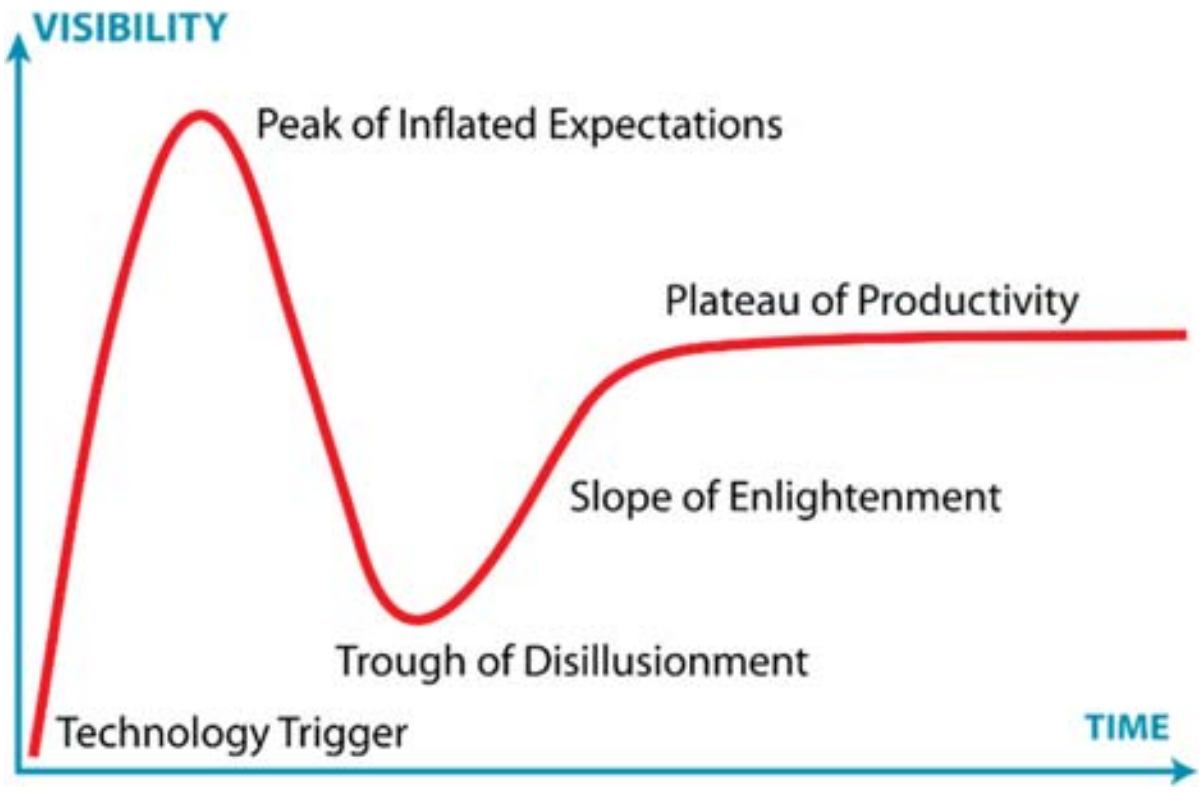

**Figure 34- The Gartner "hype cycle" [60]**

We discuss the most interesting and insightful observations in Figure 33 below:

- Trends of review volumes are changing throughout the time horizon. Often for the case of most apps under study, a large volume of mostly negative reviews has been recorded in the first few days/weeks of its first release, and then a decline has occurred. Does this imply that users have lost interest in the apps over time? This reminds us of the well-known "hype cycle" [60]. As a matter of fact, various reports have linked the uptake of contact-tracing apps to hype cycle, e.g., a report by MIT Technology Review[1] mentioned that: "*If contact-tracing apps are following Gartner's famous hype cycle, it's hard to avoid the conclusion they are now firmly in the ;trough of disillusionment'. Initial excitement that they could be a crucial part of the arsenal against covid-19 has given way to fears it could all come to nothing, despite large investments of money and time*". A news article in the American NBC News, came with this title: "*Covid apps went through the hype cycle. Now, they might be ready to work*"[2]. Trends of review volumes also show this cycle trend to some extent. But of course, we should clarify that review volumes are not necessarily the "usage" trends, but there may be some relationship between the two. In early days and weeks of an app, it is often the case that governments conduct massive publicity campaigns via TV ads, social media, and newspapers to "encourage" their citizens to install and use the apps. And, citizens indeed respond very proactively to such calls in the initial days, e.g., it was reported[3] by the UK Department (Ministry) of Health and Social Care that about six million people downloaded the NHS contact-tracing app on the first day it was launched. Out of those six million people, as Figure 33 shows, about 3,000 users left reviews in the Android version and about 900 users left reviews in the iOS version of the app.
- Some reviews have come before a given app was even released, e.g., for the Swiss app, the first review came in June 2nd; however, the app was officially lunched[4] on June 25th. By looking into that first review[5] and its author, we found that that review was written by a software engineer, and it could be guessed that he was in the development team of the app.
- The UK (NHS) app was officially "launched" in the nation to the citizens on September 24, 2020. But in its Apple app store page[6], we see that it was first released (version 3.0) on Aug. 13, 2020, more than 40 days before the official lunch date. That period was actually a "trial" (test) period[7] in the UK's Isle of Wight, in the London borough of Newham, and among NHS volunteers. The fact that the app was downloadable from the app store but not working for regular people caused many confusions among citizens and many negative reviews have come in due to that in those 40 days, e.g., "*I've been asked to download this app by my local leisure centre. The app says it is only for Isle of Wight and Newnham residents and is asking for a code. I cant use it*"[8], and "*The BBC & the NHS both say that the app is ready. The app says that's only a test for the Isle of Wight. Which is it?? FAIL!!!*"[9]

---

[1] www.technologyreview.com/2020/08/10/1006174/covid-contract-tracing-app-germany-ireland-success/
[2] www.nbcnews.com/tech/tech-news/covid-apps-went-through-hype-cycle-now-they-might-be-n1242249
[3] www.digitalhealth.net/2020/09/nhs-covid-19-app-downloaded-10-million-times-since-launch/
[4] www.bag.admin.ch/bag/en/home/das-bag/aktuell/news/news-25-06-2020.html
[5] appbot.co/apps/2348215-swisscovid/reviews/1822036574/
[6] apps.apple.com/us/app/id1520427663 3.0
[7] www.digitalhealth.net/2020/08/covid-19-new-trial-nhs-contact-tracing-app/
[8] app.appbot.co/apps/2411517-nhs-covid-19/reviews/1957894169/
[9] app.appbot.co/apps/2411517-nhs-covid-19/reviews/1953464033/

- We also noticed the review "bursts" after apps' first release and wondered about their underlying factors. For example, for the Stopp-Corona Austria app, we can notice in Figure 33 a peak in review volume on June 26th.
  - We hypothesized that bursts in the review volumes could be due to the release of new versions. But with the release of each version, we cannot notice much change in the ongoing review volumes. This could be since most mobile phones do the installation of app updates automatically and seamlessly, so layman users do not notice the new versions on their phones. As a related line of work, there have been many studies in the Mining Software Repositories (MSR) community, analyzing similar software artifacts, e.g., bugs and feature requests, in the time horizon. For example, e.g., a paper by Garousi [61] analyzed the "arrival" pattern of when issues (bugs and feature requests) have been submitted and then "closed" (addressed) by the development team. We thus find out that these trends are *quite different* than conventional software systems, for which with new versions, sometimes more bug reports are entered in their bug repositories [61].
  - As a second possible factor, we hypothesized that changes (especially jumps) in the review volumes could be due to media coverage of the apps, i.e., if citizens hear more about the apps in the news, they would install/use them more on a given day, and then possibly leave reviews for the app in the app stores. We found that this hypothesis hold for a few cases. For example, for the Austrian app, by doing a Google search for "*stopp corona app österreich 26th June*", we immediately found a news article: "*Corona app: The update for the automatic handshake is here*"[1] (the article is in German). The news article discussed an important issue: "*A new and improved version of the Stop Corona app of the Austrian Red Cross is now available in the stores. After the update, the automatic digital handshake now works on all devices with the mobile operating systems iOS (Apple) and Android (Google)*". By reviewing some of the reviews recorded on that day, we clearly saw that many (34 reviews) were recorded on that single day, some of which were:
    - *"Thanks for today's update. A very useful and good app.*
    - *The official interface is now used, so an anonymous handshake can be carried out in the background, finally! Many thanks to the developers.*
    - *All of our neighbors [countries] also rely on the API from Apple & Google and look almost identical with the exception of a few design differences. The design is really successful and intuitive."*

**Lesson learned**: Bursts in the review volumes seem to be not strongly correlated with new releases (versions) of the apps, but instead with more news coverage of the apps.

- For the cases of the two OS versions of the Ireland app, we notice slight differences in review volume "peaks" (bursts) between them (the timeline around early to mid-August). A noteworthy situation developed in that time-frame and was widely covered in the media, as we discuss next.
  - According to an Irish news article[2], "*From August 7 to August 12, more than 150,000 uninstalls [of the COVID Tracker Ireland app] had been reported [mainly due to high battery usage of the app]*". According to another article[3], "*a Google Play Services update caused the app to rapidly drain handset batteries for a two-day period earlier this month [August]*". On August 10th, another news article[4] reported that Google would "*launch fix for battery drain affecting Covid Tracker Ireland app*". Proactive communication was made to the public by the Irish health ministry on this, e.g., see the Tweet shows in Figure 35.

[1] www.roteskreuz.at/news/datum/2020/06/26/stopp-corona-app-das-update-fuer-den-automatischen-handshake-ist-da/
[2] www.irishexaminer.com/news/arid-40058456.html
[3] uk.reuters.com/article/us-health-coronavirus-ireland-apps/active-irish-covid-19-tracing-app-users-drop-on-battery-problem-hse-idUKKBN25N1PA
[4] www.siliconrepublic.com/enterprise/battery-drain-covid-tracker-ireland-app-google

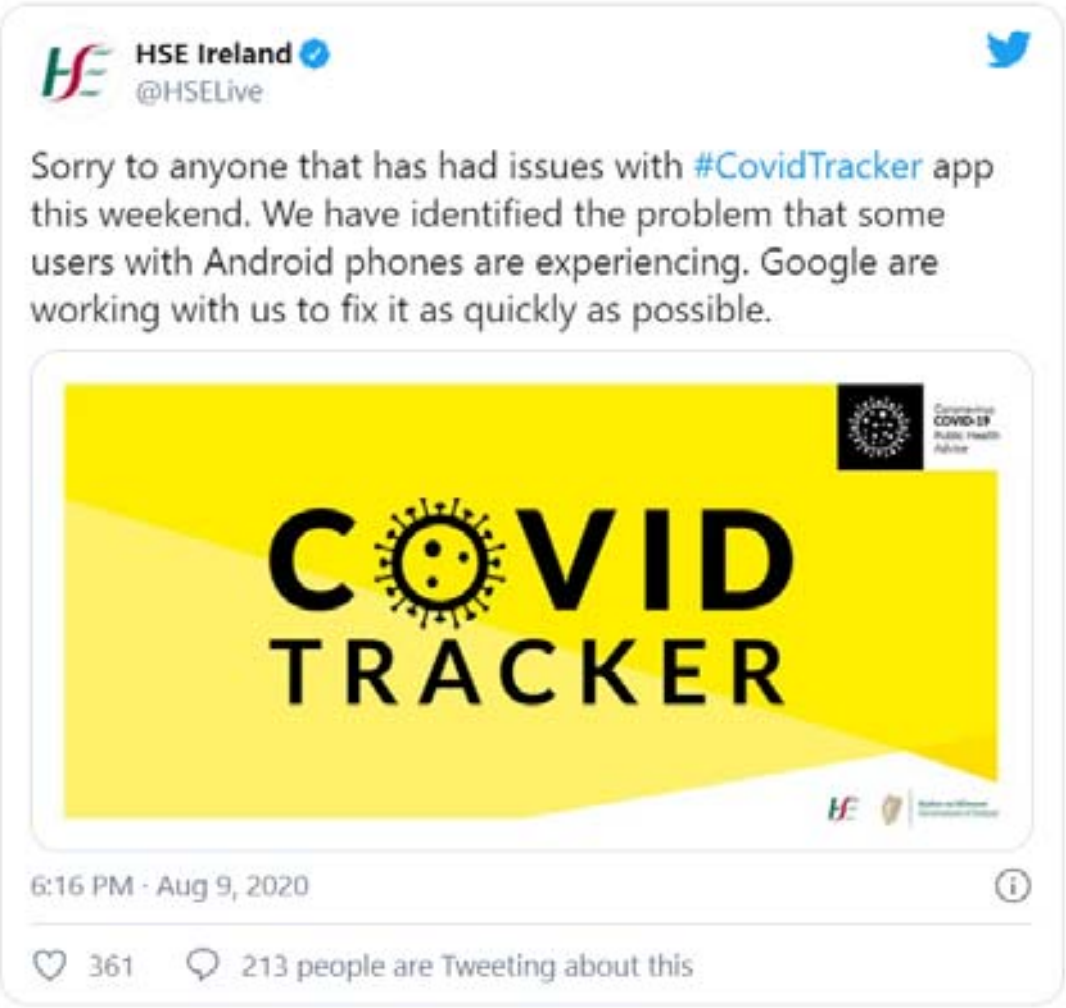

**Figure 35- A tweet[1] by the Irish health ministry about an issue in their app**

- A few days later, the health ministry publicized that the update has been installed on the phones by mentioning that: "*Google have informed us that the majority of Android phones in Ireland have been updated but it may take a day or so for every Android phone to receive the update. We would encourage anyone that uninstalled the app at the weekend to download it again over the next few days.*"[2]
- As per our analysis of review texts, most of the reviews submitted in the few days around August 10th, for the Android version of this app, are related to the high battery drainage of the app.
- A software-engineering observations (take-away message) from this issue is that even if the app itself was not modified in that time period, but an automated software update of the Google Play Services did impact the performance and battery usage of the app. This reminds us of the highly complex nature of developing these apps and the need for proactive analysis of updates and their dependencies.

**Lesson learned/recommendations**: There could be unexpected inter-dependencies among the apps and various aspects of the mobile OS that they are running on. Updates to the OS could adversely impact a given app and could easily cause major dissatisfaction by the app users. Thus, the development team should work with OS vendors (in the above example case, Google being behind the Android OS) to prevent such chaotic situations.

## 5 Implications and Discussions

Now that we have presented extensively the results of our analysis in Section 4, we take a step back and look at our original goal in this study, which has been to gain insights into the user reviews of contact-tracing apps to find out what end users think of COVID contact-tracing apps, and the main problems that users have reported. As a reminder, our focus has been to assess the "software in society" aspects of the apps based on their users reviews, and thus we need to clarify again the scope of our work, and the many related issues that are important but outside this paper's scope: public health aspects of the app, behavioral science [4], and epidemiology.

We present next the implications of our results. Afterwards, we discuss the limitations of our work.

### 5.1 Implications for Various Stakeholders

As discussed in Section 1, our study and its results could provide implications, benefits and insights to various stakeholders: (1) software engineering teams of the apps, (2) decision-makers, and public health experts who manage the development and adoption of those apps in the society, (3) developers of app-review mining tools, and (4) other SE researchers.

Also, let us recall from Section 3.2 that our research method in this paper has been "exploratory" case study [7], whose goals are defined as: "*finding out what is happening, seeking new insights and generating ideas and hypotheses for new research*", and those have been the goals of our study. Thus, throughout our work, we gathered and derived the implications, benefits and insights, as we highlighted throughout Section 4.

---

[1] twitter.com/HSELive/status/1292510112184229889

[2] www.irishtimes.com/business/technology/hse-says-fix-for-covid-tracker-app-rolled-out-to-all-android-users-1.4326646

To “package” the lesson learned, recommendations, and implications that we discussed and derived in different parts of Section 4 into one single picture, we synthesize and present them in Table 4. For full traceability, we also provide the section number from which each implication and benefit has been derived so that readers can go back and read the details.

As the table shows, most of the evidence-based implications/benefits are for the software engineering teams of the apps, but we believe there are valuable suggestions to other stakeholders too.

**Table 4- Evidence-based lessons learned, recommendations and implications derived by our study for various stakeholders**

| Lesson learned/ recommendations / implications | Based on discussions in section number… | Stakeholders: Software engineering teams of the apps | Decision-makers, behavioral scientists and public health experts | Developers (vendors) of app-review analytics / mining tools | SE researchers |
|---|---|---|---|---|---|
| The users are generally dissatisfied with the apps under study, except the Scottish app. This issue is perhaps the most clear and the most important message of our study, which should be investigated by stakeholders. | 4.1 | x | x | | |
| Future studies could look into what factors has made the Scottish app be different than others in the pool of apps under study. That could a research question (RQ) to be studied by researchers in future works. | 4.1 | | | | x |
| Contact-tracing apps should be designed to be as simple as possible to operate (for usability), as we cannot expect layperson citizens to review the online FAQ pages of the app to properly configure it, especially for a safety-critical health-related app. | 4.2 | x | | | |
| Developers of the apps can and should engage directly with reviews and reply, not only gaining insight into the most commonly raised concerns but also to answer the questions in public view. This can even provide a positive “image” of the software engineering team behind the app, in public view (in terms of accountability, transparency, responsiveness, and being open to feedback). | 4.2 | x | | | |
| Just like any other mobile app, user reviews for contact-tracing apps range from a short phrase such as “*Not working*”, often not that useful nor insightful, to detailed objective reviews, which could be useful for various stakeholders. Thus, if any stakeholder (e.g., the app’s development team) wants to benefit in a qualitative way from the reviewers, they need to filter and analyze the "informative" reviews. | 4.2 | x | x | | |
| A common issue for most apps is high battery usage (drainage). Software engineers should use various heuristics and approaches to minimize battery usage. Mobile phone users are sensitive about battery usage, and any app that uses a high amount of battery would be likely to be uninstalled by mobile users. | 4.3 | x | | | |
| For the German apps, a substantial number of reviews are about the app not working, which can be seen as bug reports. But unfortunately, since most users are non-technical people, informative and important components of a bug report (e.g., phone model/version and steps to reproduce the defect) are not included in the review. Thus, it would be quite impossible for the app’s software engineering team to utilize those reviews as bug reports. A recommendation could be that in the app itself (e.g., in its “tutorial” screens), explicit messages are given to the users, asking them that, if they wish to submit bug reports as reviews, they should include important components of a bug report (e.g., phone model/version and steps to reproduce). | 4.3.1 | x | | | |
| A large number of cross-(mobile) device issues have been reported for the German and other apps too. This denotes inadequate cross-device testing of the apps, possibly due to the rush to release the apps to the public. Given the nature | 4.3.1 | x | | | |
| of the apps, and since the apps could be installed on any mobile device model/version by any citizen, the development and testing teams should have taken extra care in cross-device development and testing of the apps. There are many sources both in the academic literature [44, 45] and also grey literature[1] on this issue, which the development and testing teams can benefit from. | | | | | |
| Certain feature of a given app did not work for many users for several days, e.g., for the German app, a functionality called "Risk assessment". Such a malfunction usually gives a negative perception to users about an app, even if the other features of the app do work properly. It is thus logical to recommend that app developers should not include a feature in the app release if they predict or see from reviews that the feature does not work for certain users or on certain times/days. | 4.3.1 | x | | | |
| We are seeing rather trivial issue in the apps, i.e., users have to "activate" multiple times, instead of just once. We would have hoped that the test teams of the apps had detected and fixed those trivial issues before release. | 4.3.2 | x | | | |
| It is important that a given app automatically switches to the home country's language since some non-English users will feel odd if they see a sudden switch from their native language to English in the app's GUI. | 4.3.2 | x | | | |
| The development team of all apps should be proactive in replying to user reviews, and filtering informative reviews and getting more information (e.g., steps to reproduce the defects/problems) from them, e.g., by direct replies to the reviews in app stores. | 4.3.3 | x | | | |
| There seems to be rather trivial usability issues with some of the apps (e.g., the case of exposure notification errors in the NI app). This raises the question of the inadequate usability testing of the apps and possibility of releasing them on a "rush". | 4.3.3 | x | | | |
| Some of reviews provide insights on software engineering issues of the apps, e.g., not enough testing has been done on all possible types of QR codes, and not enough performance (load) testing has been done. | 4.3.3 | x | | | |
| For Android phones, the update mechanism of the OS and its components (e.g., APIs) should be "seamless" (automatic), since we cannot expect all users have the "technical" skills to do such tasks properly. | 4.3.3 | x | | | |
| The apps must be clearly identifiable and findable in app stores to maximize the number of users downloading it. | 4.4.3 | x | | | |
| Where possible, some feedback (such as statistics about COVID cases in the region and also number of close-by phone IDs recorded in the past) should be provided as a feature of the app, to encourage users that the app is working to emphasize the pro-social and individual benefit it is having. | 4.4.3 | x | x | | |
| A variety of insightful feature requests has been provided by users, e.g., by user of the German app: How many encounters there were with other app users (how many people you exchanged the keys with); infection numbers and spread at district level; can the app be used without internet? As a form of "iterative" requirements engineering" (elicitation) [6, 47-53] or "crowd-based" requirements engineering [54], app's software engineering teams are encouraged to review those feature requests and select a subset to be implemented. | 4.5.1 | x | x | | |
| Given the nature of the COVID pandemic, the governmental policies and guidelines regularly change and thus the contact-tracing apps have been regularly updated/adapted to those changes. This related to the widely-discussed issue of changing/unstable software | 4.5.1 | | | | x |

[1] www.google.com/search?q=movbile+app+%22cross+device%22+testing

| | | | | | |
|---|---|---|---|---|---|
| requirements. Thus, SE researchers are encouraged to work on such issues related to contact-tracing apps. | | | | | |
| While AppBot's feature to filter reviews to see feature requests only is a useful feature, we found many example reviews which AppBot incorrectly classified them as feature requests. We realize that an NLP/AI-based algorithm has been used to do that classification and such an algorithm will have limited precision, but still there is need to improve such algorithms by developers (vendors) of App review analytics tools, such as AppBot. | 4.5.1 | | | x | |
| Many users have casted doubts on the usefulness of the apps, i.e., they do not provide most of the "right" and much-needed features that many users are looking for. Thus, using "crowd-based" requirements engineering [54] techniques for these apps are critically needed. | 4.5.2 | x | x | | |
| It would be interesting to examine the differences among the apps and also their two OS versions, at a technical level, e.g, their code-base, software architectue. | 4.6.1 | | | | x |
| The sentiment analysis of apps can provide more complex granular output compared to just the "star rating", but there seems to be an inherent negative bias especially on Android which should be further investigated in future studies to better understand the phenomenon. A possible future Research Question (RQ) would be: Why is there an inherent negative bias in Android versions of an app, compared to the iOS version? | 4.6.2 | x | x | | x |
| The semantic-overlap measures, between the two OS version of the apps ranged between 45% to 86%. Possible root causes for low or high similarity should be studied in future works. | 4.6.3 | | | | x |
| There is a moderate correlation between the number of downloads normalized by the population size and the *Trust In Public Institutions index* (TIPI). This seems to denote that the more trust a country's population, as a whole, have on their government, the higher the ratio of app downloads, and expectedly the higher the use. Behavioral scientists can possibly investigate this issue in more detail. | 4.7.1 | | x | | |
| There could be unexpected inter-dependencies among the apps and various aspects of the mobile OS that they are running on. Updates to the OS could adversely impact a given app and could easily cause major dissatisfaction by the app users. Thus, the development team should work with OS vendors (in the above example case, Google being behind the Android OS) to prevent such chaotic situations. | 4.7.2 | x | | | |

### 5.2 LIMITATIONS AND POTENTIAL THREATS TO VALIDITY

In this section, we discuss limitations and potential threats to the validity of our study and the steps we have taken to minimize or mitigate them. The threats are discussed in the context of the four types of threats to validity based on a standard checklist for validity threats presented in [62]: internal validity, construct validity, conclusion validity, and external validity.

<u>Internal validity:</u> Internal validity is a property of scientific studies that reflects the extent to which a causal conclusion based on a study and the extracted data is warranted [62]. A threat to internal validity in this study lies in the selection bias (i.e., selection of the nine apps under study). As discussed in Section 3.3, analyzing user reviews of "all" the 50+ world-wide apps would have been a major undertaking and thus, instead, we decided to sample a set of nine apps. Future studies could analyze other apps of other countries and compare the trends/findings.

<u>Construct validity:</u> Construct validity is concerned with the extent to which the objects of study truly represent theory behind the study [62]. In other words, the issue relates to whether we actually analyzed the issues that we had originally intended to assess (as per our RQs raised in Section 3.2). We defined the RQs clearly and, as discussed in Section 3.2, for data collection and measurement, we used an established approach: Goal-Question-Metric (GQM) [38]. Some research questions (RQ7 and RQ8) rely on clearly defined numbers, i.e., the number of app downloads and countries' population sizes. The other RQs rely on sentiment analysis that has been performed based on the well-established and mature tool AppBot. Furthermore, for instance in the case of the French and German app, the reviews were automatically translated, which could cause issues with respect to the construct validity. However, the author team includes researchers who speak French and German. These authors performed a review of the translated word-clouds and in case of issues with translations, which were in general rare, corrections of the translated reviews were made.

Conclusion validity: Conclusion validity of a study deals with whether correct conclusions are reached through rigorous and repeatable treatment [62]. We, as a team of three researchers together, analyzed nine apps. The conclusions for the different apps were drawn by different authors of the team and cross-checked by the other authors. Also, the analysis of the nine apps overall leads to the situation that the approach of drawing conclusions was refined and performed iteratively in several iterations, which provides an additional step of quality control.

External validity: External validity is concerned with the extent to which the results of this study can be generalized [62]. The study has clearly defined context, i.e., to analyze user reviews (feedbacks) of a subset of the COVID contact-tracing apps, both for Android and iOS. The study does not intend to generalize to other contact-tracing apps. However, we have only analyzed the data for apps from nine countries, i.e., from England and Wales, Republic of Ireland, Scotland, Northern Ireland, Germany, Switzerland, France, Austria, and Finland. In order to make the assessments more comparable, we limited the sampling to European countries by selecting the four apps developed in the British Isles and five apps from mainland Europe. Given that there are country-specific differences in mobile app user behavior [34], this is a threat to generalizability over arbitrary countries (about 50 countries and regions have, so far, developed COVID contact-tracing apps). However, we think the results provide interesting insights across different countries in Europe, and our research approach can in the future be applied to further countries.

## 6 Conclusions and future work

The initial exploratory analysis of COVID contact-tracing app reviews reported in this paper is only a starting point. As the COVID pandemic has paralyzed most of the life and businesses around the globe, contact-tracing apps, if managed well, may have the potential to help bring the COVID outbreak under control. It is vital for governments and health authorities to develop and offer effective apps that all citizens can use.

An important issue that we realized during our analysis is the need to compare the different features of the apps. It is fair to say that two main basic use cases for a contact-tracing app are: (1) scan for nearby phones which are located within 2 meters for at least 15 minutes and record their phone tokens (keys), (2) if the phone user notifies the app that s/he is COVID positive, the app should notify the recorded phones via the stored keys. Although a detailed analysis of different apps' features is outside the scope of our work, we realized that some apps have much more features beyond those two basic features, e.g., a recent version of the German app includes a feature[1] showing the number of tokens (phone IDs) the app has collected. At least for the Northern Ireland (NI) app that the first two authors are familiar with, there is no such a feature. The number of features of a given app and how well they work, of course, could impact the widespread usage and popularity of the app and also to help bring the COVID outbreak under control.

As we analyzed and reported in this paper, mining user reviews of contact-tracing apps seem like a useful analysis towards providing insights to various stakeholders, e.g., app developers, decision-makers, public health experts, researchers, and the public. Of course, such an analysis and software engineering aspects can only provide some pieces of the "big picture". Therefore, as discussed in Section 2, sharing the data and collaborations with other important disciplines, including public health, and behavioral science [4] shall be conducted. As discussed in Section 1, the first author has been a member of the Expert Advisory Committee for the StopCOVID NI app, in which he has discussed and will continue discussing the results of this study with public health and behavioral science experts.

Furthermore, more collaboration between various stakeholders of the apps (e.g., software engineering teams of the app, decision-makers, and public health experts) is needed. Further work is required in the areas of software engineering, requirements engineering, public health, and behavioral science to improve the public adoption, software quality, and public perception of these apps.

Based on our study, we can see several promising directions for future research in this area:

- First, it would be of interest to reverse engineer the list of features in each app and compare them with each other. Beyond the analysis performed with AppBot, we see the potential for thematic analysis (or qualitative coding) to group feature-request-reporting reviews into a list of suggested features by reviewers. The results can support requirements engineering of refined / better COVID contact-tracing apps or apps for other pandemics in the future.

[1] github.com/corona-warn-app/cwa-wishlist/issues/5

- As discussed in Section 4.6.2, our data and analysis showed Android app reviews were slightly more "positive" than iOS app reviews. We could not find any discussions or reported evidence in the academic or grey literature about this phenomenon and think it is worthwhile to investigate it.
- Future studies could analyze other apps of other countries and compare the trends/findings.

## REFERENCES

[1] M. Scudellari, "COVID-19 Digital contact tracing: Apple and Google work together as MIT tests validity," *IEEE Spectrum,* vol. 13, 2020.

[2] C. Martuscelli and M. Heikkilä, "Scientists cast doubt on effectiveness of coronavirus contact-tracing apps," https://www.politico.eu/article/scientists-cast-doubt-on-the-effectiveness-of-contact-tracing-apps/, Last accessed: Oct. 2020.

[3] I. Braithwaite, T. Callender, M. Bullock, and R. W. Aldridge, "Automated and partly automated contact tracing: a systematic review to inform the control of COVID-19," *The Lancet Digital Health,* 2020.

[4] The British Psychological Society, "Behavioural science and success of the proposed UK digital contact tracing application for Covid-19," *Technical report,* June 2020.

[5] N. Genc-Nayebi and A. Abran, "A systematic literature review: Opinion mining studies from mobile app store user reviews," *Journal of Systems and Software,* vol. 125, pp. 207-219, 2017.

[6] C. Iacob and R. Harrison, "Retrieving and analyzing mobile apps feature requests from online reviews," in *Working conference on mining software repositories*, 2013, pp. 41-44.

[7] P. Runeson and M. Höst, "Guidelines for conducting and reporting case study research in software engineering," *Empirical Software Engineering,* vol. 14, no. 2, pp. 131-164, 2009.

[8] R. Kazman and L. Pasquale, "Software engineering in society," *IEEE Software,* vol. 37, no. 1, pp. 7-9, 2019.

[9] J. Budd *et al.*, "Digital technologies in the public-health response to COVID-19," *Nature medicine,* pp. 1-10, 2020.

[10] N. Ahmed *et al.*, "A survey of covid-19 contact tracing apps," *IEEE Access,* vol. 8, pp. 134577-134601, 2020.

[11] C. Gomez, J. Oller, and J. Paradells, "Overview and evaluation of bluetooth low energy: An emerging low-power wireless technology," *Sensors,* vol. 12, no. 9, pp. 11734-11753, 2012.

[12] H. Cho, D. Ippolito, and Y. W. Yu, "Contact tracing mobile apps for COVID-19: Privacy considerations and related trade-offs," *arXiv preprint arXiv:2003.11511,* 2020.

[13] S. Trang, M. Trenz, W. H. Weiger, M. Tarafdar, and C. M. Cheung, "One app to trace them all? Examining app specifications for mass acceptance of contact-tracing apps," *European Journal of Information Systems,* pp. 1-14, 2020.

[14] E. Rizzo, "COVID-19 contact tracing apps: the 'elderly paradox'," *Public health,* 2020.

[15] Q. Zhao, H. Wen, Z. Lin, D. Xuan, and N. Shroff, "On the accuracy of measured proximity of bluetooth-based contact tracing apps," in *International Conference on Security and Privacy in Communication Networks*, 2020.

[16] R. Sun, W. Wang, M. Xue, G. Tyson, S. Camtepe, and D. Ranasinghe, "Vetting Security and Privacy of Global COVID-19 Contact Tracing Applications," *arXiv preprint arXiv:2006.10933,* 2020.

[17] J. Li and X. Guo, "COVID-19 Contact-tracing Apps: A Survey on the Global Deployment and Challenges," *arXiv preprint arXiv:2005.03599,* 2020.

[18] K. Rekanar *et al.*, "Sentiment Analysis of User Feedback on the HSE Contact Tracing App," *Pre-print,* 2020.

[19] D. J. Leith and S. Farrell, "Measurement-Based Evaluation Of Google/Apple Exposure Notification API For Proximity Detection In A Commuter Bus," *Technical report, Trinity College Dublin,* June 2020.

[20] R. Davidson, "'Searching for Mary, Glasgow': Contact Tracing for Sexually Transmitted Diseases in Twentieth-Century Scotland," *Social history of medicine,* vol. 9, no. 2, pp. 195-214, 1996.

[21] C. Browne, H. Gulbudak, and G. Webb, "Modeling contact tracing in outbreaks with application to Ebola," *Journal of theoretical biology,* vol. 384, pp. 33-49, 2015.

[22] J. A. Sacks *et al.*, "Introduction of mobile health tools to support Ebola surveillance and contact tracing in Guinea," *Global Health: Science and Practice,* vol. 3, no. 4, pp. 646-659, 2015.

[23] L. O. Danquah *et al.*, "Use of a mobile application for Ebola contact tracing and monitoring in northern Sierra Leone: a proof-of-concept study," *BMC infectious diseases,* vol. 19, no. 1, p. 810, 2019.

[24] C. Farronato, M. Iansiti, M. Bartosiak, S. Denicolai, L. Ferretti, and R. Fontana, "How to get people to actually use contact-tracing apps," *Harvard Business Review Digital Articles,* 2020.

[25] A. G. Blom *et al.*, "Barriers to the Large-Scale Adoption of the COVID-19 Contact-Tracing App in Germany."

[26] M. Walrave, C. Waeterloos, and K. Ponnet, "Adoption of a Contact Tracing App for Containing COVID-19: A Health Belief Model Approach," *JMIR Public Health and Surveillance,* vol. 6, no. 3, p. e20572, 2020.

[27] S. Nicholas, C. Armitage, T. Tampe, and K. Dienes, "Public attitudes towards COVID-19 contact tracing apps: a UK-based focus group study," 2020.

[28] S. Altmann *et al.*, "Acceptability of app-based contact tracing for COVID-19: Cross-country survey study," *JMIR mHealth and uHealth,* vol. 8, no. 8, p. e19857, 2020.

[29] E. M. Redmiles, "User Concerns & Tradeoffs in Technology-Facilitated Contact Tracing," *arXiv preprint arXiv:2004.13219,* 2020.

[30] L. Kukuk, "Analyzing adoption of contact tracing apps using UTAUT," University of Twente, 2020.

[31] I. Morales-Ramirez, A. Perini, and R. S. Guizzardi, "An ontology of online user feedback in software engineering," *Applied Ontology,* vol. 10, no. 3-4, pp. 297-330, 2015.

[32] H. Khalid, E. Shihab, M. Nagappan, and A. E. Hassan, "What do mobile app users complain about?," *IEEE software,* vol. 32, no. 3, pp. 70-77, 2014.

[33] H. Hu, S. Wang, C.-P. Bezemer, and A. E. Hassan, "Studying the consistency of star ratings and reviews of popular free hybrid Android and iOS apps," *Empirical Software Engineering,* vol. 24, no. 1, pp. 7-32, 2019.

[34] S. L. Lim, P. J. Bentley, N. Kanakam, F. Ishikawa, and S. Honiden, "Investigating country differences in mobile app user behavior and challenges for software engineering," *IEEE Transactions on Software Engineering,* vol. 41, no. 1, pp. 40-64, 2014.

[35] S. A. Scherr, F. Elberzhager, and S. Meyer, "Listen to Your Users–Quality Improvement of Mobile Apps Through Lightweight Feedback Analyses," in *International Conference on Software Quality*, 2019: Springer, pp. 45-56.

[36] E. Guzman and W. Maalej, "How do users like this feature? a fine grained sentiment analysis of app reviews," in *IEEE international requirements engineering conference*, 2014, pp. 153-162.

[37] S. R. Stoyanov, L. Hides, D. J. Kavanagh, O. Zelenko, D. Tjondronegoro, and M. Mani, "Mobile app rating scale: a new tool for assessing the quality of health mobile apps," *JMIR mHealth and uHealth,* vol. 3, no. 1, p. e27, 2015.

[38] V. R. Basili, "Software modeling and measurement: the Goal/Question/Metric paradigm," Technical Report, University of Maryland at College Park, 1992.

[39] W. Maalej and H. Nabil, "Bug report, feature request, or simply praise? on automatically classifying app reviews," in *IEEE international requirements engineering conference*, 2015: IEEE, pp. 116-125.

[40] D. M. Fernández, D. Graziotin, S. Wagner, and H. Seibold, "Open science in software engineering," *arXiv preprint arXiv:1904.06499,* 2019.

[41] B. Liu, "Sentiment analysis and opinion mining," *Synthesis lectures on human language technologies,* vol. 5, no. 1, pp. 1-167, 2012.

[42] K. Potter, H. Hagen, A. Kerren, and P. Dannenmann, "Methods for presenting statistical information: The box plot," *Visualization of large and unstructured data sets,* vol. 4, pp. 97-106, 2006.

[43] M. E. Joorabchi, A. Mesbah, and P. Kruchten, "Real challenges in mobile app development," in *ACM/IEEE International Symposium on Empirical Software Engineering and Measurement*, 2013, pp. 15-24.

[44] M. Husmann, M. Spiegel, A. Murolo, and M. C. Norrie, "UI testing cross-device applications," in *Proceedings of the 2016 ACM International Conference on Interactive Surfaces and Spaces*, 2016, pp. 179-188.

[45] M. Nebeling, M. Husmann, C. Zimmerli, G. Valente, and M. C. Norrie, "XDSession: integrated development and testing of cross-device applications," in *Proceedings of the 7th ACM SIGCHI Symposium on Engineering Interactive Computing Systems*, 2015, pp. 22-27.

[46] V. Guttal, S. Krishna, and R. Siddharthan, "Risk assessment via layered mobile contact tracing for epidemiological intervention," *medRxiv,* 2020.

[47] N. Jha and A. Mahmoud, "Mining non-functional requirements from App store reviews," *Empirical Software Engineering,* vol. 24, no. 6, pp. 3659-3695, 2019.

[48] N. Jha and A. Mahmoud, "Mining user requirements from application store reviews using frame semantics," in *International working conference on requirements engineering: Foundation for software quality*, 2017: Springer, pp. 273-287.

[49] G. Williams and A. Mahmoud, "Mining twitter feeds for software user requirements," in *2017 IEEE 25th International Requirements Engineering Conference (RE)*, 2017: IEEE, pp. 1-10.

[50] E. Guzman, M. Ibrahim, and M. Glinz, "A little bird told me: Mining tweets for requirements and software evolution," in *2017 IEEE 25th International Requirements Engineering Conference (RE)*, 2017: IEEE, pp. 11-20.

[51] M. Lu and P. Liang, "Automatic classification of non-functional requirements from augmented app user reviews," in *Proceedings of the 21st International Conference on Evaluation and Assessment in Software Engineering*, 2017, pp. 344-353.

[52] W. Maalej, M. Nayebi, and G. Ruhe, "Data-Driven Requirements Engineering-An Update," in *2019 IEEE/ACM 41st International Conference on Software Engineering: Software Engineering in Practice (ICSE-SEIP)*, 2019: IEEE, pp. 289-290.

[53] M. Nayebi, M. Marbouti, R. Quapp, F. Maurer, and G. Ruhe, "Crowdsourced exploration of mobile app features: A case study of the Fort McMurray wildfire," in *IEEE/ACM International Conference on Software Engineering: Software Engineering in Society Track*, 2017: IEEE, pp. 57-66.

[54] E. C. Groen, J. Doerr, and S. Adam, "Towards crowd-based requirements engineering a research preview," in *International Working Conference on Requirements Engineering: Foundation for Software Quality*, 2015: Springer, pp. 247-253.

[55] H. M. Wallach, "Topic modeling: beyond bag-of-words," in *Proceedings of the international conference on Machine learning*, 2006, pp. 977-984.

[56] F. D. S. Webber, "Semantic folding theory and its application in semantic fingerprinting," *arXiv preprint arXiv:1511.08855,* 2015.

[57] F. G. Hayden, *Policymaking for a good society: the social fabric matrix approach to policy analysis and program evaluation*. Springer Science & Business Media, 2006.

[58] A. Aslam, "Research ideas: Correlation does not imply causation," *British dental journal,* vol. 219, no. 2, pp. 49-49, 2015.

[59] C. Ksir and C. L. Hart, "Correlation still does not imply causation," *The Lancet Psychiatry,* vol. 3, no. 5, p. 401, 2016.

[60] J. Fenn and M. Raskino, *Mastering the hype cycle: how to choose the right innovation at the right time*. Harvard Business Press, 2008.

[61] V. Garousi, "Evidence-based Insights about Issue Management Processes: An Exploratory Study," in *Proceedings of the International Conference on Software Process (ICSP)*, 2009, pp. 112-123.

[62] C. Wohlin, P. Runeson, M. Höst, M. C. Ohlsson, B. Regnell, and A. Wesslén, *Experimentation in Software Engineering: An Introduction*. Kluwer Academic Publishers, 2000.